\documentclass{iopart}  

\usepackage{iopams}  
 \usepackage{graphicx}
 \usepackage{epstopdf}

 \newcommand{\lya}{Ly$\alpha$ }
 
\newcommand{\lyb}{Ly$\beta$ } 
\newcommand{\lybns}{Ly$\beta$} 
 \newcommand{\lyn}{Ly$n$ }
 \newcommand{\deriv}{{\rm d}}
  \newcommand{\ud}{{\rm d}}
    \newcommand{\ue}{{\rm e}}
  \newcommand{\eV}{{\rm eV}}
  \newcommand{\keV}{{\rm keV}}
  \newcommand{\iMpc}{{\rm Mpc^{-1}}}

    \newcommand{\K}{{\rm K}}   
\begin{document}

\title[21-cm Cosmology]{21-cm cosmology in the 21st Century}

\author{Jonathan R. Pritchard and Abraham Loeb} \address{Institute for
Theory \& Computation, Harvard University, 60 Garden St., Cambridge,
MA 02138, USA} \ead{jpritchard@cfa.harvard.edu; aloeb@cfa.harvard.edu}

\begin{abstract}


Imaging the Universe during the first hundreds of millions of years
remains one of the exciting challenges facing modern cosmology.
Observations of the redshifted 21 cm line of atomic hydrogen offer the
potential of opening a new window into this epoch.  This will
transform our understanding of the formation of the first stars and
galaxies and of the thermal history of the Universe.  A new generation
of radio telescopes is being constructed for this purpose with the
first results starting to trickle in.  In this review, we detail the
physics that governs the 21 cm signal and describe what might be
learnt from upcoming observations. We also generalize our discussion
to intensity mapping of other atomic and molecular lines.
\end{abstract}


\tableofcontents

\maketitle

\section{Introduction}

Our understanding of cosmology has matured significantly over the last
twenty years.  In that time, observations of the Universe from its
infancy, 400,000 years after the Big Bang, through to the present day,
some 13.7 billion years later, have given us a basic picture of how
the Universe came to be the way it is today.  Despite this progress
much of the first billion years of the Universe, a period when the
first stars and galaxies formed, is still an unobserved mystery.

Astronomers have an advantage over archaeologists in that the finite
speed of light gives them a way of looking into the past.  The further
away an object is located the longer the light that it emits takes to
reach an observer today.  The image recorded at a telescope is
therefore a picture of the object long ago when the light was first
emitted.  The construction of telescopes both on the Earth, such as
Keck, Subaru and VLT, and in space, such as the Hubble Space Telescope, has
enabled astronomers to directly observe galaxies out to distances
corresponding to a time when the Universe was a billion years old.

Added to this, observations at microwave frequencies reveal the
cooling afterglow of the big bang.  This cosmic microwave background
(CMB) decoupled from the cosmic gas 400,000 years after the Big Bang
when the Universe cooled sufficiently for protons and electrons to
combine to form neutral hydrogen.  Radiation from this time is able to
reach us directly, providing a snapshot of the primordial Universe.

Despite current progress, connecting these two periods represents a
considerable challenge.  Our understanding of structure is
based upon the observation of small perturbations in the temperature
maps of the CMB.  These indicate that the early Universe was
inhomogeneous at the level of 1 part in 100,000.  Over time the action
of gravity causes the growth of these small perturbations into larger
non-linear structures, which collapse to form sheets, filaments, and
halos.  These non-linear structures provide the framework within which
galaxies form via the collapse and cooling of gas until the density
required for star formation is reached.

The theoretical picture is well established, but the middle phase is
largely untested by observations.  To improve on this astronomers are
pursuing two main avenues of attack.  The first is to extend existing
techniques by building larger, more sensitive, telescopes at a variety
of wavelengths.  On the ground, there are plans for optical telescopes
with an aperture diameter of 24--39 meters - the {\em Giant Magellan
Telescope} (GMT), the {\em Thirty Meter Telescope} (TMT), and the {\em
European Extremely Large Telescope} (E-ELT) - that would be able to
detect an individual galaxy out to redshifts $z>10$.  In space, the
{\em James Webb Space Telescope} (JWST) will operate at infrared
wavelengths and potentially image some of the first galaxies at $z\sim
10$--$15$.  Other efforts involve the {\em Atacama Large
Millimeter/submillimeter Array} (ALMA), which will observe the
molecular gas that fuels star formation in galaxies during
reionization ($z=8-10$).  These efforts target individual galaxies
although the objects of interest are far enough away that only the
brightest sources may be seen.


This review focuses on an alternative approach based upon making
observations of the red-shifted 21 cm line of neutral hydrogen.  This
21 cm line is produced by the hyperfine splitting caused by the
interaction between electron and proton magnetic moments.  Hydrogen is
ubiquitous in the Universe, amounting to $\sim$75\% of the gas mass
present in the intergalactic medium (IGM).  As such, it provides a
convenient tracer of the properties of that gas and of major
milestones in the first billion years of the Universe's
history.

The 21 cm line from gas during the first billion years after the Big
Bang redshifts to radio frequencies 30-200 MHz making it a prime
target for a new generation of radio interferometers currently being
built.  These instruments, such as {\em Murchison Widefield Array}
(MWA), the {\em LOw Frequency ARray} (LOFAR), the Precision Array to
Probe the Epoch of Reionization (PAPER), the {\em 21 cm Array} (21CMA), and the Giant Meter-wave
Radio Telescope (GMRT), seek to detect the radio fluctuations in the redshifted 21 cm background arising from variations in the amount of neutral hydrogen.  Next generation instruments (e.g. SKA) will be able to go further and might make detailed maps of the ionized regions during reionization and measure properties of hydrogen out to $z=30$.  These observations constrain the properties of the intergalactic medium and by extension the cumulative impact of light from all galaxies, not just the brightest ones.  In combination with direct observations of the sources they provide a powerful tool for learning about the first stars and galaxies.  They will also provide information about active galactic nuclei (AGN), such as quasars, by observing the ionized bubbles surrounding individual AGN.


In addition to learning about galaxies and reionization, 21 cm
observations have the potential to inform us about fundamental physics too.  Part of the signal traces the density field giving information about neutrino masses and the initial conditions from the early epoch of cosmic inflation in the form of the power spectrum.  However spin temperature fluctuations driven by astrophysics also contribute to the signal.  Getting at this cosmology is a challenge, since the astrophysical effects must be understood before cosmology can be disentangled.  One possibility is to exploit the effect of redshift space distortions, which also produce 21 cm fluctuations but directly trace the density field.  In the long term, 21 cm cosmology may allow precision measurements of cosmological parameters by opening up large volumes of the Universe to observation.

\begin{figure}[htbp]
\begin{center}
\includegraphics[scale=0.2]{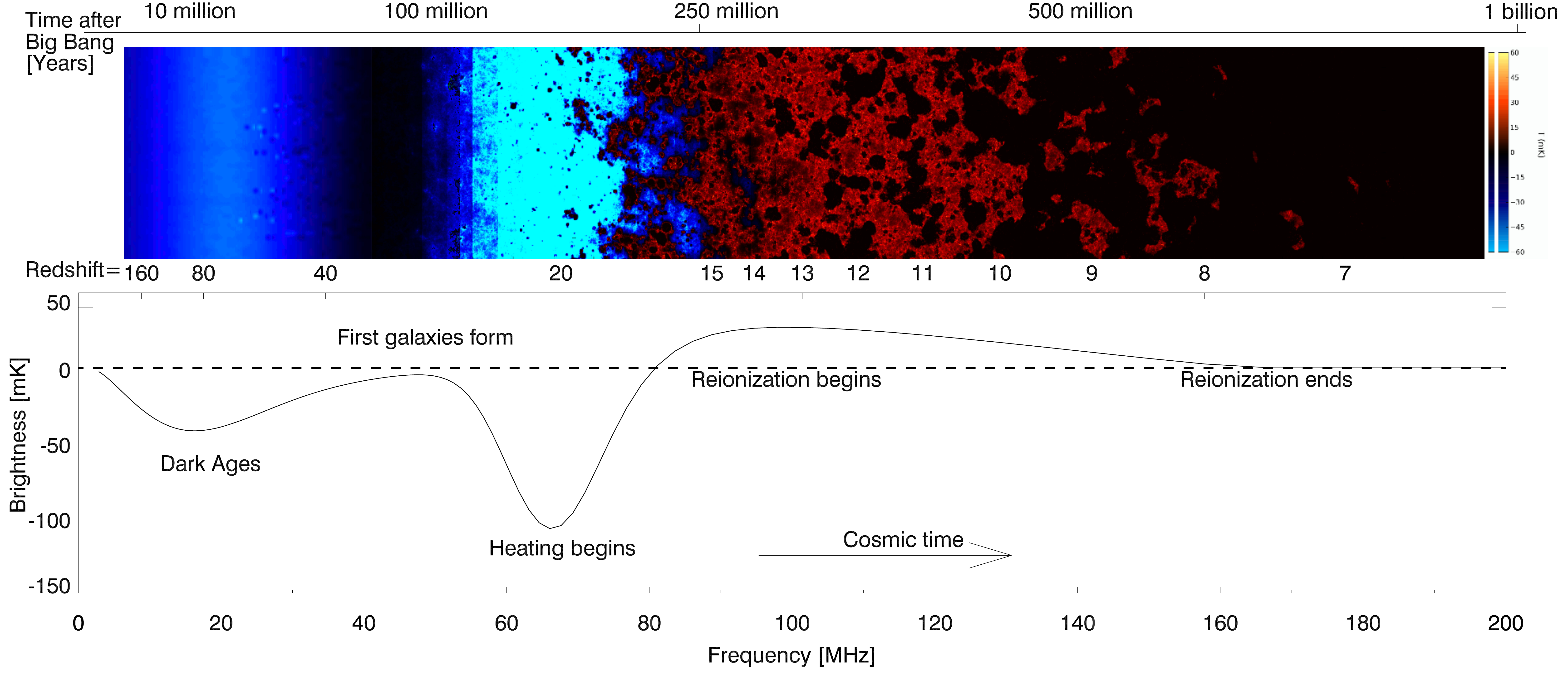}
\caption{The 21-centimeter cosmic hydrogen signal. (a) Time evolution of fluctuations in the 21-cm brightness from just before the first stars formed through to the end of the reionization epoch. This evolution is pieced together from redshift slices through a simulated cosmic volume \cite{santos2008}. Coloration indicates the strength of the 21-cm
brightness as it evolves through two absorption phases (purple and blue), separated by a period (black) where the excitation temperature of the 21-cm hydrogen transition decouples from the temperature of the hydrogen gas, before it transitions to emission (red) and finally disappears (black) owing to the ionization of the hydrogen gas. (b)
Expected evolution of the sky-averaged 21-cm brightness from the ``dark ages'' at redshift 200 to the end of reionization, sometime before redshift 6 (solid curve indicates the signal; dashed curve indicates $T_b=0$). The frequency structure within this redshift range is driven by several physical processes, including the formation of the first galaxies and the heating and ionization of the hydrogen gas. There is considerable uncertainty in the exact form of this signal, arising from the unknown properties of the first galaxies. }
\label{fig:NVplot}
\end{center}
\end{figure}

The goal of this review is to summarise the physics that determines
the 21 cm signal, along with a comprehensive overview of related
astrophysics.  Figure \ref{fig:NVplot} provides a summary of the 21 cm
signal showing the key features of the signal with the relevant cosmic
time, frequency, and redshift scales indicated.  The earliest period
of the signal arises in the period after thermal decoupling of the
ordinary matter (baryons) from the CMB, so that the gas is able to
cool adiabatically with the expansion of the Universe.  In these
cosmic ``Dark Ages", before the first stars have formed, the first
structures begin to grow from the seed inhomogeneties thought to be produced by
quantum fluctuations during inflation.  The cold gas can be seen in a
21 cm absorption signal, which has both a mean value (shown in the
bottom panel) and fluctuations arising from variation in density
(shown in the top panel).  Once the first stars and galaxies form,
their light radically alters the properties of the gas.  Scattering of
Ly$\alpha$ photons leads to a strong coupling between the excitation
of the 21 cm line spin states and the gas temperature.  Initially,
this leads to a strong absorption signal that is spatially varying due
to the strong clustering of the rare first generation of galaxies.
Next, the X-ray emission from these galaxies heats the gas leading to
a 21 cm emission signal.  Finally, ultraviolet photons ionize the gas
producing dark holes in the 21 cm signal within regions of ionized
bubbles surrounding groups of galaxies.  Eventually all of the
hydrogen gas, except for that in a few dense pockets, is ionized.

Throughout this review, we will make reference to parameters describing
the standard $\Lambda$CDM cosmology.  These describe the mass
densities in non-relativistic matter $\Omega_m=0.26$, dark energy
$\Omega_\Lambda=0.74$, and baryons $\Omega_b=0.044$ as a fraction of
the critical mass density.  We further parametrise the Hubble
parameter $H_0=100h\,\rm{km\,s^{-1}\,Mpc^{-1}}$ with $h=0.74$.
Finally, the spectrum of fluctuations is described by a logarithmic
slope or ``tilt" $n_S=0.95$, and the variance of matter fluctuations
today smoothed on a scale of 8$h^{-1}$ Mpc is $\sigma_8=0.8$.  The
values quoted are indicative of those found by the latest measurements
\cite{komatsu2011}.

The layout of this review is as follows.  We first discuss the basic atomic physics of the 21 cm line in \S\ref{sec:physics}.  In \S \ref{sec:global}, we turn to the evolution of the sky averaged 21 cm signal and the feasibility of observing it.  In \S
\ref{sec:fluctuations} we describe three-dimensional 21 cm
fluctuations, including predictions from analytical and numerical calculations.  After reionization, most of the 21 cm signal originates from cold gas in galaxies (which is self-shielded from the background of ionizing radiation). In \S \ref{sec:intensity} we describe the prospects for intensity mapping of this signal as well as using the same technique to map the cumulative emission of other atomic and molecular lines from galaxies without resolving the galaxies individually.  The 21 cm forest that is expected against radio bright sources is described in \S \ref{sec:forest}. Finally, we conclude with an outlook for the future in \S \ref{sec:conclusions}.

We direct interested readers to a number of other worthy reviews on the subject.  Ref. \cite{fob} provides a comprehensive overview of the entire field, and Ref.  \cite{morales2010} takes a more observationally orientated approach focussing on the near term observations of reionization.

\section{Physics of the 21 cm line of atomic hydrogen}
\label{sec:physics}

\subsection{Basic 21 cm physics}

As the most common atomic species present in the Universe, hydrogen is a useful
tracer of local properties of the gas.  The simplicity of its
structure - a proton and electron - belies the richness of the
associated physics.  In this review, we will be focusing on the 21 cm
line of hydrogen, which arises from the hyperfine splitting of the
$1S$ ground state due to the interaction of the magnetic moments of
the proton and the electron.  This splitting leads to two distinct
energy levels separated by $\Delta E=5.9\times10^{-6}\eV$,
corresponding to a wavelength of 21.1 cm and a frequency of 1420 MHz.
This frequency is one of the most precisely known quantities in
astrophysics having been measured to great accuracy from studies of
hydrogen masers \cite{goldenberg1960}.

The 21 cm line was theoretically predicted by van de Hulst in 1942
\cite{vandehulst1945} and has been used as a probe of astrophysics
since it was first detected by Ewen \& Purcell in 1951
\cite{ewen1951}.  Radio telescopes look for emission by warm hydrogen gas within galaxies.  Since the line is narrow with a well measured rest frame frequency it can be used in the local Universe as a probe of the velocity distribution of gas within our galaxy and other nearby galaxies.  21 cm rotation curves are often used to trace galactic dynamics.  Traditional techniques for observing 21 cm emission have only detected the line in relatively local galaxies, although the 21 cm line has been seen in absorption against radio loud background sources from individual systems at redshifts $z\lesssim3$  \cite{kanekar2007,srianand2010}.  A new generation of radio telescopes offers the exciting prospect of using the 21 cm line as a probe of cosmology. 

In passing, we note that other atomic species show hyperfine
transitions that may be useful in probing cosmology.  Of particular
interest are the 8.7 GHz hyperfine transition of $^3$He$^+$
\cite{bagla2009,mcquinn2009}, which could provide a probe of Helium
reionization, and the 92 cm deuterium analogue of the 21 cm line
\cite{sigurdson2006}.  The much lower abundance of deuterium and $^3$He compared to neutral hydrogen makes it more difficult to take advantage of these transitions.

In cosmological contexts the 21 cm line has been used as a probe of
gas along the line of sight to some background radio source.  The
detailed signal depends upon the radiative transfer through gas along
the line of sight.  We recall the basic equation of radiative transfer
for the specific intensity $I_\nu$ (per unit frequency $\nu$) in the absence of scattering along a
path described by coordinate $s$ \cite{rybicki1986}
\begin{equation}\label{RTequation}
\frac{\ud I_\nu}{\ud s}= -\alpha_\nu I_\nu+j_\nu
\end{equation}
where absorption and emission by gas along the path are described by
the coefficients $\alpha_\nu$ and $j_\nu$, respectively.

To simplify the discussion, we will work in the Rayleigh-Jeans limit,
appropriate here since the relevant photon frequencies $\nu$ are much
smaller than the peak frequency of the CMB blackbody.  This allows us
to relate the intensity $I_\nu$ to a brightness temperature $T$ by the
relation $I_\nu=2k_B T\nu^2/c^2$, where $c$ is the speed of light and
$k_B$ is Boltzmann's constant.  We will also make use of the standard
definition of the optical depth $\tau=\int\ud s\,\alpha_\nu(s)$.  With
this we may rewrite \eref{RTequation} to give the radiative transfer
for light from a background radio source of brightness temperature
$T_R$ along the line of sight through a cloud of optical depth
$\tau_\nu$ and uniform excitation temperature $T_{ex}$ so that the
observed temperature $T_b^{\rm obs}$ at a frequency $\nu$ is given by
\begin{equation}
T_b^{\rm obs}=T_{ex}(1-\ue^{-\tau_\nu})+T_R(\nu)\ue^{-\tau_\nu}.
\end{equation}

The excitation temperature of the 21 cm line is known as the spin
temperature $T_S$.  It is defined through the ratio between the number
densities $n_i$ of hydrogen atoms in the two hyperfine levels (which
we label with a subscript 0 and 1 for the $1S$ singlet and $1S$
triplet levels, respectively)
\begin{equation}
n_1/n_0=(g_1/g_0)\exp(-T_\star/T_S),
\end{equation}
 where $(g_1/g_0)=3$ is the ratio of the statistical degeneracy
factors of the two levels, and $T_\star\equiv hc/k\lambda_{21
\rm{cm}}=0.068\,\rm{K}$.

With this definition, the optical depth of a cloud of hydrogen is then
\begin{equation}
\tau_\nu=\int\ud s\,[1-\exp(-E_{10}/k_B T_S)]\sigma_0\phi(\nu)n_0,
\end{equation}
where $n_0=n_H/4$ with $n_H$ being the hydrogen density, and we have
denoted the 21 cm cross-section as $\sigma(\nu)=\sigma_0\phi(\nu)$,
with $\sigma_0\equiv3c^2A_{10}/8\pi\nu^2$, where
$A_{10}=2.85\times10^{-15}\,\rm{s}^{-1}$ is the spontaneous decay rate
of the spin-flip transition, and the line profile is normalised so
that $\int\phi(\nu)\ud\nu=1$.  To evaluate this expression we need to
find the column length as a function of frequency $s(\nu)$ to
determine the range of frequencies $\ud\nu$ over the path $\ud s$ that
correspond to a fixed observed frequency $\nu_{\rm obs}$.  This can be
done in one of two ways: by relating the path length to the
cosmological expansion $\ud s=-c\,\ud z/(1+z)H(z)$ and the redshifting
of light to relate the observed and emitted frequencies $\nu_{\rm
obs}=\nu_{\rm em}/(1+z)$ or assuming a linear velocity profile locally
$v=(\ud v/\ud s)s$ (the well known Sobolev approximation
\cite{sobolev1957}) and using the Doppler law $\nu_{\rm obs}=\nu_{\rm
em}(1-v/c)$ self-consistently to ${\cal{O}}(v/c)$.  Since the latter
case describes the well known Hubble law in the absence of peculiar
velocities these two approaches give identical results for the optical
depth.  The latter picture brings out the effect of peculiar
velocities that modify the local velocity-frequency conversion.

The optical depth of this transition is small at all relevant redshifts,
yielding a differential brightness temperature
\begin{eqnarray}
\delta T_b&=&\frac{T_S-T_R}{1+z}(1-\ue^{-\tau_\nu})\\
&\approx&\frac{T_S-T_R}{1+z}\tau\\
&\approx&27  x_{\rm{HI}}\left(1+\delta_b\right)\left(\frac{\Omega_bh^2}{0.023}\right)\left(\frac{0.15}{\Omega_mh^2}\frac{1+z}{10}\right)^{1/2}\nonumber\\
&&\times\left(\frac{T_S-T_R}{T_S}\right)\left[\frac{\partial_r v_r}{(1+z)H(z)}\right]\,\rm{mK}\label{brightnessT},
\end{eqnarray}
Here $x_{\rm{HI}}$ is the neutral fraction of hydrogen, $\delta_b$ is
the fractional overdensity in baryons, and the final term arises from
the velocity gradient along the line of sight $\partial_r v_r$.

The key to the detectability of the 21 cm signal hinges on the spin
temperature.  Only if this temperature deviates from the background
temperature, will a signal be observable.  Much of this review will
focus on the physics that determines the spin temperature and how
spatial variation in the spin temperature conveys information about
astrophysical sources.

Three processes determine the spin temperature: {\it (i)}
absorption/emission of 21 cm photons from/to the radio background,
primarily the CMB; {\it (ii)} collisions with other hydrogen atoms and
with electrons; and {\it (iii)} resonant scattering of Ly$\alpha$
photons that cause a spin flip via an intermediate excited state.  The
rate of these processes is fast compared to the de-excitation time of
the line, so that to a very good approximation the spin temperature is
given by the equilibrium balance of these effects.  In this limit, the
spin temperature is given by \cite{field1958}
\begin{equation}
T_S^{-1}=\frac{T_\gamma^{-1}+x_\alpha T_\alpha^{-1}+x_c T_K^{-1}}{1+x_\alpha+x_c},
\end{equation}
where $T_\gamma$ is the temperature of the surrounding bath of radio
photons, typically set by the CMB so that $T_\gamma=T_{\rm CMB}$;
$T_\alpha$ is the color temperature of the \lya radiation field at the
\lya frequency and is closely coupled to the gas kinetic temperature
$T_K$ by recoil during repeated scattering, and $x_c, x_\alpha$ are
the coupling coefficients due to atomic collisions and scattering of
Ly$\alpha$ photons, respectively.  The spin temperature becomes
strongly coupled to the gas temperature when $x_{\rm{tot}}\equiv
x_c+x_\alpha\gtrsim1$ and relaxes to $T_\gamma$ when $x_{\rm
tot}\ll1$.

Two types of background radio sources are important for the 21 cm line as a probe of astrophysics.  Firstly, we may use the CMB as a radio background source.  In this case, $T_R=T_{\rm CMB}$ and the 21 cm feature is seen as a spectral distortion to the CMB blackbody at appropriate radio frequencies (since fluctuations in the CMB temperature are small $\delta T_{\rm CMB}\sim10^{-5}$ the CMB is effectively a source of uniform brightness).  The distortion forms a diffuse background that can be studied across the whole sky in a similar way to CMB anisotropies.  Observations at different frequencies probe different spherical shells of the observable Universe, so that a 3D map can be constructed.  This is the main subject of \S\ref{sec:fluctuations}.

The second situation uses a radio loud point source, for example a radio loud quasar, as the background.  In this case, the source will always be much brighter than the weak emission from diffuse hydrogen gas, $T_R\gg T_S$, so that the gas is seen in absorption against the source.  The appearance of lines from regions of neutral gas at different distances to the source leads to a ``forest" of lines known
as the ``21 cm forest" in analogy to the \lya forest.  The high brightness of the background source allows the 21 cm forest to be studied with high frequency resolution so probing small scale structures ($\sim kpc$) in the IGM.  For useful
statistics, many lines of sight to different radio sources are
required, making the discovery of high redshift radio sources a
priority.  We leave discussion of the 21 cm forest to
\S\ref{sec:forest}.

Note that we have a number of different quantities with units of
temperature, many of which are not true thermodynamic temperatures.
$T_R$ and $\delta T_b$ are measures of a radio intensity.  $T_S$
measures the relative occupation numbers of the two hyperfine levels.
$T_\alpha$ is a colour temperature describing the photon distribution
in the vicinity of the \lya transition. Only the CMB blackbody
temperature $T_{\rm CMB}$ and $T_K$ are genuine thermodynamic
temperatures.

\subsection{Collisional coupling}

Collisions between different particles may induce spin-flips in a
hydrogen atom and dominate the coupling in the early Universe where
the gas density is high.  Three main channels are available:
collisions between two hydrogen atoms and collisions between a
hydrogen atom and an electron or a proton.  The collisional coupling
for a species $i$ is \cite{field1958,fob}
\begin{equation}
x_c^i\equiv\frac{C_{10}}{A_{10}}\frac{T_\star}{T_\gamma}=\frac{n_i\kappa_{10}^i}{A_{10}}\frac{T_\star}{T_\gamma},
\end{equation}
where $C_{10}$ is the collisional excitation rate, $\kappa_{10}^i$ is
the specific rate coefficient for spin deexcitation by collisions with
species $i$ (in units of cm$^3$ s$^{-1}$).

The total collisional coupling coefficient can be written as
\begin{eqnarray}
x_c&=&x_c^{HH}+x_c^{eH}+x_c^{pH}\nonumber\\&=&\frac{T_\star}{A_{10}T_\gamma}\left[\kappa^{HH}_{1-0}(T_k)n_H +\kappa^{eH}_{1-0}(T_k)n_e+\kappa^{pH}_{1-0}(T_k)n_p\right],
\end{eqnarray}
where $\kappa^{HH}_{1-0}$ is the scattering rate between hydrogen
atoms, $\kappa^{eH}_{1-0}$ is the scattering rate between electrons
and hydrogen atoms, and $\kappa^{pH}_{1-0}$ is the scattering rate
between protons and hydrogen atoms.

The collisional rates require a quantum mechanical calculation.
Values for $\kappa^{HH}_{1-0}$ have been tabulated as a function of
$T_k$ \cite{allison1969,zygelman2005}, the scattering rate between
electrons and hydrogen atoms $\kappa^{eH}_{1-0}$ was considered in
Ref. \cite{furlanettobros}, and the scattering rate between protons
and hydrogen atoms $\kappa^{pH}_{1-0}$ was considered in
Ref. \cite{furlanettobros_pH}.  Useful fitting functions exist for
these scattering rates: the HH scattering rate is well fit in the
range $10\K<T_K<10^3\K$ by
$\kappa^{HH}_{1-0}(T_K)\approx3.1\times10^{-11}T_K^{0.357}\exp(-32/T_K){\rm\,cm^3\,s^{-1}}$
\cite{kuhlen2006}; and the e-H scattering rate is well fit by
$\log(\kappa^{eH}_{1-0}/{\rm cm^3\,s^{-1}})=-9.607+0.5\log
T_K\times\exp[-(\log T_K)^{4.5}/1800]$ for $T\le10^4\K$ and
$\kappa^{eH}_{1-0}(T_K>10^4\K)=\kappa^{eH}_{1-0}(10^4\K)$
\cite{liszt2001}.

During the cosmic dark ages, where the coupling is dominated by
collisional coupling the details of the process become important.  For
example, the above calculations make use of the assumption that the
collisional cross-sections are independent of velocity; the actual
velocity dependance leads to a non-thermal distribution for the
hyperfine occupation \cite{hirata2006col}. This effect can lead to a
suppression of the 21 cm signal at the level of 5\%, which although
small is still important from the perspective of using the 21 cm
signal from the dark ages for precision cosmology.

\subsection{Wouthuysen-Field effect}

\begin{figure}[htbp]
\begin{center}
\includegraphics[scale=0.3]{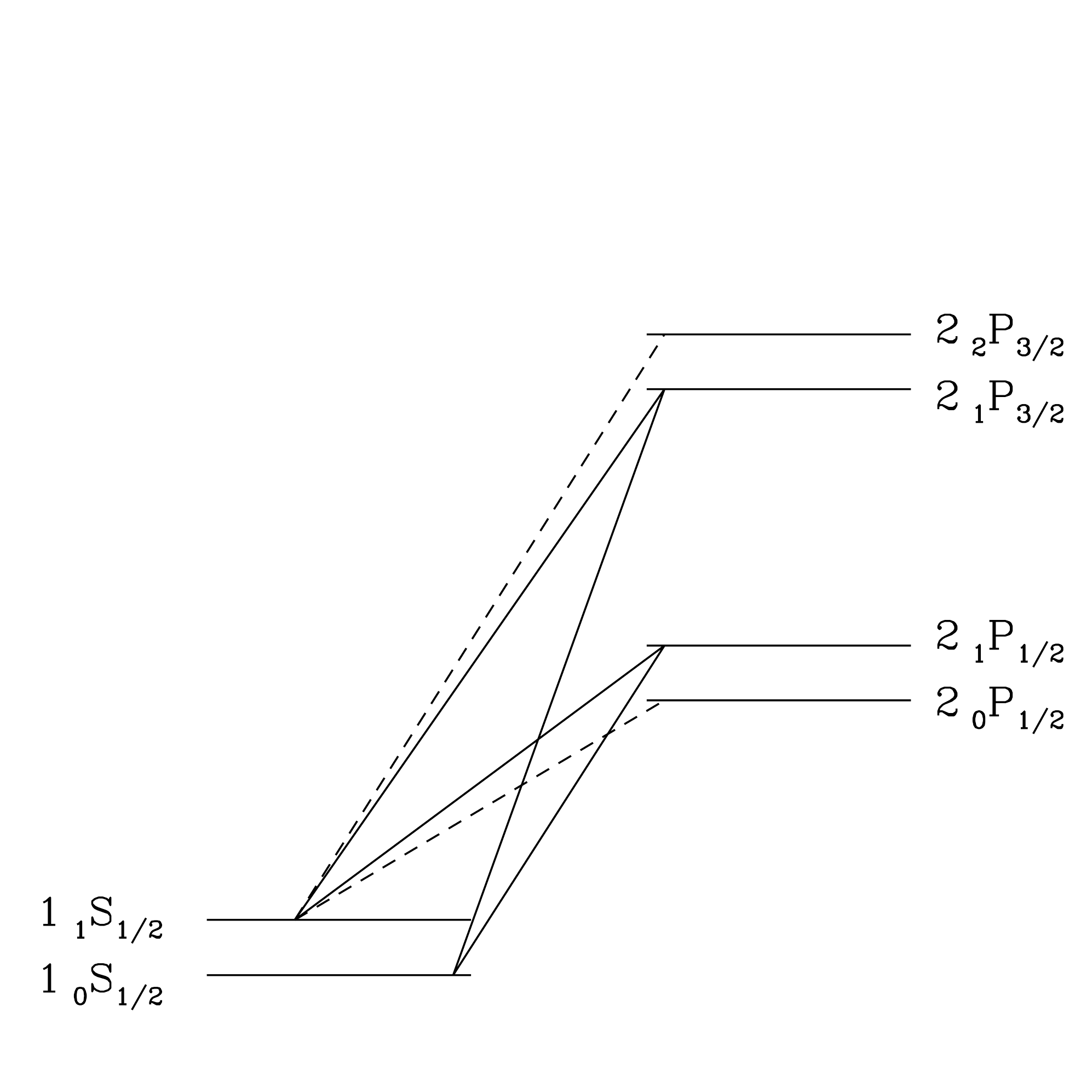}
\includegraphics[scale=0.25]{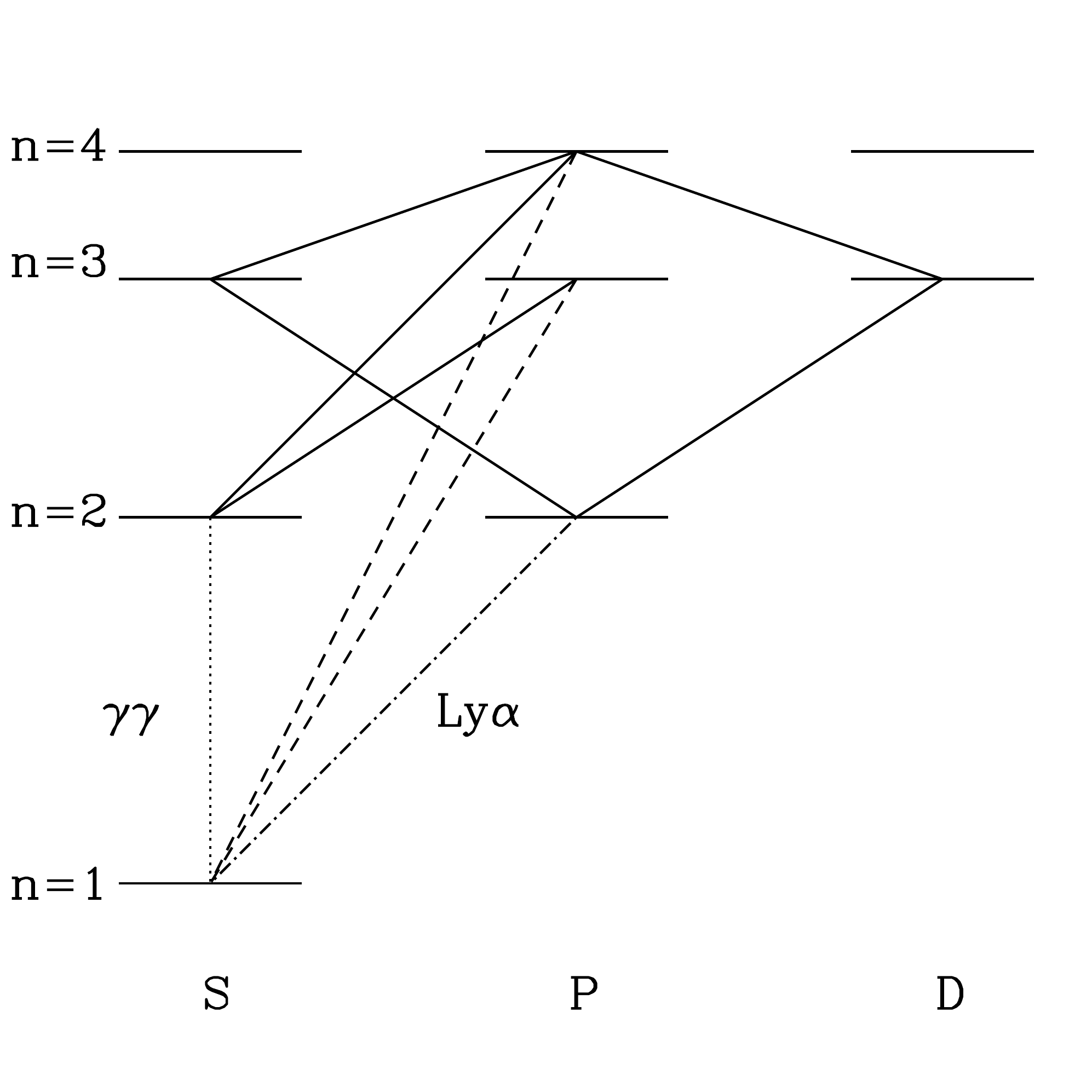}
\caption{{\em Left panel: }Hyperfine structure of the hydrogen atom
and the transitions relevant for the Wouthuysen-Field effect
\cite{pritchard2006}.  Solid line transitions allow spin flips, while
dashed transitions are allowed but do not contribute to spin
flips.{\em Right panel: }Illustration of how atomic cascades convert
\lyn photons into \lya photons.  }
\label{fig:wf_diagram}
\end{center}
\end{figure}

For most of the redshifts that are likely to be observationally probed
in the near future collisional coupling of the 21 cm line is
inefficient.  However, once star formation begins, resonant scattering
of Ly$\alpha$ photons provides a second channel for coupling.  This
process is generally known as the Wouthuysen-Field effect
\cite{wouth1952,field1958} and is illustrated in Figure
\ref{fig:wf_diagram}, which shows the hyperfine structure of the
hydrogen 1S and 2P levels.  Suppose that hydrogen is initially in the
hyperfine singlet state.  Absorption of a \lya photon will excite the
atom into either of the central 2P hyperfine states (the dipole
selection rules $\Delta F=0,1$ and no $F=0\rightarrow0$ transitions
make the other two hyperfine levels inaccessible).  From here emission
of a \lya photon can relax the atom to either of the two ground state
hyperfine levels.  If relaxation takes the atom to the ground level
triplet state then a spin-flip has occurred.  Hence, resonant
scattering of \lya photons can produce a spin-flip.

The physics of the Wouthuysen-Field effect is considerably more subtle
than this simple description would suggest.  We may write the coupling
as
\begin{equation}
x_\alpha=\frac{4P_\alpha}{27A_{10}}\frac{T_\star}{T_\gamma},
\end{equation}
where $P_\alpha$ is the scattering rate of \lya photons.  Here we have
related the scattering rate between the two hyperfine levels to
$P_\alpha$ using the relation $P_{01}=4P_\alpha/27$, which results
from the atomic physics of the hyperfine lines and assumes that the
radiation field is constant across them \cite{meiksin2000}.

The rate at which \lya photons scatter from a hydrogen atom is given by
\begin{equation}
P_\alpha=4\pi\chi_\alpha\int\ud\nu\,J_\nu(\nu)\phi_\alpha(\nu),
\end{equation}
where $\sigma_\nu\equiv\chi_\alpha\phi_\alpha(\nu)$ is the local
absorption cross section, $\chi_\alpha\equiv(\pi e^2/m_e c)f_\alpha$
is the oscillation strength of the \lya transition, $\phi_\alpha(\nu)$
is the \lya absorption profile, and $J_\nu(\nu)$ is the angle-averaged
specific intensity of the background radiation field (by number).

Making use of this expression, we can express the coupling as
\begin{equation}\label{xalpha}
x_\alpha=\frac{16\pi^2T_\star e^2 f_\alpha}{27A_{10}T_\gamma m_e c}S_\alpha J_\alpha,
\end{equation}
where $J_\alpha$ is the specific flux evaluated at the \lya frequency.
Here we have introduced $S_\alpha\equiv\int\ud
x\phi_\alpha(x)J_\nu(x)/J_\infty$, with $J_\infty$ being the flux away
from the absorption feature, as a correction factor of order unity to
describe the detailed structure of the photon distribution in the
neighborhood of the \lya resonance.

Equation \eref{xalpha} can be used to calculate the critical flux
required to produce $x_\alpha=S_\alpha$.  We rewrite \eref{xalpha} as
$x_\alpha=S_\alpha J_\alpha/J_\alpha^C$ where
$J_\alpha^C\equiv1.165\times10^{10}[(1+z)/20]{\rm
\,cm^{-2}\,s^{-1}\,Hz^{-1}\,sr^{-1}}$.  The critical flux can also be
expressed in terms of the number of \lya photons per hydrogen atom
$J_\alpha^C/n_H=0.0767[(1+z)/20]^{-2}$.  In practice, this condition
is easy to satisfy once star formation begins.

The above physics couples the spin temperature to the colour
temperature of the radiation field, which is a measure of the shape of
the radiation field as a function of frequency in the neighbourhood of
the \lya line defined by \cite{rybicki2006}
\begin{equation}
\frac{h}{k_B T_c}=-\frac{\ud\log n_\nu}{\ud \nu},
\end{equation}
where $n_\nu=c^2 J_\nu/2\nu^2$ is the photon occupation number.  Some
care must to taken with this definition; other definitions that do
not obey detailed balance can be found in the literature.

Typically, $T_C\approx T_K$, because in most cases of interest the
optical depth to \lya scattering is very large leading to a large
number of scatterings of \lya photons that bring the radiation field
and the gas into local equilibrium for frequencies near the line
center \cite{field1959relax}.  At the level of microphysics this
relation occurs through the process of scattering \lya photons in the
neighbourhood of the \lya resonance, which leads to a distinct feature
in the frequency distribution of photons.  Without going into the
details, one can understand the formation of this feature in terms of
the ``flow" of photons in frequency.  Redshifting with the cosmic
expansion leads to a flow of photons from high to low frequency at a
fixed rate.  As photons flow into the \lya resonance they may scatter
to larger or smaller frequencies.  Since the cross-section is
symmetric, one would expect the net flow rate to be preserved.
However, each time a \lya photon scatters from a hydrogen atom it will
lose a fraction of its energy $h\nu/m_pc^2$ due to the recoil of the
atom.  This loss of energy increases the flow to lower energy and
leads to a deficit of photons close to line center.  As this feature
develops scattering redistributes photons leading to an asymmetry
about the line.  This asymmetry is exactly that required to bring the
distribution into local thermal equilibrium with $T_C\approx T_K$.

The shape of this feature determines $S_\alpha$ and, since recoils
source an absorption feature, ensures $S_\alpha\le1$.  At low
temperatures, recoils have more of an effect and the suppression of
the Wouthuysen-Field effect is most pronounced.  If the IGM is warm
then this suppression becomes negligible
\cite{chen2004,hirata2006lya,chuzhoy2006heat,furlanetto2006heat}.  The
above discussion has neglected processes whereby the distribution of
photons is changed by spin-exchanges.  Including this complicates the
determination of $T_S$ and $T_C$ considerably since they must then be
iterated to find a self-consistent solution for the level- and
photon-populations \cite{hirata2006lya}.  However, the effect of
spin-flips on the photon distribution is small $\lesssim10\%$.

A useful approximation for $S_\alpha$ is outlined in
Ref. \cite{chuzhoy2006heat}: $S_\alpha\approx\exp(-1.79\alpha)$, where
$\alpha\equiv\eta(3a/2\pi\gamma)^{1/3}$, $a=\Gamma/(4\pi\Delta\nu_D)$,
$\Gamma$ the inverse lifetime of the upper 21 cm level,
$\Delta\nu_D/\nu_0=(2k_BT_K/mc^2)^{1/2}$ is the Doppler parameter,
$\nu_0$ the line center frequency, $\gamma=\tau_{\rm GP}^{-1}$, and
$\eta=(h\nu_0^2)/(mc^2\Delta\nu_D)$ is the mean frequency drift per
scattering due to recoil.  which is accurate at the 5\% level provided
that $T_K\gtrsim1{\rm\,K}$ and the Gunn-Peterson optical depth
$\tau_{\rm GP}$ is large.

In the astrophysical context, we will primarily be interested in
photons redshifting into the \lya resonance from frequencies below the \lyb resonance.  In addition, \lya photons can be produced by atomic cascades from photons redshifting into higher Lyman series resonances. These atomic cascades are illustrated in Figure \ref{fig:wf_diagram}, where the probability of converting a \lyn photon into a \lya photon is set by the atomic rate coefficients and can be found in tabular form in Refs. \cite{hirata2006lya,pritchard2006}.  For large $n$, approximately 30\% conversion is typical.  These photons are injected into the \lya line rather than being redshifted from outside of the line.  This changes their contribution to the Wouthuysen-Field coupling since the photon distribution is now one-sided.  Similar processes to those described above apply to the redistribution of these photons, and they can lead to an important amplification of the \lya flux.

This discussion gives a sense of some of the subtleties that go into to determining the strength of the \lya coupling.  These effects can modify the 21 cm signal at the $\sim$10\% level, which will be important as observations begin to detect 21 cm fluctuations.  At this stage, it appears that the underlying atomic physics is understood, although the details of \lya radiative transfer still requires some work.

\section{Global 21 cm signature}
\label{sec:global}

\subsection{Outline}

Next we examine the cosmological context of the 21 cm signal.  We may
express the 21 cm brightness temperature as a function of four
variables $T_b=T_b(T_K,x_i,J_\alpha,n_H)$, where $x_i$ is the
volume-averaged ionized fraction of hydrogen.  In calculating the 21
cm signal, we require a model for the global evolution of and
fluctuations in these quantities.  Before looking at the evolution of
the signal quantitatively, we will first outline the basic picture to
delineate the most important phases.

\begin{figure}[htbp]
\begin{center}
\includegraphics[scale=0.4]{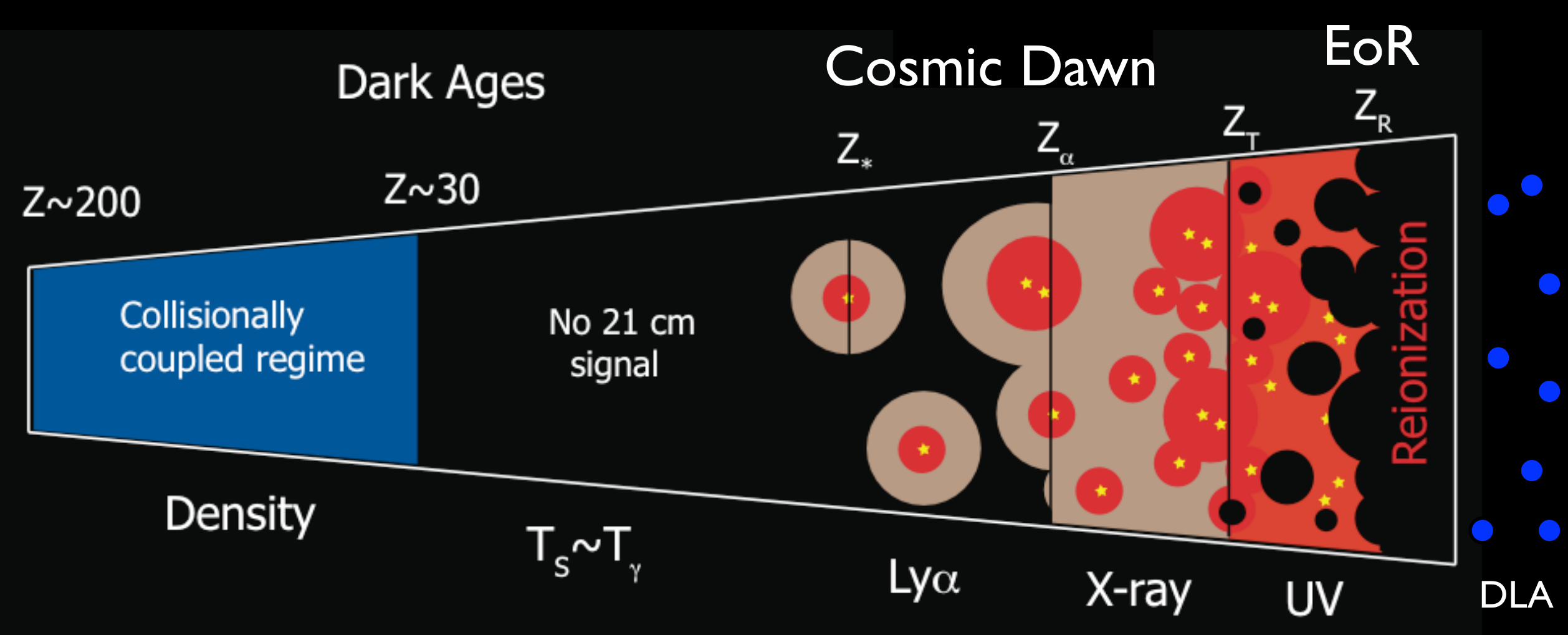}
\caption{Cartoon of the different phases of the 21 cm signal. The
signal transitions from an early phase of collisional coupling to a
later phase of \lya coupling through a short period where there is
little signal.  Fluctuations after this phase are dominated
successively by spatial variation in the \lya, X-ray, and ionizing UV
radiation backgrounds.  After reionization is complete there is a
residual signal from neutral hydrogen in galaxies. }
\label{fig:21cm_cartoon}
\end{center}
\end{figure}

An important feature of $T_b$ is that its dependence on each of these
quantities saturates at some point, for example once the \lya flux is
high enough the spin and kinetic gas temperatures become tightly
coupled and further variation in $J_\alpha$ becomes irrelevant to the
details of the signal.  This leads to conceptually separate regimes
where variation in only one of the variables dominating fluctuations
in the signal.  These different regimes can be seen in Figure
\ref{fig:NVplot} and are shown in schematic form in Figure
\ref{fig:21cm_cartoon} for clarity. We now discuss each of these
phases in turn.

\begin{itemize}

\item $\mathbf{200\lesssim z\lesssim1100}$: The residual free electron fraction left after recombination allows Compton scattering to maintain thermal coupling of the gas to the CMB, setting $T_K=T_\gamma$.  The high gas density leads to effective collisional coupling so that $T_S=T_\gamma$ and we expect $\bar{T}_b=0$ and no detectable 21 cm signal.

\item $\mathbf{40\lesssim z\lesssim200}$: In this regime, the gas cools adiabatically so that $T_K\propto(1+z)^2$ leading to $T_K<T_\gamma$ and
collisional coupling sets $T_S<T_\gamma$, leading to $\bar{T}_b<0$ and
an early absorption signal.  At this time, $T_b$ fluctuations are
sourced by density fluctuations, potentially allowing the initial
conditions to be probed \cite{loeb_zald2004,hirata2006col}.

\item $\mathbf{z_\star\lesssim z \lesssim 40}$: As the expansion
continues, decreasing the gas density, collisional coupling becomes
ineffective and radiative coupling to the CMB sets $T_S=T_\gamma$, and
there is no detectable 21 cm signal.

\item $\mathbf{z_\alpha\lesssim z \lesssim z_\star}$: Once the first sources switch on at $z_\star$, they emit both \lya photons and X-rays.  In general, the emissivity required for \lya coupling is significantly less than that for heating $T_K$ above $T_\gamma$. We therefore expect a regime where the spin temperature is coupled to cold gas so that $T_S\sim T_K<T_\gamma$ and there is an absorption signal.  Fluctuations are dominated by density fluctuations and variation in the \lya flux \cite{bl2005detect,pritchard2006,chen2006}.  As further star formation occurs the \lya
coupling will eventually saturate ($x_\alpha\gg1$), so that by a redshift $z_\alpha$ the gas will everywhere be strongly coupled.

\item $\mathbf{z_h \lesssim z \lesssim z_\alpha}$: After \lya coupling saturates, fluctuations in the \lya flux no longer affect the 21 cm signal.  By this point, heating becomes significant and gas temperature fluctuations source $T_b$ fluctuations. While $T_K$ remains below $T_\gamma$ we see a 21 cm signal in absorption, but as $T_K$ approaches $T_\gamma$ hotter regions may begin to be seen in emission.  Eventually by a redshift $z_h$ the gas will be heated everywhere so that $\bar{T}_K=T_\gamma$.

\item $\mathbf{z_T \lesssim z \lesssim z_h}$: After the heating
transition, $T_K>T_\gamma$ and we expect to see a 21 cm signal in
emission.  The 21 cm brightness temperature is not yet saturated,
which occurs at $z_T$, when $T_S\sim T_K\gg T_\gamma$.  By this time,
the ionization fraction has likely risen above the percent level.
Brightness temperature fluctuations are sourced by a mixture of
fluctuations in ionization, density and gas temperature.

\item $\mathbf{z_r\lesssim z \lesssim z_T}$: Continued heating drives
$T_K\gg T_\gamma$ at $z_T$ and temperature fluctuations become
unimportant.  $T_S\sim T_K\gg T_\gamma$ and the dependence on $T_S$
may be neglected in equation \eref{brightnessT}, which greatly
simplifies analysis of the 21 cm power spectrum \cite{santos2006}.  By
this point, the filling fraction of HII regions probably becomes
significant and ionization fluctuations begin to dominate the 21 cm
signal \cite{fzh2004}.

\item $\mathbf{z\lesssim z_r}$: After reionization, any remaining 21
cm signal originates primarily from collapsed islands of neutral
hydrogen (damped \lya systems).
\end{itemize}

Most of these epochs are not sharply defined, and so there could be
considerable overlap between them.  In fact, our ignorance of early
sources is such that we can not definitively be sure of the sequence
of events.  The above sequence of events seems most likely and can be
justified on the basis of the relative energetics of the different
processes and the probable properties of the sources.  We will discuss
this in more detail as we quantify the evolution of the sources.

Perhaps the largest uncertainty lies in the ordering of $z_\alpha$ and
$z_h$.  Ref. \cite{nasser2005} explores the possibility that
$z_h>z_\alpha$, so that X-ray preheating allows collisional coupling
to be important before the \lya flux becomes significant.  Simulations
of the very first mini-quasar \cite{kuhlen2005,kuhlen2006} also probe
this regime and show that the first luminous X-ray sources can have a
great impact on their surrounding environment.  We note that these
studies ignored \lya coupling, and that an X-ray background may
generate significant \lya photons \cite{chen2006}, as we discuss in
\S\ref{ssec:lyaflux}.  Additionally, while these authors looked at the
case where the production of \lya photons was inefficient, one can
consider the case where heating is much more efficient.  This can be
the case where weak shocks raise the IGM temperature very early on \cite{furlanetto2004}
or if exotic particle physics mechanisms such as dark matter
annihilation are important.  Clearly, there is still considerable
uncertainty in the exact evolution of the signal making the potential
implications of measuring the 21 cm signal very exciting.

\subsection{Evolution of global signal}

Having outlined the evolution of the signal qualitatively, we will
turn to the details of making quantitative predictions.  In
calculating the 21 cm signal it will help us to treat the IGM as a two phase medium.  Initially, the IGM is composed of a single mostly neutral phase left over after recombination.  This phase is characterised by a gas temperature $T_K$ and a small fraction of free electrons $x_e$.  This is the phase that generates the 21 cm signal.

Once galaxy formation begins, energetic UV photons ionize HII regions surrounding, first individual galaxies and then clusters of galaxies. These UV photons have a very short mean free path in a neutral medium leading to the ionized HII regions having a very sharp boundary (although the boundary can be softened if the ionizing photons are particularly hard \cite{geil2009}).  We may therefore treat the ionized HII bubbles as a second phase in the IGM characterised by a
volume filling fraction $x_i$ (provided that the free electron fraction is small $x_i$ is approximately the mean ionization fraction).  We will assume that these bubbles are fully ionized and that the temperature inside the bubbles is fixed at $T_{\rm HII}=10^4{\,\rm K}$ determining the collisional recombination
rate inside these bubbles.  Since the photons that redshift into the \lya resonance initially have long mean free paths, we may treat the \lya flux, $J_\alpha$ as being the same in both phases (although in practice, since there is no 21 cm signal from the fully ionized bubbles, it is only the \lya flux in the mostly neutral phase that matters).  To determine the 21 cm signal at a given redshift, we must calculate the four quantities $x_i$, $x_e$, $T_K$, and $J_\alpha$.  We begin by describing the evolution of the gas temperature $T_K$,
\begin{eqnarray}
\frac{\deriv T_K}{\deriv t}&=&\frac{2T_K}{3n}\frac{\deriv n}{\deriv t}+\frac{2}{3 k_B}\sum_j \frac{\epsilon_j}{n}.\label{thistory}
\end{eqnarray}
Here, the first term accounts for adiabatic cooling of the gas due to
the cosmic expansion while the second term accounts for other sources
of heating/cooling $j$ with $\epsilon_j$ the heating rate per unit
volume for the process $j$.

Next, we consider the volume filling fraction $x_i$ and the ionization
of the neutral IGM $x_e$
\begin{eqnarray}
\frac{\deriv x_i}{\deriv t}&=&(1-x_e)\Lambda_i-\alpha_ACx_i^2n_H,\label{xihistory}\\
\frac{\deriv x_e}{\deriv t}&=&(1-x_e)\Lambda_e-\alpha_B(T)x_e^2n_H.\label{xehistory}
\end{eqnarray}
In these expressions, we define $\Lambda_i$ to be the rate of
production of ionizing photons per unit time per baryon applied to HII
regions, $\Lambda_e$ is the equivalent quantity in the bulk of the
IGM, $\alpha_A=4.2\times10^{-13}\rm{cm^3\,s^{-1}}$ is the case-A
recombination coefficient at $T=10^4\,\rm{K}$, $\alpha_B(T)$ is the
case-B recombination rate (whose temperature dependence can be
obtained from \cite{seager1999}), and $C\equiv\langle
n_e^2\rangle/\langle n_e \rangle^2$ is the clumping factor.

Superficially these look the same, since in each case the ionization
rate is a balance between ionizations and recombinations.  The main
distinction lies in the manner in which we treat the recombinations.
In the fully ionized bubbles, recombinations occur in those dense
clumps of material capable of self-shielding against ionizing
radiation.  These overdense regions will have a locally enhanced
recombination rate, making it important to account for the
inhomogeneous distribution of matter through the clumping factor $C$.
Since recombinations will occur on the edge of these neutral clumps,
secondary photons produced by the recombinations will likely be
absorbed inside the clumps rather than in the mean IGM, justifying the
use of case-A recombination \cite{me2000}.  In contrast,
recombinations in the bulk of the neutral IGM will occur at close to
mean density in gas with temperature $T_K$.  Here recombination
radiation will be absorbed in the IGM, so we must use case-B
recombination.  By keeping track of this carefully our evolution
matches onto that of RECFAST \cite{seager1999}.

This two phase approximation will eventually break down should $x_e$
become close to unity, indicating that most of the IGM has been ionized
and that there is no clear distinction between ionized bubbles and a
neutral bulk IGM.  In most of our models, $x_e$ remains small until the end
of reionization making this a reasonable approximation.

\subsection{Growth of HII regions}

The growth of ionized HII regions is governed by the interplay between
ionization and recombination, both of which contain considerable
uncertainties.  We may write the ionization rate per hydrogen atom as
\begin{equation}\label{lambdai_v1}
\Lambda_i=A_{\rm He}f_{\rm esc}N_{\rm ion} \dot{\rho}_{\star}(z),
\end{equation}
with $N_{\rm{ion}}$ being the number of ionizing photons per baryon
produced in stars, $f_{\rm{esc}}$ the fraction of ionizing photons
that escape the host halo, and $A_{\rm{He}}$ a correction factor for
the presence of Helium. Here $\dot{\rho}_{\star}(z)$ is the star
formation rate density as a function of redshift, which is still
poorly known observationally. For a present-day initial mass function
of stars, $N_{\rm ion}\sim 4\times 10^3$, whereas for very massive
($>10^2M_\odot$) stars of primordial composition, $N_{\rm ion}\sim
10^5$ \cite{bromm2001,zackrisson2011}.

We model the star formation rate as tracking the collapse of matter,
so that we may write the star formation rate per (comoving) unit
volume
\begin{equation}\label{sfrd}
\dot{\rho}_{\star}(z)=\bar{\rho}^0_b f_\star\frac{\deriv }{\deriv t}f_{\rm{coll}}(z).
\end{equation}
where $\bar{\rho}^0_b$ is the cosmic mean baryon density today and
$f_\star$ is the fraction of baryons converted into stars.  This
formalism is appropriate for $z\gtrsim10$, as at later times star
formation as a result of mergers becomes important.

With these assumptions, we may rewrite the ionization rate per hydrogen atom as
\begin{equation}\label{lambdai}
\Lambda_i=\zeta(z)\frac{\deriv f_{\rm{coll}}}{\deriv t},
\end{equation}
where $f_{\rm{coll}}(z)$ is the fraction of gas inside collapsed objects at $z$ and the ionization efficiency parameter $\zeta$ is given by 
\begin{equation}
\zeta=A_{\rm{He}}f_\star f_{\rm{esc}}N_{\rm{ion}}.
\end{equation}

This model for $x_i$ is motivated by a picture of HII regions
expanding into neutral hydrogen \cite{bl2001}.  In calculating
$f_{\rm{coll}}$, we use the Sheth-Tormen \cite{sheth1999mfn} mass
function $\deriv n/\deriv m$ and determine a minimum mass
$m_{\rm{min}}$ for collapse by requiring the virial temperature
$T_{\rm{vir}}\ge10^4\,\rm{K}$, appropriate for cooling by atomic
hydrogen. Decreasing this minimum galaxy mass, say to the virial
temperature corresponding to molecular hydrogen cooling ($\sim 300$
K), will allow star formation to occur at earlier times, shifting the
features that we describe in redshift.

The sources of ionizing photons in the early Universe are believed to
have been primarily galaxies.  However, the properties of these
galaxies are currently only poorly constrained.  Recent observations
with the Hubble Space Telescope provide some of the best constraints
on early galaxy formation.  Faint galaxies are identified as being at
high redshift using a ``\lya dropout technique" where a naturally
occurring break in the galaxy spectrum at the \lya wavelength 1216\AA\
is seen in different colour filters as a galaxy is redshifted.  So
far, galaxies at redshifts up to $z\sim10$ have been found providing
information on the sources of reionization.  There are unfortunately
considerable limitations on the existing surveys owing to their small
sky coverage, which makes it unclear whether those galaxies seen are
properly representative, and the limited frequency coverage.  Even
more problematic for our purposes is that the optical frequencies at
which the galaxies are seen do not correspond to the UV photons that
ionize the IGM.  Our limited understanding of the mass distribution of
the emitting stars introduces an uncertainty in the number of ionizing
photons per baryon $N_{\rm ion}$ is emitted by galaxies.  There is
also considerable uncertainty in the fraction of ionizing photons
$f_{\rm esc}$ that escape the host galaxy to ionize the IGM.

The recombination rate is primarily important at late times once a
significant fraction of the volume has already been ionized.  At this
stage, dense clumps within an ionized bubble can act as sinks of
ionizing photons slowing or even stalling further expansion of the
bubble.  The degree to which gas resides in these dense clumps is an
important uncertainty in modelling reionization.  Important
hydrodynamic effects, such as the evaporation of gas from a halo as a
result of photoionization heating \cite{pawlik2009}, can significantly
modify the clumping factor.

A simple model for the clumping factor \cite{me2000} assumes that the
Universe will be fully ionized up to some critical overdensity
$\Delta_c$.  If the probability distribution for the gas density
$P_V(\Delta)$ is specified, we may then write the clumping factor as
\begin{equation}
C\equiv\int^{\Delta_c}\ud\Delta\,\Delta^2P_V(\Delta).
\end{equation}
The quantity $P_V(\Delta)$ can be modeled analytically starting from a consideration of behaviour of low density voids and accounting for Gaussian initial conditions.  The analytic form resembles that of a Gaussian with a power law tail
\cite{me2000} and can be measured from simulations \cite{bolton2009}.  To accurately capture the clumping one should self-consistently perform a full hydrodynamical simulation of reionization, since thermal feedback can modify the gas density distribution \cite{pawlik2009}.

To set the critical density $\Delta_c$, we account for the patchy
nature of reionization, which proceeds via the expansion and overlap
of ionized bubbles \cite{furlanetto2005tax}.  The size of a bubbles
will become limited if the mean free path of ionized photons becomes
shorter than the size of the bubble, for example if the bubble
contains many small self-shielded absorbers.  The mean free path of
ionizing photons can be related to the underlying density field as
\begin{equation}
\lambda_i=\lambda_0\left[1-F_V(\Delta_i)\right]^{-2/3}.
\end{equation}
Here $\lambda_0$ is an unknown normalisation constant that was found by Ref. \cite{me2000} in the context of simulations at $z=2-4$ to be well fit by $\lambda_0 H(z)=60{\rm\,km\,s^{-1}}$.  This scaling
relationship is likely to be very approximate, but we make use of it
for convenience.  With this we can fix $\Delta_i$ within an ionized
bubble by setting the relevant $R_b=\lambda_i(\Delta_c)$.  We then
average the clumping factor over the distribution of bubble sizes
(discussed in more details later) to get the mean clumping factor.

\subsection{Heating and ionization}

To determine the heating rate, we must integrate equation
\eref{thistory} and therefore we must specify which heating mechanisms
are important.  At high redshifts, the dominant mechanism is Compton
heating of the gas arising from the scattering of CMB photons from the
small residual free electron fraction.  Since these free electrons
scatter readily from the surrounding baryons this transfers energy
from the CMB to the gas.  Compton heating serves to couple $T_K$ to
$T_\gamma$ at redshifts $z\gtrsim 150$, but becomes ineffective below
that redshift.  In our context, it serves to set the initial
conditions before star formation begins.  The heating rate per
particle for Compton heating is given by \cite{naoz2005}
\begin{equation}\label{compton}
\frac{2}{3}\frac{\epsilon_{\rm{compton}}}{k_Bn}=\frac{x_e}{1+f_{\rm{He}}+x_e}\frac{T_\gamma-T_K}{t_\gamma}\frac{u_\gamma}{\bar{u}_\gamma}(1+z)^{4},
\end{equation}
where $f_{\rm{He}}$ is the helium fraction (by number), $u_\gamma$ is the energy density of the CMB, $\sigma_T=6.65\times10^{-25}\rm{cm^2}$ is the Thomson cross-section, and we define
\begin{equation}
t_\gamma^{-1}=\frac{8 \bar{u}_\gamma\sigma_T}{3m_e c}=8.55\times 10^{-13}\,\rm{yr}^{-1}.
\end{equation}

At lower redshifts, the growth of non-linear structures leads to other
possible sources of heat.  Shocks associated with large scale
structure occur as gas separates from the Hubble flow and undergoes
turnaround before collapsing onto a central overdensity.  After
turnaround different fluid elements may cross and shock due to the
differential accelerations.  Such turnaround shocks could provide
considerable heating of the gas at late times \cite{furlanetto2004}.

Another source of heating is the scattering of \lya photons off
hydrogen atoms, which leads to a slight recoil of the nucleus that
saps energy from the photon.  It was initially believed that this
would provide a strong source of heating sufficient to prevent the
possibility of seeing the 21 cm signal in absorption.  Early
calculations showed that by the time the scattering rate required for
\lya photons to couple the spin and gas temperatures was reached, the
gas would have been heated well above the CMB temperature
\cite{mmr1997}.  These early estimates, however, did not account for
the way the distribution of \lya photon energies was changed by
scattering.  This spectral distortion is a part of the photons coming
into equilibrium with the gas and serves to greatly reduce the heating
rate \cite{mmr1997,chen2004,chuzhoy2006heat,furlanetto2006heat}.
While \lya heating can be important it typically requires very large
\lya fluxes and so is most relevant at late times and may be
insufficient to heat the gas to the CMB temperature alone.

The most important source of energy injection into the IGM is likely
via X-ray heating of the gas
\cite{venkatesan2001,chen2004,pritchard2007xray,zaroubi2007} .  While
shock heating dominates the thermal balance in the present day
Universe, during the epoch we are considering they heat the gas only
slightly before X-ray heating dominates. For sensible source
populations, \lya heating is mostly negligible compared to X-ray
heating \cite{furlanetto2006heat,ciardi2010}.

Since X-ray photons have a long mean free path, they are able to heat
the gas far from the source, and can be produced in large quantities
once compact objects are formed. The comoving mean free path of an
X-ray with energy $E$ is \cite{fob}
\begin{equation}
\lambda_X\approx4.9 \bar{x}_{\rm{HI}}^{-1/3}\left(\frac{1+z}{15}\right)^{-2}\left(\frac{E}{300 \eV}\right)^3 {\rm\,Mpc} .
\end{equation}
Thus, the Universe will be optically thick over a Hubble length to
all photons with energy below
$E\sim2[(1+z)/15]^{1/2}\bar{x}_{\rm{HI}}^{1/3}\keV$.  The $E^{-3}$
dependence of the cross-section means that heating is dominated by
soft X-rays, which fluctuate on small scales.  In addition, though,
there will be a uniform component to the heating from harder X-rays.

X-rays heat the gas primarily through photo-ionization of HI and HeI:
this generates energetic photo-electrons, which dissipate their energy
into heating, secondary ionizations, and atomic excitation.  With this
in mind, we calculate the total rate of energy deposition per unit
volume as
\begin{equation}
\epsilon_X=4\pi \sum_i n_i\int \deriv \nu\, \sigma_{\nu,i}J_\nu (h\nu-h\nu_{{\rm th},i}),
\end{equation}
where we sum over the species $i=$HI, HeI, and HeII, $n_i$ is the
number density of species $i$, $h\nu_{\rm{th}}=E_{\rm{th}}$ is the
threshold energy for ionization, $\sigma_{\nu,i}$ is the cross-section
for photoionization, and $J_\nu$ is the number flux of photons of
frequency $\nu$.

We may divide this energy into heating, ionization, and excitation by
inserting the factor $f_i(\nu,x_e)$, defined as the fraction of energy
converted into form $i$ at a specific frequency. This allows us to
calculate the contribution of X-rays to both the heating and the
partial ionization of the bulk IGM.  The relevant division of the
X-ray energy depends on both the X-ray energy $E$ and the free
electron fraction $x_e$ and can be calculated by Monte-Carlo methods.
This partitioning of X-ray energy in this way was first calculated by
Ref. \cite{SVS1985} and subsequently updated
\cite{valdes2008,furlanetto2010cascade}.  In the following
calculations, we make use of fitting formula for the $f_i(\nu)$
calculated by Ref. \cite{SVS1985}, which are approximately independent
of $\nu$ for $h\nu\gtrsim100\eV$, so that the ionization rate is
related to the heating rate by a factor
$f_{\rm{ion}}/(f_{\rm{heat}}E_{\rm{th}})$.

The X-ray number flux is found from
\begin{eqnarray}\label{jxflux}
J_X(z)&=&\int_{\nu_{\rm{th}}}^{\infty}\deriv \nu\,J_X(\nu,z),\\
&=&\int_{\nu_{\rm{th}}}^{\infty}\deriv \nu\int_z^{z_{\star}} \deriv z'\,\frac{(1+z)^2}{4\pi}\frac{c}{H(z')}\hat{\epsilon}_X(\nu',z')e^{-\tau},\nonumber
\end{eqnarray}
where $\hat{\epsilon}_X(\nu,z)$ is the comoving photon emissivity for
X-ray sources, and $\nu'$ is the emission frequency at $z'$
corresponding to an X-ray frequency $\nu$ at $z$
\begin{equation}\label{nup}
\nu'=\nu\frac{(1+z')}{(1+z)}.
\end{equation}
The optical depth is given by
\begin{equation}
\tau(\nu,z,z')=\int_z^{z'} \frac{\deriv l}{\ud z''}\ud z''\,[ n_{\rm{HI}}\sigma_{\rm{HI}}(\nu'')+n_{\rm{HeI}}\sigma_{\rm{HeI}}(\nu'')\\+n_{\rm{HeII}}\sigma_{\rm{HeII}}(\nu'')],
\end{equation}
where we calculate the cross-sections using the fits of
\cite{verner1996}.  Care must be taken here, as the cross-sections
have a strong frequency dependence and the X-ray photon frequency can
redshift considerably between emission and absorption. In practice,
the abundance of HeII may be neglected \cite{venkatesan2001}.

X-rays may be produced by a variety of different sources with three
main candidates at high redshifts being identified as starburst
galaxies, supernova remnants (SNR), and miniquasars
\cite{oh2001,glover2003,furlanetto2006}.  Galaxies with high rates of
star formation produce copious numbers of X-ray binaries, whose total
X-ray luminosity can be considerable.  Two populations of X-ray
binaries may be identified in the local Universe distinguished by the
mass of the donor star which feeds its black hole companion - low-mass
X-ray binaries (LMXB) and high-mass X-ray binaries (HMXB).  The short
life time of HMXBs ($t_{\rm HMXB}\sim10^7{\rm\,yr}$) leads the X-ray
luminosity $L^{\rm HMXB}_X$ to track the star formation rate.  At the
same time, the longer lived LMXB ($t_{\rm LMXB}\sim10^{10}{\rm\,yr}$)
tracks the total mass of stars formed.  Since we will focus on the
early Universe and on the first billion years of evolution, when few LMXB are expected to have formed, the
dominant contribution to $L_X$ in galaxies is likely to be from HMXB \cite{grimm2003}. This has conventionally been defined in terms of a parameter $f_X$ such that the emissivity per unit (comoving) volume per unit frequency
\begin{equation}\label{ehatX}
\hat{\epsilon}_X(z,\nu)=\hat{\epsilon}_X(\nu)\left(\frac{\dot{\rho}_{\star}(z)}{\rm{M}_\odot \,\rm{yr}^{-1}\,Mpc^{-3}}\right),
\end{equation}
where $\dot{\rho}_{\star}$ is the star formation rate density, and the spectral distribution function is a power law with index $\alpha_S$
\begin{equation}\label{emissivityX}
\hat{\epsilon}_X(\nu)=\frac{L_0}{h\nu_0} \left(\frac{\nu}{\nu_0}\right)^{-\alpha_S-1},
\end{equation}
and the pivot energy $h\nu_0=1\,\rm{keV}$.  We assume emission within
the band 0.2 -- 30 keV, and set $L_0=3.4\times10^{40}
f_X\,\rm{erg\,s^{-1}\,Mpc^{-3}}$, where $f_X$ is a highly uncertain
constant factor \cite{furlanetto2006}.  For a spectral index $\alpha_S=1.5$, roughly
corresponding to that for starburst galaxies, $f_X=1$ corresponds to
the emission of approximately $560\eV$ for every baryon converted into
stars.

This normalisation was chosen so that, with $f_X=1$, the total X-ray luminosity per unit star formation rate (SFR) is consistent with that observed in starburst galaxies at the present epoch \cite{grimm2003,gilfanov2004}.  Since then improved observations have revised this figure owing to better separation of the contribution from LMXB and HMXB.  While the data is still as of yet fairly patchy and shows considerable scatter $f_X\approx0.2$ seems a better fit to the most recent data in the local universe \cite{lehmer2010,mineo2010}.
Extrapolating observations from the present day to high redshift is fraught with uncertainty, and we note that this normalisation is very uncertain and probably evolves with redshift \cite{dijkstra2011}. In particular, the metallicity evolution of galaxies with redshift is likely to impact the ratio of black holes to neutron stars that form the compact object in the HMXB and with it the efficiency of X-ray production.  Additionally, the fraction of stars in binaries may evolve with redshift and is only poorly constrained at high redshifts \cite{mirabel2011}.

Other sources of X-rays are inverse Compton scattering of CMB photons from the energetic electrons in supernova remnants. Estimates of the luminosity of such sources is again highly uncertain, but of a similar order of magnitude as from HMXB \cite{oh2001}.  Like HMXB, the X-ray luminosity from supernovae remnants is expected to track the star formation rate.  Finally, miniquasars - accretion onto black holes with intermediate masses in the range $10^{1-5} M_\odot$ - can produce significant levels of X-rays.  Since the early formation of black holes depends sensitively on the source of seed black holes and their subsequent merger history there is again considerable uncertainty.  For simplicity, we will assume
that miniquasars similarly track the star formation rate (SFR). In reality, of course, their evolution could be considerably more complex \cite{madau2004,ricotti2004}.

The total X-ray luminosity at high redshift is constrained by
observations of the present day unresolved soft X-ray background (SXRB).  An early population of X-ray sources would produce hard X-rays that would redshift to lower energies contributing to this background.  Since there will be faint X-ray sources at lower redshift that also contribute to this background, the SXRB can be used to place a conservative upper limit on the amount of X-ray production at early times.  This rules out complete reionization by X-rays but allows considerable latitude for heating \cite{dijkstra2004}.  

Since heating requires considerably less energy than ionization, $f_X$ is still relatively unconstrained with values as high as $f_X\lesssim10^3$ possible without violating constraints from the CMB polarisation anisotropies on the optical depth for electron scattering.  Constraining this parameter will mark a step forward in our understanding of the thermal history of the IGM and the population of X-ray sources at high redshifts.

\subsection{Coupling}
\label{ssec:lyaflux}

Finally, we need to specify the evolution of the \lya flux.  This is
produced by stellar emission ($J_{\alpha,\star}$) and by X-ray
excitation of HI ($J_{\alpha,X}$).  Photons emitted by stars, between
\lya and the Lyman limit, will redshift until they enter a Lyman
series resonance.  Subsequently, they may generate \lya photons via
atomic cascades \cite{pritchard2006,hirata2006lya}.  The \lya flux
from stars $J_{\alpha,\star}$ arises from a sum over the Ly$n$ levels,
with the maximum $n$ that contributes $n_{\rm{max}}\approx23$
determined by the size of the HII region of a typical (isolated)
galaxy (see \cite{bl2005detect} for details).  The average \lya
background is then
\begin{eqnarray}\label{jaflux}
J_{\alpha,\star}(z)&=&\sum_{n=2}^{n_{\rm{max}}}J^{(n)}_\alpha(z),\\
&=&\sum_{n=2}^{n_{\rm{max}}}f_{\rm{recycle}}(n)\int_z^{z_{\rm{max}}(n)} \deriv z'\,\frac{(1+z)^2}{4\pi}\frac{c}{H(z')}\hat{\epsilon}_\star(\nu'_n,z'),\nonumber
\end{eqnarray}
where $z_{\rm max}(n)$ is the maximum redshift from which emitted
photons will redshift into the level $n$ Lyman resonance, $\nu'_n$ is
the emission frequency at $z'$ corresponding to absorption by the
level $n$ at $z$, $f_{\rm{recycle}}(n)$ is the probability of
producing a \lya photon by cascade from level $n$, and
$\hat{\epsilon}_\star(\nu,z)$ is the comoving photon number emissivity
for stellar sources.  We connect $\hat{\epsilon}_\star(\nu,z)$ to the
star formation rate in the same way as for X-rays in equation
\eref{emissivityX}, and define $\hat{\epsilon}_\star(\nu)$ to be the
spectral distribution function of the stellar sources.

Stellar sources typically have a spectrum that falls rapidly above the
\lyb transition.  We consider models with present-day (Pop. I \& II)
and very massive (Pop. III) stars.  In each case, we take
$\hat{\epsilon}_\star(\nu)$ to be a broken power law with one index
describing emission between \lya and \lybns, and a second describing
emission between \lyb and the Lyman limit (see \cite{pritchard2006}
for details).  The details of the signal depend primarily on the total
\lya emissivity and not on the shape of the spectrum (but see
\cite{chuzhoy2006,pritchard2007xray} for details of how precision
measurements of the 21 cm fluctuations might say something about the
source spectrum).

For convenience, we define a parameter controlling the normalisation
of the \lya emissivity $f_\alpha$ by setting the total number of \lya
photons emitted per baryon converted into stars as $N_\alpha=f_\alpha
N_{\rm \alpha,ref}$ where we take the reference values appropriate for
normal (so-called, {\it Population I \& II}) stars $N_{\rm
\alpha,ref}=6590$ \cite{leitherer1999,bl2005detect}.  For comparison, in this notation, the very massive
({\it Population III}) stars have \cite{bromm2001}, $N_\alpha=3030$
($f_\alpha=0.46$), when the contribution from higher Lyman series
photons is included.  We expect the value of $f_\alpha$ to be close to unity, since stellar properties are relatively well understood.

Photoionization of HI or HeI by X-rays may also lead to the production
of \lya photons.  In this case, some of the primary photo-electron's
energy ends up in excitations of HI \cite{SVS1985}, which on
relaxation may generate \lya photons
\cite{mmr1997,chen2006,chuzhoy2006}.  The rate at which \lya photons
are produced $\epsilon_{X,\alpha}$ can be calculated in the same way
as the contribution to heating by X-rays, but with the appropriate
change in the fraction of the total X-ray energy that goes into
excitations rather than heating.

Calculating the fraction of energy that goes into producing \lya
photons is a problem that has been considered by a number of authors.
Early work,\cite{shull1979,SVS1985} focused on the amount of
energy that went into atomic excitations as a whole, but we require
only the fraction that leads to \lya production.  Although excitations
to the 2P level will always generate \lya photons, only some fraction
of excitations to other levels will lead to \lya generating cascades.
The rest will end with two photon decay from the 2S level. This has
been addressed by Monte-Carlo simulation of the X-ray scattering
process using up to date cross-sections in
\cite{valdes2008,furlanetto2010cascade}.  These simulations find that
around $p_\alpha\approx0.7$ of the total energy that goes into
excitation ends up as \lya photons, consistent with simple estimates
based on the atomic cross-sections \cite{pritchard2007xray}.

This \lya flux $J_{\alpha,X}$ produced by X-ray excitation may be
found by balancing the rate at which \lya photons are produced via
cascades with the rate at which photons redshift out of the \lya
resonance \cite{chen2006}, giving
\begin{equation}
J_{\alpha,X}=\frac{c}{4\pi}\frac{\epsilon_{X,\alpha}}{h\nu_\alpha}\frac{1}{H\nu_\alpha}.
\end{equation}

The relative importance of \lya photons from X-rays or directly
produced by stars is highly dependent upon the nature of the sources
that existed at high redshifts.  Furthermore, it can vary
significantly from place to place.  In general, X-rays with their long
mean free path seem likely to dominate the \lya flux far from sources
while the contribution from stellar sources dominates closer in
\cite{chen2006}.

\subsection{Astrophysical sources and histories}
In the above sections, we have outlined the mathematical formalism for describing the 21 cm signal and have omitted a detailed discussion of the sources.  This was deliberate; although we have a reasonable understanding of the physical processes involved, our knowledge of the properties of early sources of radiation is highly uncertain.

Many models of galaxy formation assume that the first stars to form from the collapse of primordial gas are very massive ($\sim10-100{\rm\,M_\odot}$) population III stars \cite{bromm2001}.  This is predicated on the inference that the absence of coolants more efficient than molecular hydrogen leads to monolithic collapse into a single massive star rather than fragmentation into many lower mass stars.  This assumption has recently begun to be challenged by new numerical simulations that use ``sink particles" to better follow the collapsing gas for many dynamical times.  Such simulations show that fragmentation into many $\sim0.1-1{\rm\,M_\odot}$ stars may be the preferred channel of star formation \cite{clark2011}.  This would naturally explain tentative observations of low mass metal-free stars \cite{caffau2011} and could lead to a much higher fraction of early X-ray binaries \cite{mirabel2011}.  Once earlier generations of star formation has enriched the IGM with metals low mass population II stars will begin to form due to more efficient gas cooling \cite{bromm2003}.  Different predictions for the mode of star formation will lead to quite different IGM histories.

We have three radiation backgrounds to account for - ionizing UV, X-ray, and Lyman series photons (identified as those photons with energy $10.2\,\eV\le E<13.6\,\eV$).  For each of these radiation fields we must specify a single parameter: the ionization efficiency
$\zeta$, the X-ray emissivity $f_X$, and the \lya emissivity
$f_\alpha$.  These parameters enter our model as a factor multiplying
the star formation rate and are therefore individually degenerate with
the star formation efficiency $f_\star$.  This split provides a
natural separation between the physics of the sources and the star
formation rate and, in practice, one might imagine using observations
of the star formation rate by other means as a way of breaking the
degeneracy between them.  In addition to these parameters, we must
specify the minimum mass halo in which galaxies form $M_{\rm min}$ and
make use of the Sheth-Tormen mass function of dark matter halos.

We now show results for the 21 cm global signal that explore this
parameter space to give a sense of how the signal depends on these
astrophysical parameters.  Model A uses ($N_{\rm ion,IGM}$,
$f_\alpha$, $f_X$, $f_*$) = (200,1,1,0.1) giving $z_{\rm reion}=6.47$
and $\tau=0.063$. Model B uses ($N_{\rm ion,IGM}$, $f_\alpha$, $f_X$,
$f_*$) = (600,1,0.1,0.2) giving $z_{\rm reion}=9.76$ and $\tau=0.094$.
Model C uses ($N_{\rm ion,IGM}$, $f_\alpha$, $f_X$, $f_*$) =
(3000,0.46,1,0.15) giving $z_{\rm reion}=11.76$ and $\tau=0.115$.

%
%
%
%
%
%

\begin{figure}[htbp]
\begin{center}
\includegraphics[scale=0.4]{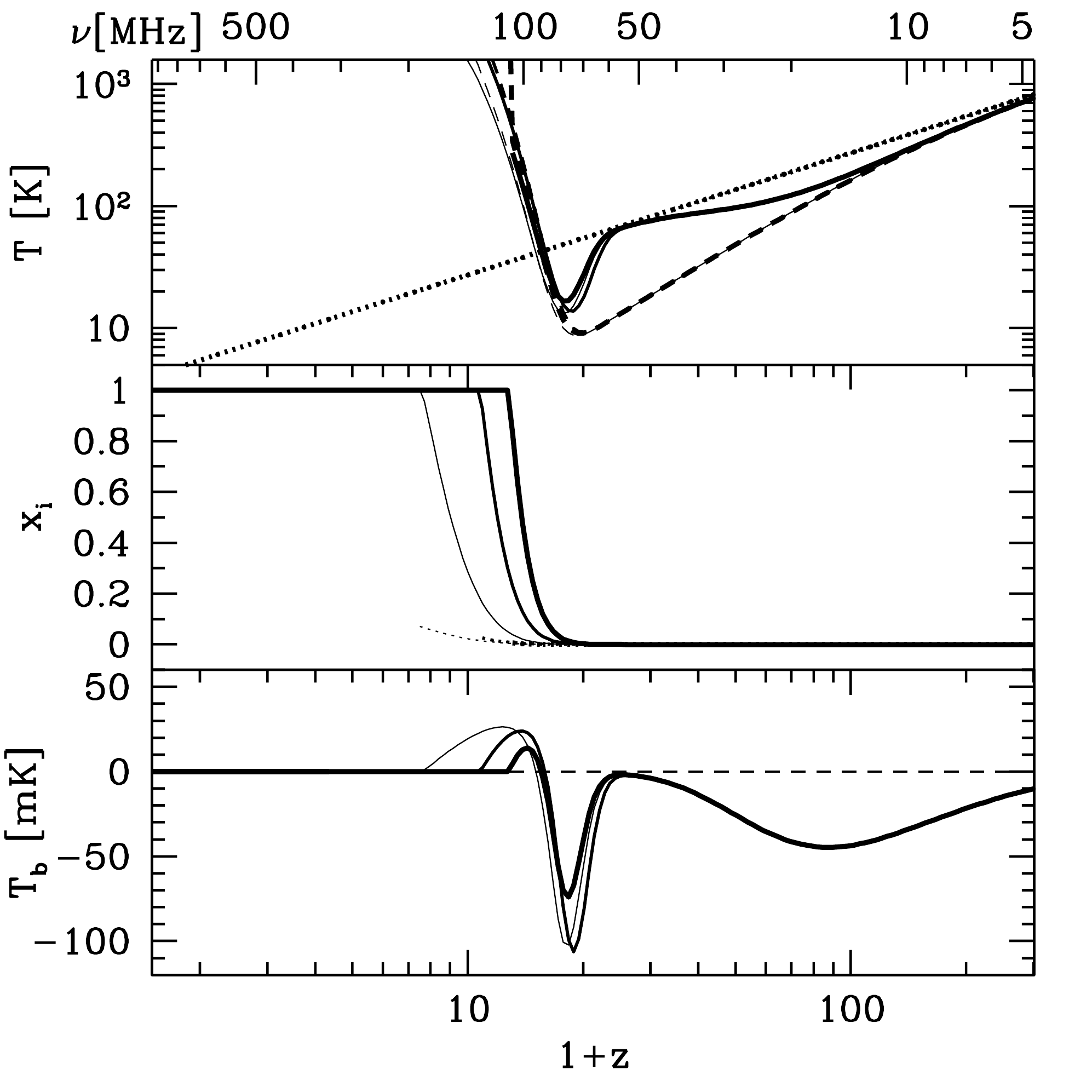}
\caption{{\em Top panel: }Evolution of the CMB temperature $T_{\rm CMB}$
(dotted curve),the gas kinetic temperature $T_K$ (dashed curve), and the
spin temperature $T_S$ (solid curve).  {\em Middle panel: }Evolution of the
gas fraction in ionized regions $x_i$ (solid curve) and the ionized
fraction outside these regions (due to diffuse X-rays) $x_e$ (dotted
curve). {\em Bottom panel: } Evolution of mean 21 cm brightness temperature
$T_b$.  In each panel we plot curves for model A (thin curves), model B
(medium curves), and model C (thick curves). \cite{pritchard2008}}
\label{fig:pl_global}
\end{center}
\end{figure}
Figure \ref{fig:pl_global} shows several examples of the global 21 cm
signal and the associated evolution in the neutral fraction and gas
temperatures.  While the details of the models may vary considerably,
all show similar basic properties.  At high redshift, $10\gg z\lesssim
200$, the gas temperature cools adiabatically faster than the CMB
(since the residual fraction of free electrons is insufficient to
couple the two temperatures).  At the same time, collisional coupling
is effective at coupling spin and gas temperatures leading to the
absorption trough seen at the right of the lower panel.  The details
of this trough are fixed by cosmology and therefore may be predicted
relatively robustly.  The minima of this trough corresponds to the
point at which collisional coupling starts to become relatively
ineffective.

Once star formation begins, the spin and gas temperatures again become
tightly coupled leading to a second, potentially deeper, absorption
trough.  The minimum of this trough corresponds to the point when
X-ray heating switches on heating the gas above the CMB temperature
leading to an emission signal.  The signal then reaches the curve for
a saturated signal ($T_S\gg T_{\rm CMB}$) briefly before the
ionization of neutral hydrogen diminishes it.

The ordering of these events is determined primarily by the energetics
of the processes involved and by the basic properties of the
reasonable source spectra.  For example, ionization requires at least
one ionizing photon with energy $E\ge13.6\,\eV$ per baryon while
depositing only $\sim10\%$ of that energy per baryon would heat the
gas to $T_K\gtrsim10^4\,\K$.  However, the details of the shape of the
curve after star formation begins are highly uncertain - in our model
we have neglected any possible redshift evolution in the various
photon emissivity parameters - but the basic structure of one emission
feature and two absorption troughs are likely to be robust.  By
determining the positions of the various turning points in the signal
one could hope to constrain the underlying astrophysics and learn
about the first stars and galaxies.

\subsection{Exotic heating}

One of the key points to take away from this discussion is that the 21
cm global signal plays the role of a very sensitive calorimeter of the
IGM gas temperature.  Provided that the coupling is saturated and that
the IGM is close to neutral there is a direct connection between the
21 cm brightness temperature and the IGM temperature.  Many models of
physics beyond the standard model make concrete predictions for exotic
heating of the IGM.  For example, dark matter annihilation in the
early Universe can act as a source of X-rays leading to heating.  In
this subsection, we consider some of the possibilities that have been
advanced for exotic heating mechanisms and discuss the possibility of
constraining them.

Perhaps the most commonly considered source of heating in the dark
ages is that of dark matter annihilation
\cite{chen2004dm,furlanetto2006dm,valdes2007,belikov2009,slatyer2009}.
Dark matter is widely assumed to explain the observed galaxy rotation
curves as well as the detailed features of the CMB acoustic peaks.
Simple models of dark matter production and freeze out in the early
Universe lead to a prediction for the annihilation cross-section
required to leave a freeze out abundance corresponding to the measured
value of the dark matter density parameter, $\Omega_{\rm DM}$.

Depending on the dark matter mass, which for fixed $\Omega_{\rm DM}$
determines the dark matter number density $n_{\rm DM}$, annihilation
of dark matter in the later Universe may be an important source of
heating.  It is important to note that there are two regimes in which
dark matter annihilation may be important.  Since the annihilation
rate scales as $n_{\rm DM}^2$ the rate may be large at early times
where $n_{\rm DM}$ has yet to be diluted by the cosmic expansion.
Alternatively, dark matter annihilation can become important once
significant numbers of collapsed dark matter halos form, leadin to a
local enhancement in the dark matter density \cite{cumberbatch2008}.

Alternatives to dark matter annihilation include dark matter decaying
into standard model particles or photons leading to the deposition of
energy in the IGM \cite{furlanetto2006dm,valdes2007}.  The different
density dependence of dark matter decay and annihilation might lead to
distinguishable redshift evolution of the heating.  Further, models
where dark matter contains an internal excited state that relaxes to
the ground state releasing energy have been proposed
\cite{finkbeiner2008}.

Many other scenarios for exotic heating of the IGM have been put
forward, emphasising the interest in a new technique for
distinguishing models of new physics.  Primordial black holes produced
in the early Universe may evaporate after recombination if their
masses lie in the range of $10^{14}-10^{17}{\rm\,g}$ \cite{mack2008}
and the Hawking radiation given off could provide a strong heating
source \cite{ricotti2008}.  Moving cosmic strings produce wakes that
stir the IGM imparting heat into the gas.  These were originally put
forward as a source of density fluctuations for seeding the growth of
structure.  While ruled out for this purpose, cosmic strings might be
further constrained via their heating effect on the IGM
\cite{brandenberger2010}.

\begin{figure}[htbp]
\begin{center}
\includegraphics[scale=0.8]{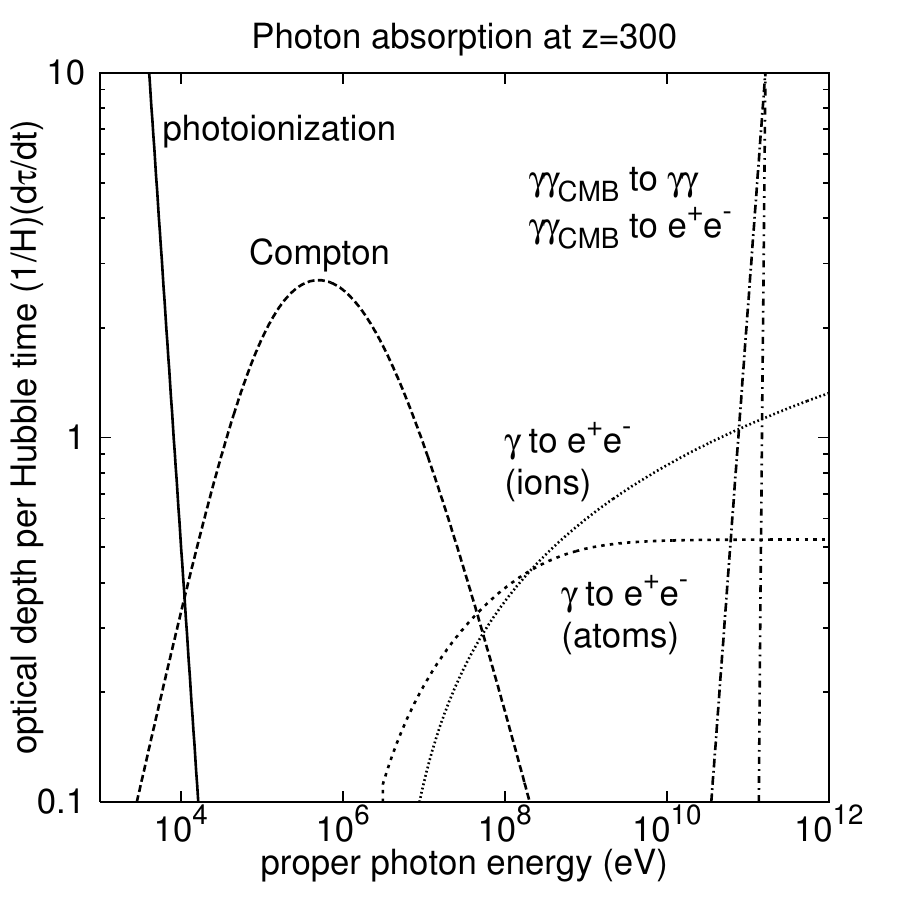}
\caption{Optical depths per time for various photon-IGM processes, in
units of the Hubble time, at z = 300, assuming a neutral IGM. These
include processes which deposit energy directly into the IGM (pair
production and photoionization), processes which redistribute photons
(2$\gamma\rightarrow2\gamma$) and ones that do both (Compton). At very
low energies, photoionization is the dominant process; at very high
energies, e$\pm$ pair production dominates \cite{mack2008}.}
\label{fig:mack_scattering}
\end{center}
\end{figure}

Incorporating the heating effect arising from exotic sources requires
a knowledge of the energy spectrum of photons produced by the source,
which must then be carefully processed to determine how much of the
radiative energy is ultimately deposited into the IGM.  The
cross-section for photon absorption has a number of minima, which
reflect windows at which the IGM is transparent to photons so that
rather than being absorbed they may propagate to the present as a
diffuse background.  For most scenarios, this consideration greatly
constrains the amount of energy deposited as heat in the IGM.  Figure
\ref{fig:mack_scattering} illustrates the various processes that
dominate the loss of energy from an energetic photon at $z=300$.

\subsection{Detectability of the global signal with small numbers of dipoles}

The global 21 cm signal could potentially be measured by absolute
temperature measurements as a function of frequency, averaged over the
sky.  Since the global signal is constant over different large patches
of the sky, experimental efforts to measure it do not need high
angular resolution and can be carried out with just a single dipole.
The attempted measurement is complicated however by the need to remove
galactic foregrounds, which are much larger than the desired signal.
Foreground removal is predicated on the assumption of spectral
smoothness of the foregrounds in contrast to the frequency structure
of the signal.  This should allow removal of the foregrounds by, for
example, fitting a low order polynomial to the foregrounds leaving the
21 cm signal in the residuals.  This methodology requires very precise
calibration of the instrumental frequency response, which could
otherwise become confused with the foregrounds.

The first experimental efforts to detect the 21 cm global signal have
been carried out by the COsmological Reionization Experiment (CORE)
\cite{chippendale2005} and the Experiment to Detect the Reionization
Step (EDGES) \cite{bowman2008}.  These have been analysed using a tanh
model of reionization that depends upon the redshift of reionization
$z_r$ and its duration $\Delta z$.  EDGES is presently able to rule
out the most rapid models of reionization that occur over a redshift
interval as short as $\Delta z<0.06$ \cite{bowman2010}.  These first
experimental efforts should be seen as the first steps along a road that may lead to considerably better constraints.  Other experiments using different experimental approaches are underway. Some of these use individual dipoles, such as the Shaped Antenna measurement of the RAdio Spectrum (SARAS) (R. Subrahmanyan,
private communications) and the Broadband Instrument for the Global HydrOgen ReionizatioN Signal (BIGHORNS) (S. Tingay, private communications), while others are exploring ways of using many dipoles as with the Large-aperture Experiment to Detect the Dark Ages (LEDA) (L. Greenhill, private communications).

Theoretical estimates for the ability of a single dipole experiment to
constrain models of the 21 cm signal can be made via the Fisher matrix
formalism \cite{EHT99}.  For a single dipole experiment, the Fisher
matrix may be written as \cite{pritchard2010}
\begin{equation}
F_{ij}=\sum_{n=1}^{N_{\rm channel}}\left(2+Bt_{\rm int}\right)\frac{\ud\log T_{\rm sky}(\nu_n)}{\ud p_i}\frac{\ud\log T_{\rm sky}(\nu_n)}{\ud p_j},
\end{equation}
where $t_{int}$ is the total integration time (before systematics
limit the performance), and we divide the total bandwidth $B$ into
$N_{\rm channel}$ frequency bins \{$\nu_n$\} running between
[$\nu_{\rm min}$, $\nu_{\rm max}$].  For the 21 cm global signature,
our observable is the antennae temperature $T_{\rm sky}(\nu)=T_{\rm
fg}(\nu)+T_{b}(\nu)$, where we assume the dipole sees the full sky so
that spatial variations can be ignored.  Best case errors on the
parameters $\{p_i\}$, which include both foreground and signal model
parameters, are then given by $\sigma_i\le\sqrt{F^{-1}_{ii}}$.

\begin{figure}[htbp]
\begin{center}
\includegraphics[scale=0.5]{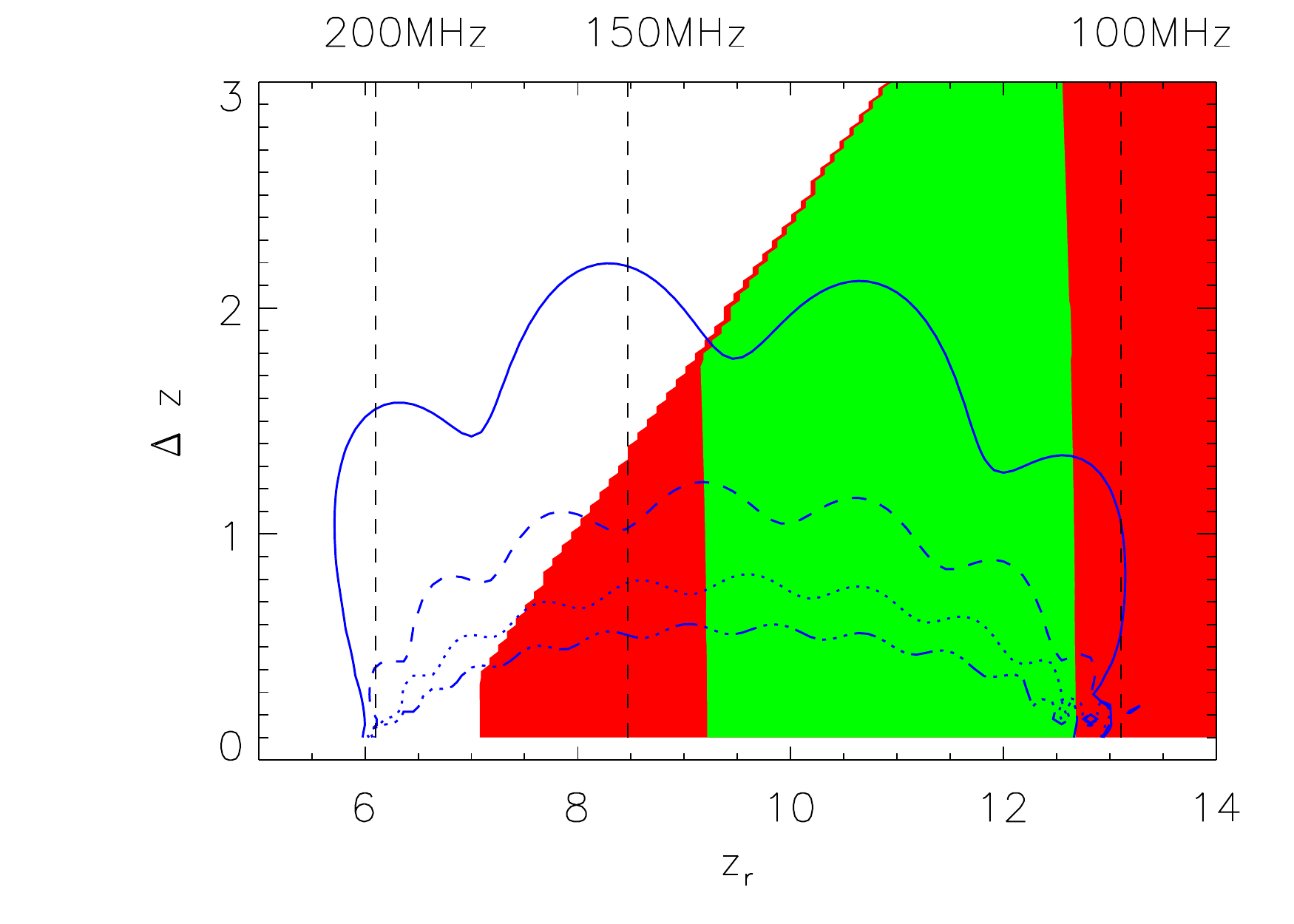}
\caption{95\% constraint region on a ``{\it tanh}'' reionization model
$T_b(z)=T_{21}\tanh[(z-z_r)/\Delta z]$ of the end of reionization for
an EDGES-like experiment assuming $N_{\rm poly}=3$ (solid curve), 6
(dashed curve), 9 (dotted curve), and 12 (dot-dashed curve). Also
plotted are the 68 and 95\% contours for the WMAP5 electron scattering
optical depth constraint combined with a prior that $x_i(z=6.5)>0.95$
(green and red coloured regions). \cite{pritchard2010}}
\label{fig:zr_dz}
\end{center}
\end{figure}
Such estimates show that global 21 cm experiments should be able to
constrain realistic reionization models with $\Delta z\lesssim2$
\cite{pritchard2010,morandi2011}.  The results of integrating for 500
hours between 100-200 MHz with a single dipole are shown in Figure
\ref{fig:zr_dz}, where the reionization history has been parametrized
with a {\it tanh} function.

\begin{figure}[htbp]
\begin{center}
\includegraphics[scale=0.4]{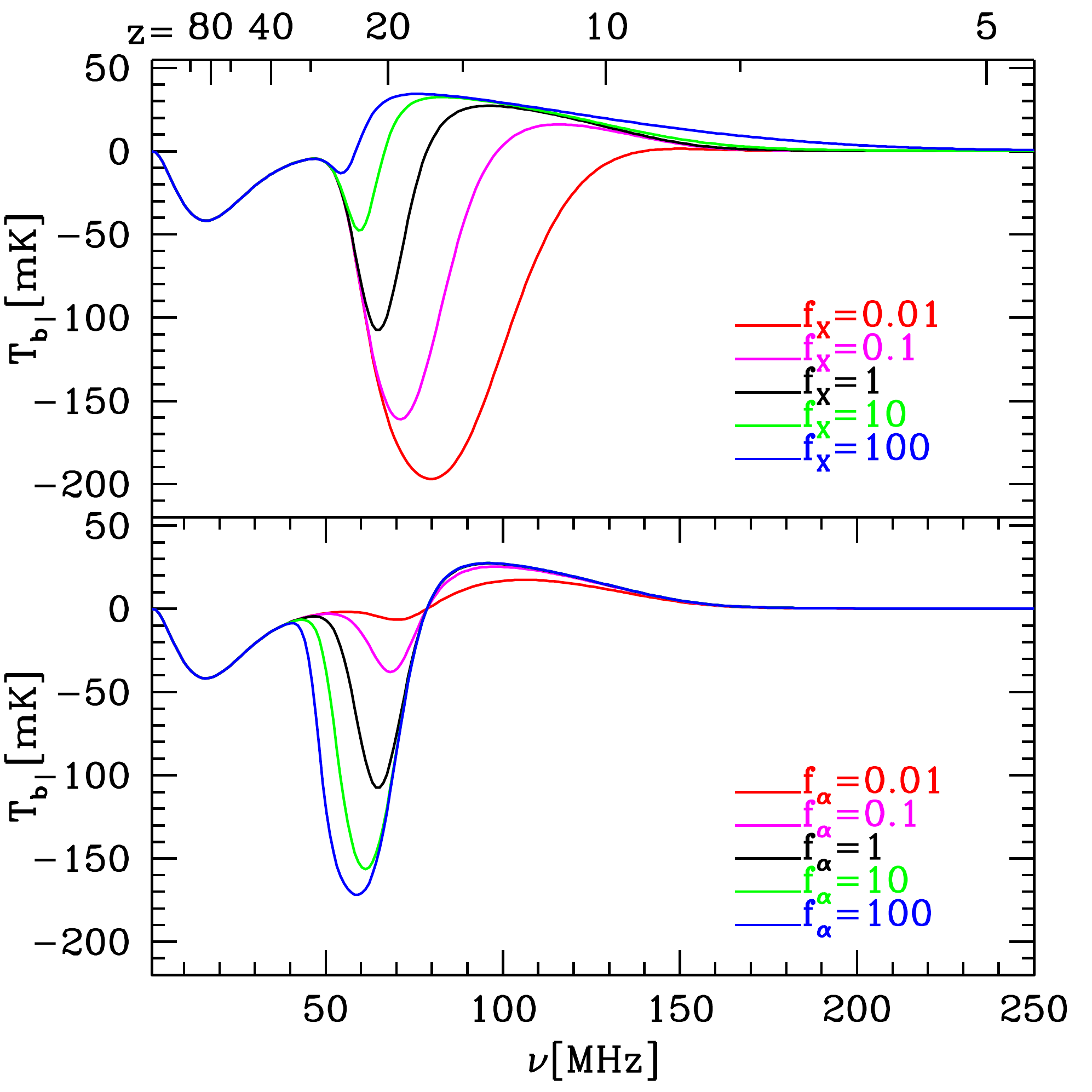}
\caption{Dependence of 21 cm signal on the X-ray (top panel) and \lya (bottom panel) emissivity.  In each case, we consider examples with the emissivity reduced or increased by a factor of up to 100.  Note that in our model $f_X$ and $f\alpha$ are really the product of the emissivity and the star formation efficiency.}
\label{fig:global_fx_flya}
\end{center}
\end{figure}

In addition to constraining reionization, global 21 cm experiments
might be used to probe the thermal evolution of the IGM at redshifts
$z>12$.  Such high redshifts are very difficult to probe via the 21 cm
fluctuations (discussed later) since they require very large
collecting areas.  Global experiments bypass this requirement, but
still suffer from the larger foregrounds at lower frequencies.  By
going to high redshifts such experiments could place constraints on
X-ray heating and \lya coupling giving information about when the
first black holes and galaxies form, respectively.  The absorption
feature resulting from this physics can potentially be larger
($\sim100$ mK) making it a good target for observations.  Figure \ref{fig:global_fx_flya} shows how the global 21 cm signal can vary with different values of $f_X$ and $f_\alpha$.  Measuring the global signal would offer a useful avenue for distinguishing these models although there is some degeneracy between the two parameters.  

EDGES type experiments at frequencies $\nu<100{\rm\,MHz}$ are underway
from the ground and a lunar orbiting
dipole experiment - the Dark Ages Radio Experiment (DARE)
\cite{burns2011} - has also been proposed.  Lunar orbit offers a number of advantages including being shielded from terrestrial radio frequency interference (RFI) while on the far side of the moon and the ability to use the moon to
chop the beam aiding instrumental calibration \cite{harker2011}.  DARE would be targeted at the 40-120 MHz range ideal for measuring the deep absorption feature and determining $f_X$ and $f_\alpha$.

\section{21 cm tomography}
\label{sec:fluctuations}

The 21 cm global signal can be viewed as a zeroth order approximation
to the full 21 cm signal, as it is averaged over large angular scales.
The full 3D signal will be highly inhomogeneous as a result of the
spatial variation in the different radiation fields and properties of
the IGM.  In this section, we consider the physics underlying 21 cm
brightness fluctuations in 3D and detail the existing techniques for
calculating the statistical properties of the signal.

Ultimately, one might wish to make use of fully numerical simulations
of the relevant physics and so produce detailed maps of the 21 cm
signal along thelight cone.  At present, the large dynamic range
required and the computational cost make this a dream for the future.
For the moment, it is important to make use of a variety of analytic,
semi-numerical, and numerical techniques to calculate the expected 21
cm signal.  These different methods complement one another in speed,
accuracy, and detail as we describe below.

Fluctuations in the 21 cm signal may be expanded to linear order \cite{fob}
\begin{equation}\label{deltaTb}
\delta_{T_b}=\beta_b\delta_b+\beta_x\delta_x+\beta_\alpha\delta_\alpha+\beta_T\delta_T-\delta_{\partial v},
\end{equation}
where each $\delta_i$ describes the fractional variation in the
quantity $i$ and we include fluctuations in the baryon density (b),
neutral fraction (x), \lya coupling coefficient ($\alpha$), gas
temperature (T), and line-of-sight peculiar velocity gradient
($\partial v$). The expansion coefficients are given by
\begin{eqnarray}
\beta_b&=&1+\frac{x_c}{x_{\rm{tot}}(1+x_{\rm{tot}})},\\
\beta_x&=&1+\frac{x_c^{HH}-x_c^{eH}}{x_{\rm{tot}}(1+x_{\rm{tot}})},\nonumber\\
\beta_\alpha&=&\frac{x_\alpha}{x_{\rm{tot}}(1+x_{\rm{tot}})},\nonumber\\
\beta_T&=&\frac{T_\gamma}{T_K-T_\gamma}
+\frac{1}{x_{\rm{tot}}(1+x_{\rm{tot}})}\left(x_c^{eH}\frac{\deriv \log\kappa_{10}^{eH}}{\deriv \log T_K}+x_c^{HH}\frac{\deriv \log\kappa_{10}^{HH}}{\deriv \log T_K}\right)\nonumber.
\end{eqnarray}
In this expression, we have treated all the terms as being of a
similar size, but it is important to realise that fluctuations in
$x_H$ can be of order unity.  This means that terms in higher order of
$\delta_x$, which one might naively think to be small, can still
contribute at a significant level to the power spectrum.

In general, homogeneity and isotropy of the Universe suggest that the power spectrum of brightness temperature fluctuations should be spherically symmetric in Fourier space i.e. it should only depend on $k=|\mathbf{k}|$ for a wavevector $\mathbf{k}$ of a given Fourier mode.  However, redshift space distortions induced by peculiar velocities break this symmetry since the direction to the observer becomes important and so only a cylindrical symmetry is preserved.  This symmetry may be useful in separating signal from foregrounds, which typically do not share this symmetry (see e.g. Ref \cite{morales_hewitt2004}). In Fourier space and at linear order, we may write the peculiar velocity term $\delta_{\partial
v}=-\mu^2\delta$ \cite{bharadwaj2004}, where $\mu$ is cosine of the
angle between the line of sight and the wavevector $\mathbf{k}$ of the
Fourier mode.  With this, we may use \eref{deltaTb} to form the power spectrum
\begin{eqnarray}\label{Pexpansion}
P_{T_b}(k,\mu)&=&P_{bb}+P_{xx}+P_{\alpha\alpha}+P_{TT}+2P_{bx}\nonumber\\
&&+2P_{b\alpha}+2P_{bT}+2P_{x\alpha}+2P_{xT}+2P_{\alpha T}\nonumber\\
&&+P_{x\delta x\delta}+ {\rm other\,quartic\,terms}\nonumber\\
&+&2\mu^2\left(P_{b\delta}+P_{x\delta}+P_{\alpha\delta}+P_{T\delta}\right) \nonumber\\
&+&\mu^4P_{\delta\delta}\nonumber\\
&+&2P_{x\delta\delta_{\partial v} x}+P_{x\delta_{\partial v}\delta_{\partial v} x}\nonumber\\ 
&&+{\rm other\,quartic\,terms\, with\, }\delta_{\partial v}.
\end{eqnarray}
Here we note that all quartic terms must be quadratic in $x_H$ and separate them depending upon whether they contain powers of $\delta_{\partial v}$ or not.  Those that contain powers of $\delta_{\partial v}$ will not be isotropic and will lead to the angular dependence of $P_{T_b}$ (see Ref. \cite{mcquinn2005} for further discussion).  

We may rewrite Eq. \eref{Pexpansion} in more compact form
\begin{equation}
P_{T_b}(k,\mu)=P_{\mu^0}(k)+\mu^2P_{\mu^2}(k)+\mu^4 P_{\mu^4}(k)+P_{f(k,\mu)}(k,\mu),
\end{equation}
where we have grouped those quartic terms with anomalous $\mu$
dependence into the term $P_{f(k,\mu)}(k,\mu)$.  In principle, high
precision measurements of the 3D power spectrum will allow the
separation of $P_{T_b}(k,\mu)$ into these four terms by their angular
dependence on powers of $\mu^2$ \cite{bl2005sep}.  The contribution of
the $P_{f(k,\mu)}(k,\mu)$ term, with its more complicated angular
dependence, threatens this decomposition \cite{mcquinn2005}.  Since
this term is only important during the final stages of reionization,
we will not discuss it in detail in this paper noting only that the
angular decomposition by powers of $\mu^2$ may not be possible when
ionization fluctuations are important.

It is unclear whether the first generation of 21 cm experiments will
be able to achieve the high signal-to-noise required for this
separation \cite{mcquinn2005}.  Instead, they might measure the angle
averaged quantity
\begin{equation}\label{Pangle_avg}
\bar{P}_{T_b}(k)=P_{\mu^0}(k)+P_{\mu^2}(k)/3+P_{\mu^4}(k)/5 ,
\end{equation}
(where we neglect the $P_{f(k,\mu)}(k,\mu)$ term). One typically
plots the power per logarithmic interval $\Delta=[k^3
P(k)/2\pi^2]^{1/2}$.

\subsection{Redshift space distortions}

Peculiar velocity effects can have a significant effect on the 21 cm
signal.  At linear order, the effects of peculiar velocities are well
understood \cite{bharadwaj2004,bl2005sep,wang2006hu} since the
$\delta_{\partial v}$ term is simply related to the total density
field.  However, as the density field evolves and non-linear
corrections to the velocity field become important the picture can
change in ways that are not yet well understood.

Redshift space effects become important because our observations are
made in frequency space, while theory makes predictions most directly
in coordinate space.  The conversion between the two is affected by
the local bulk velocity of the gas.  We may write the comoving distance
to an object as
\begin{equation}
\chi_z=\int_0^z\frac{c\, \ud z'}{(1+z')\mathcal{H}(z')},
\end{equation}
where we introduce $\mathcal{H}$ as the comoving Hubble parameter.
Using $\mathcal{H}$ makes the notion compact and comes from
introducing the conformal time coordinate $\eta$ related to the proper
time $t$ via $\ud\eta=\ud t/a(t)$, where $a(t)$ is the usual scale
factor, so that the comoving Hubble parameter $\mathcal{H}=(1/a)\ud
a(\eta)/\ud\eta$.

Including the effects of peculiar velocity, the true coordinate
distance to an object with measured redshift $z$ is
\begin{equation}
\chi=\chi_z-\mathbf{v}(\mathbf{x})\cdot\hat{\mathbf{n}}/\mathcal{H}|_z.
\end{equation}
Writing our coordinates as $\mathbf{x}=\chi\hat{\mathbf{n}}$ in real
space and $\mathbf{s}=\chi_z\hat{\mathbf{n}}$ in redshift space gives
the mapping between the two as
\begin{equation}
\mathbf{s}=\mathbf{x}+[\mathbf{v}(\mathbf{x})\cdot\hat{\mathbf{n}}/\mathcal{H}]\hat{\mathbf{n}}
\end{equation}
Accounting for this difference in the coordinate systems leads to the
redshift space distortions.  In linear theory, we have
\begin{equation}
\nabla\cdot\mathbf{v}(\mathbf{x})= -\mathcal{H}\delta(\mathbf{x}),
\end{equation}
where we are assuming the Universe to be matter dominated so that the
growth factor $f\equiv\ud\log D_+/\ud\log a=1$.  The Fourier transform
of this result introduces the factor of $\mu^2$.  More generally,
large peculiar velocities can lead to the so-called ``finger of god''
effects from virialised structures and greatly complicate efforts to
separate components via the angular structure of the power spectrum
\cite{mellema2006,shaw2008,mao2011}.  Mistakenly attempting to use the
form \eref{Pangle_avg} would lead to significantly biased results
\cite{shaw2008}, and so new estimators calibrated by simulations are
needed.

\subsection{Ionization fluctuations} 
\label{ssec:fluctuate:ion}

Reionization is a complicated process involving the balancing of
ionizing photons originating in highly clustered collections of
galaxies and recombinations in dense clumps of matter.  It is perhaps
surprising then that one can produce remarkably robust models of the
topology of reionization by simply counting the number of ionizing
photons.  This basic insight lies at the center of the analytic
calculation of ionization fluctuations \cite{fzh2004,barkana2004}.

Imagine a spherical region of gas containing a total mass of ionized
gas $m_{\rm ion}$.  As an ansatz for determining whether this region
of gas will be ionized, we can ask whether the region contains a
quantity of galaxies sufficient to ionize it.  Connecting to the
language we set up earlier when discussing the 21 cm global signal, we
ask whether the following condition is satisfied
\begin{equation}
m_{\rm ion}\ge \zeta m_{\rm gal},
\end{equation}
where $\zeta$ is the ionizing efficiency and $m_{\rm gal}$ is the
total mass in galaxies.  Note this is essentially the condition that
the galaxies produce enough ionizing photons to have ionized all the
gas within the region.

This condition equates to asking if the collapse fraction exceeds a
critical value $f_{\rm coll}\ge f_x\equiv\zeta^{-1}$.  From the
definition of the collapse fraction (making use of the Press-Schechter \cite{ps1974mfn}
mass fraction for analytic simplicity), these can be translated into a
condition on the mass overdensity if a region is to self ionize
\begin{equation}
\delta_m\ge\delta_x(m,z)\equiv\delta_c(z)-\sqrt{2}K(\zeta)\left[\sigma_{\rm min}^2-\sigma^2(m)\right]^{1/2},
\end{equation}
where $K(\zeta)={\rm erf^{-1}}(1-\zeta^{-1})$.  This condition to
self-ionization can be used to calculate the probability distribution
of ionized regions or bubble sizes $n_{\rm bub}(m)$ by reference to
the excursion set formalism \cite{bond1991}.  Smoothing a Gaussian
density field on decreasing mass scales corresponds to a random walk
in overdensity.  Once the overdensity for a region crosses the
ionization threshold, the mass enclosed will be ionized.  This is
similar to the mass function calculations of Press-Schechter or
Sheth-Tormen, except with a mass dependent rather than constant
barrier.  The distribution of bubbles sizes found from this analytic
calculation has been shown to be a good match to numerical
reionization simulations at the same neutral fraction
\cite{zahn2006,zahn2011}.

To connect the bubble distribution (a one-point statistic) to the
power spectrum (a two-point statistic) requires extra thought.  It is
possible to generalise the basic excursion set formalism to keep track
of two correlated random walks corresponding to two spatially
separated locations.  This very directly gives the two-point
correlation function for the ionization (or density) field
\cite{scannapieco2002,barkana2007}.  Unfortunately the resulting
expressions are somewhat complicated to deal with and simpler more
approximate calculations can be more useful.  We follow
Ref. \cite{fzh2004}, which incorporates some simple ansatzes for the
form of $P_{xx}$ and $P_{x\delta}$ based upon the expected clustering
properties of the bubbles.

\subsection{Fluctuations in the coupling} 
\label{sec:fluctuate_alpha}

Next we consider fluctuations in the \lya coupling
\cite{bl2005detect,pritchard2006}.  Provided that we neglect the mild
temperature dependence of $S_\alpha$ \cite{furlanetto2006heat},
the fluctuations in the coupling are simply sourced by fluctuations in
the flux and we may write $\delta_\alpha=\delta_{J_\alpha}$.

Density perturbations at redshift $z'$ source fluctuations in
$J_\alpha$ seen by a gas element at redshift $z$ via three effects.
First, the number of galaxies traces, but is biased with respect to,
the underlying density field.  As a result an overdense region will
contain a factor $[1+b(z')\delta]$ more sources, where $b(z')$ is the
(mass-averaged) bias, and will emit more strongly.  Next, photon
trajectories near an overdense region are modified by gravitational
lensing, increasing the effective area by a factor $(1+2 \delta/3)$.
Finally, peculiar velocities associated with gas flowing into
overdense regions establish an anisotropic redshift distortion, which
modifies the width of the region contributing to a given observed
frequency.  Given these three effects, we can write
$\delta_{\alpha}=\delta_{J_\alpha}=W_\alpha(k)\delta$, where we
compute the window function $W_{\alpha,\star}(k)$ for a gas element at
$z$ by adding the coupling due to \lya flux from each of the \lyn
resonances and integrating over radial shells \cite{bl2005detect},

\begin{eqnarray}
\label{wk}
W_{\alpha,\star}(k)&=&\frac{1}{J_{\alpha,\star}}\sum_{n=2}^{n_{\rm{max}}}\int^{z_{\rm{max}}(n)}_z \deriv z' \frac{\deriv J^{(n)}_\alpha}{\deriv z'}\nonumber\\
&&\times\frac{D(z')}{D(z)} \left\{[1+b(z')]j_0(kr)-\frac{2}{3}j_2(kr)\right\},
\end{eqnarray}
where $D(z)$ is the linear growth function, $r=r(z,z')$ is the
distance to the source, and the $j_l(x)$ are spherical Bessel
functions of order $l$.  The first term in brackets accounts for
galaxy bias while the second describes velocity effects. The ratio
$D(z')/D(z)$ accounts for the growth of perturbations between $z'$ and
$z$. Each resonance contributes a differential comoving \lya flux $\deriv J_\alpha^{(n)}/\deriv z'$, calculated from equation
\eref{jaflux}.  This analytic prescription has been bound to match later simulations fairly well \cite{santos2011}.  At present, simulations generally do not account for the full \lya radiative transfer so this agreement is not unexpected and comparisons to future, more detailed simulations will be needed.

On large scales, $W_{\alpha,\star}(k)$ approaches the average bias of
sources, while on small scales it dies away rapidly encoding the
property that the \lya flux becomes more uniform.  In addition to the
fluctuations in $J_{\alpha,\star}$, there will be fluctuations in
$J_{\alpha,X}$.  We discuss these below, but note in passing that the
effective value of $W_\alpha$ is the weighted average $W_\alpha=\sum_i
W_{\alpha,i}(J_{\alpha,i}/J_\alpha)$ of the contribution from stars
and X-rays.

This description of the coupling fluctuations neglects one important
piece of physics.  Namely, it assumes that UV photons redshift until
they reach the line center of a Lyman series resonance and only then
they scatter.  At redshifts before reionization the Universe is filled
with neutral hydrogen leading to an optical depth for scattering of
\lya photons that is very large $\tau_\alpha\gtrsim10^5$.  While this
is reduced for higher series transitions, the Universe is optically
thick for all but the highest $n$ transitions.  This has the
consequence that photons will preferentially scatter in the wings of
the line and only a few will make it to line center.

The related implications have been investigated in the literature
\cite{chuzhoy2007,semelin2007,naoz2008}.  Since UV photons scatter in
the wings of the line they will travel a significantly reduced
distance from the source before scattering.  This reduces the size of
the coupled region around a source and acts to steepen the flux
profile surrounding a source.  These effects increase the variance in
the \lya flux on small scales.  This can be incorporated into our
analytic formalism as a modification to the $\ud J_\alpha^{(n)}/\ud
z'$ term in \eref{wk} to account for the modified flux profile around
individual sources.  For our purposes, we will neglect this since it
adds considerable numerical complication without modifying the
qualitative features of the power spectrum.

Once the effect of the optical depth is taken into account, another
limitation of this approach becomes apparent.  We have assumed that
the photons propagate through a uniform IGM with mean properties.
Density inhomogeneities and especially velocity flows may modify the
propagation and scattering of UV photons leading to extra fluctuations
that have not been accounted for here.

This becomes especially apparent when one recalls that the sources are
placed in ionized bubbles.  This means that on small scales there are
regions where there is no coupling because there is no neutral
hydrogen.  A simple way to account for this is by changing the lower
limit of \eref{wk} to $z_{\rm HII}$, the typical size of an ionized
region \cite{naoz2008}.  Clearly, if the patch of gas being considered
is closer to a source than the HII region that surrounds that source
then the patch will be ionized and there will be no signal.  This
leads to a reduction in power on small scales.  For the sharp cutoff
of \cite{naoz2008} one numerically sees oscillation in the power
spectrum, but in reality averaging over a distribution of ionized
bubble sizes would remove these oscillations yielding a smooth
reduction in power.

\subsection{Formalism for temperature and ionization fluctuations from X-rays} 
\label{sec:fluctuate}

We next turn to fluctuations in the gas temperature and the free
electron fraction in the IGM.  Following on from our discussion of the
global history, we will assume that the evolution of $T_K$ and $x_e$
is driven by the effect of X-rays and will consider the fluctuations
$\delta_T$, and $\delta_e$ that arise from clustering of the X-ray
sources.  In doing so, we will follow the approach of
Ref. \cite{pritchard2007xray}.

Before plunging into the calculation, it is worth detailing some of the issues that face radiative transfer of X-ray photons.  The
cross-section for X-ray photoionization depends sensitively on photon energy $E$, $\sigma\sim E^{-3}$, so that we must keep track of separate energies.  Most importantly, this energy dependence means that the IGM is optically thick for soft X-rays ($E\sim20\eV$), but optically thin for hard X-rays ($E\gtrsim1\keV$).  Moreover, heating is a continuous process and the temperature of a gas element depends on its past history.  This is different from UV coupling where only the \lya flux at a given redshift is important.  We must therefore be careful to track the change in fluctuations in the heating and integrate these to get the temperature fluctuations at a later time. Insight can be gained by looking at the results of 1D numerical simulations of X-ray radiative transfer \cite{venkatesan2001,zaroubi2007}.  

The prescription we adopt here describes X-ray fluctuations produced by the clustering of X-ray sources.  We neglect the possibility of Poisson fluctuations in the distribution of X-ray sources, since it is not clear how to calculate this.  Poisson fluctuations in the \lya flux were considered in \cite{bl2005detect}, but the formalism is not readily applicable to heating.  As a result, this formalism is most appropriate in scenarios where X-ray sources are relatively common and would not be appropriate for describing heating by extremely rare quasars \cite{alvarez2010}.

We begin by forming equations for the evolution of $\delta_T$ and
$\delta_{e}$ (the fractional fluctuation in $x_e$) by perturbing
equations \eref{thistory} and \eref{xehistory} (see also
\cite{bl2005infall,naoz2005}).  This gives
\begin{equation}
\frac{\deriv \delta_T}{\deriv t}-\frac{2}{3}\frac{\deriv \delta}{\deriv t}=\sum_i\frac{2\bar{\Lambda}_{{\rm heat},i}}{3k_B\bar{T}_K} [\delta_{\Lambda_{{\rm heat},i}}-\delta_T],\label{evolve_deltaT}
\end{equation}
\begin{equation}
\frac{\deriv \delta_e}{\deriv t}=\frac{(1-\bar{x}_e)}{\bar{x}_e}\bar{\Lambda}_e[\delta_{\Lambda_e}-\delta_e]-\alpha_A C \bar{x}_e\bar{n}_H[\delta_e+\delta],\label{evolve_deltax}
\end{equation}
where an overbar denotes the mean value of that quantity, and
$\Lambda=\epsilon/n$ is the ionization or heating rate per baryon.  We
note that the fluctuation in the neutral fraction $\delta_x$ is simply
related to the fluctuation in the free electron fraction by
$\delta_x=-x_e/(1-x_e)\delta_{e}$.

Our challenge then is to fill in the right hand side of these
equations by calculating the fluctuations in the heating and ionizing
rates.  We will focus here on Compton and X-ray heating processes.
Perturbing equation \eref{compton} we find that the contribution of
Compton scattering to the right hand side of equation
\eref{evolve_deltaT} becomes \cite{naoz2005}
\begin{eqnarray}
\label{compton_perturb}
\frac{2\bar{\Lambda}_{{\rm heat,C}}}{3k_B\bar{T}_K}[\delta_{\Lambda_{{\rm heat,C}}}-\delta_T]&=&\frac{\bar{x}_e}{1+f_{\rm{He}}+\bar{x}_e}\frac{a^{-4}}{t_\gamma}  \nonumber \\
&&\times \left[
4\left(\frac{\bar{T}_\gamma}{\bar{T}_K}-1\right)\delta_{T_\gamma}
+\frac{\bar{T}_\gamma}{\bar{T}_K}(\delta_{T_\gamma}-\delta_T)
\right],
\end{eqnarray}
where $\delta_{T_\gamma}$ is the fractional fluctuation in the CMB
temperature, and we have ignored the effect of ionization variations
in the neutral fraction outside of the ionized bubbles, which are
small.  Before recombination, tight coupling sets $T_K=T_\gamma$ and
$\delta_T=\delta_{T_\gamma}$.  This coupling leaves a scale dependent
imprint in the temperature fluctuations, which slowly decreases in
time.  We will ignore this effect, as it is small ($\sim10\%$) below
$z=20$ and once X-ray heating becomes effective any memory of these
early temperature fluctuations is erased.  At low $z$, the amplitude
of fluctuations in the CMB $\delta_{T_\gamma}$ becomes negligible, and
equation \eref{compton_perturb} simplifies.  Our main challenge then
is to calculate the fluctuations in the X-ray heating.  We shall
achieve this by paralleling the approach we took to calculating
fluctuations in the \lya flux from a population of stellar sources.

First, note that for X-rays
$\delta_{\Lambda_{\rm{ion}}}=\delta_{\Lambda_{\rm{heat}}}=\delta_{\Lambda_\alpha}=\delta_{\Lambda_X}$,
as the rate of heating, ionization, and production of \lya photons
differ only by constant multiplicative factors (provided that we may
neglect fluctuations in $x_e$, which are small, and focus on X-ray
energies $E\gtrsim100\eV$).  In each case, fluctuations arise from
variation in the X-ray flux.  We then write
$\delta_{\Lambda_X}=W_X(k)\delta$ and obtain,

\begin{eqnarray}
\label{wk_xray}
W_{X}(k)&=&\frac{1}{\bar{\Lambda}_{X}}\int_{E_{\rm{th}}}^{\infty} \deriv E\,\int^{z_{\star}}_z \deriv z' \frac{\deriv \Lambda_X(E)}{\deriv z'}\nonumber\\ 
&&\times\frac{D(z')}{D(z)} \left\{[1+b(z')]j_0(kr)-\frac{2}{3}j_2(kr)\right\},
\end{eqnarray}
where the contribution to the energy deposition rate by X-rays of energy $E$ emitted with energy $E'$ from between redshifts $z'$ and $z'+\deriv z'$ is given by
\begin{equation}
\frac{\deriv \Lambda_X(E)}{\deriv z'}=\frac{4\pi}{h}\sigma_{\nu}(E)\frac{\deriv J_X(E,z)}{\deriv z'}(E-E_{\rm{th}}),
\end{equation}
where $\sigma_\nu(E)$ is the cross-section for photo-ionization,
$E_{\rm th}$ is the minimum energy threshold required for
photo-ionization, and $\bar{\Lambda}_X$ is obtained by performing the
energy and redshift integrals.  Note that rather than having a sum
over discrete levels, as in the \lya case, we must integrate over the
X-ray energies.  The differential X-ray number flux is found from
equation \eref{jxflux}.

The window function $W_X(k)$ gives us a ``mask" to relate fluctuations
to the density field; its scale dependence means that it is more than
a simple bias.  The typical sphere of influence of the sources extends
out to several Mpc.  On scales smaller than this, the shape of $W_X(k)$
will be determined by the details of the X-ray source spectrum and the
heating cross-section.  On larger scales, the details of the heated
regions remain unresolved so that $W_X(k)$ will trace the density
fluctuations.

An X-ray is emitted with energy $E'$ at a redshift $z'$ and redshifts
to an energy $E$ at redshift $z$, where it is absorbed.  To calculate
$W_X$ we perform two integrals in order to capture the contribution of
all X-rays produced by sources at redshifts $z'>z$.  The integral over
$z'$ counts X-rays emitted at all redshifts $z'>z$ which redshift to
an energy $E$ at $z$; the integral over $E$ then accounts for all the
X-rays of different energies arriving at the gas element.  Together,
these integrals account for the full X-ray emission history and source
distribution.  Many of these X-rays have travelled considerable
distances before being absorbed.  The effect of the intervening gas is
accounted for by the optical depth term in $J_X$.  Soft X-rays have a
short mean free path and are absorbed close to the source; hard
X-rays will travel further, redshifting as they go, before being
absorbed.  Correctly accounting for this redshifting when calculating
the optical depth is crucial as the absorption cross-section shows
strong frequency dependence.  In our model, heating is dominated by
soft X-rays, from nearby sources, although the contribution of harder
X-rays from more distant sources can not be neglected.

Returning now to the calculation of temperature fluctuations, to
obtain solutions for equations \eref{evolve_deltaT} and
\eref{evolve_deltax}, we let $\delta_T=g_T(k,z)\delta$,
$\delta_e=g_e(k,z)\delta$, $\delta_\alpha=W_\alpha(k,z) \delta$, and
$\delta_{\Lambda_X}=W_X(k,z)\delta$ \cite{bharadwaj2004}. Since we do
not assume these quantities to be independent of scale we must solve
the resulting equations for each value of $k$.  Note that we do not
include the scale dependence induced by coupling to the CMB
\cite{naoz2005}. In the matter dominated limit, we have
$\delta\propto(1+z)^{-1}$ and so obtain
\begin{equation}
\frac{\deriv g_T}{\deriv z}=\left(\frac{g_T-2/3}{1+z}\right)-Q_X(z)[W_{X}(k)-g_T]-Q_C(z)g_T,\label{evolve_gT2}
\end{equation}
\begin{equation}
\frac{\deriv g_e}{\deriv z}=\left(\frac{g_e}{1+z}\right)-Q_I(z)[W_{X}(k)-g_e]+Q_R(z)[1+g_e],\label{evolve_ge2}
\end{equation}
where we define
\begin{equation}
Q_I(z)\equiv\frac{(1-\bar{x}_e)}{\bar{x}_e}\frac{\bar{\Lambda}_{{\rm ion},X}}{(1+z)H(z)},
\end{equation}
\begin{equation}
Q_R(z)\equiv\frac{\alpha_A C \bar{x}_e\bar{n}_H}{(1+z)H(z)},
\end{equation}
\begin{equation}
Q_C(z)\equiv\frac{\bar{x}_e}{1+f_{{\rm He}}+\bar{x}_e}\frac{(1+z)^3}{t_\gamma H(z)}  \frac{T_\gamma}{\bar{T}_K},
\end{equation}
and
\begin{equation}
Q_X(z)\equiv\frac{2\bar{\Lambda}_{{\rm heat},X}}{3k_B\bar{T}_K(1+z)H(z)}.
\end{equation}
These are defined so that $Q_R$ and $Q_I$ give the fractional change
in $x_e$ per Hubble time as a result of recombination and ionization
respectively.  Similarly, $Q_C$ and $Q_X$ give the fractional change
in $\bar{T}_K$ per Hubble time as a result of Compton and X-ray
heating.  Immediately after recombination $Q_C$ is large, but it
becomes negligible once Compton heating becomes ineffective at
$z\sim150$.  The $Q_R$ term becomes important only towards the end of
reionization, when recombinations in clumpy regions slows the
expansion of HII regions.  Only the $Q_X$ and $Q_I$ terms are relevant
immediately after sources switch on.  We must integrate these
equations to calculate the temperature and ionization fluctuations at
a given redshift and for a given value of $k$.

These equations illuminate the effect of heating.  First, consider
$g_T$, which is related to the adiabatic index of the gas $\gamma_a$
by $g_T=\gamma_a-1$, giving it a simple physical interpretation.
Adiabatic expansion and cooling tends to drive $g_T\rightarrow2/3$
(corresponding to $\gamma_a=5/3$, appropriate for a monoatomic ideal
gas), but when Compton heating is effective at high $z$, it deposits
an equal amount of heat per particle, driving the gas towards
isothermality ($g_T\rightarrow0$).  At low $z$, when X-ray heating of
the gas becomes significant, the temperature fluctuations are
dominated by spatial variation in the heating rate ($g_T\rightarrow
W_X$).  This embodies the higher temperatures closer to clustered
sources of X-ray emission.  If the heating rate is uniform
$W_X(k)\approx0$, then the spatially constant input of energy drives
the gas towards isothermality $g_T\rightarrow0$.

The behaviour of $g_e$ is similarly straightforward to interpret.  At
high redshift, when the IGM is dense and largely neutral, the
ionization fraction is dominated by the recombination rate, driving
$g_x\rightarrow-1$, because denser regions recombine more quickly.  As
the density decreases and recombination becomes ineffective, the first
term of equation \eref{evolve_ge2} gradually drives $g_x\rightarrow0$.
Again, once ionization becomes important, the ionization fraction is
driven towards tracking spatial variation in the ionization rate
($g_x\rightarrow W_X$).  Note that, because the ionization fraction in
the bulk remains less than a few percent, fluctuations in the neutral
fraction remain negligibly small at all times.

Fluctuations in the temperature $g_T$ attempt to track the heating
fluctuations $W_X(k)$, but two factors prevent this.  First, until
heating is significant, the effect of adiabatic expansion tends to
smooth out variations in $g_T$.  Second, $g_T$ responds to the
integrated history of the heating fluctuations, so that it tends to
lag $W_X$ somewhat.  When the bulk of star formation has occurred
recently, as when the star formation rate is increasing with time,
then there is little lag between $g_T$ and $W_X$.  In contrast, when
the star formation rate has reached a plateau or is decreasing the
bulk of the X-ray flux originates from noticeably higher $z$ and so
$g_T$ tends to track the value of $W_X$ at this higher redshift. On
small scales, the heating fluctuations are negligible and $g_T$
returns to the value of the (scale independent) uniform heating case.

\subsection{Evolution of the full power spectrum}
\label{ssec:full_power}

Having described the different components of the 21 cm power spectrum,
we now need to put them together.  The 21 cm power spectrum is a 3D
quantity observed as a function of scale and redshift, much like a
movie evolving with time on a 2D screen.  Displaying this information
on a static 2D paper is therefore challenging.

\begin{figure}[htbp]
\begin{center}
\includegraphics[scale=0.4]{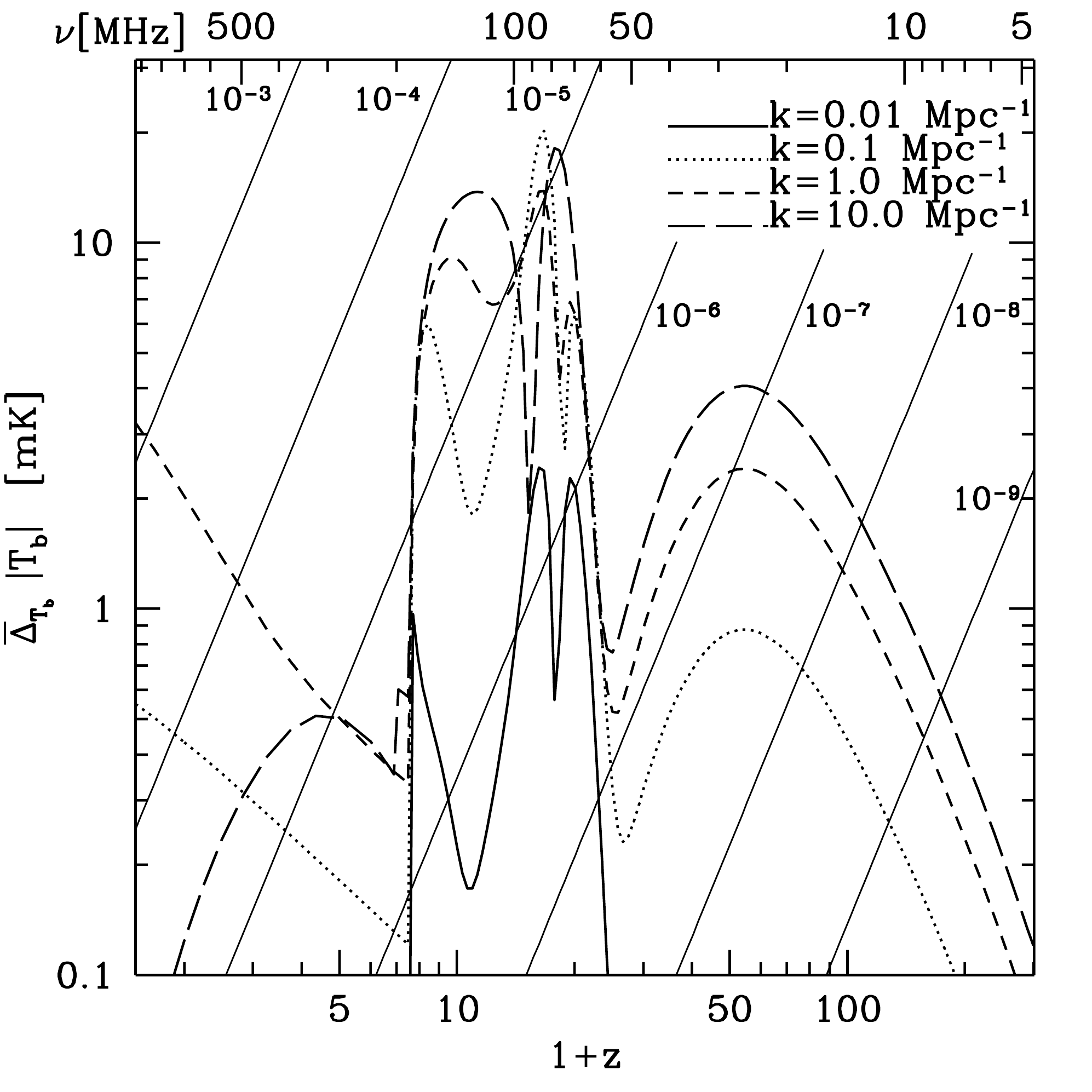}
\caption{Evolution of power spectrum fluctuations.  The different
curves show $P(k,z)$ as a function of $z$ at fixed $k$ for $k=0.01$,
0.1, 1, 10 $\iMpc$.  Diagonal lines show $\epsilon T_{\rm fg}(\nu)$,
the foreground temperature reduced by a factor $\epsilon$ ranging from
$10^{-3}-10^{-9}$ to indicate the level of foreground removal required
to detect the signal \cite{pritchard2008}.}
\label{fig:pl_fluctuations}
\end{center}
\end{figure}

Figure \ref{fig:pl_fluctuations} shows the evolution of the power
spectrum as a function of redshift for several fixed $k$-values.  Four
key epochs can be picked out.  At early times $z\gtrsim30$ before star
formation, the power spectrum rises towards a peak at $z\approx50$ and
falls thereafter as the 21 cm power spectrum tracks the density field
modulated by the mean brightness temperature.  Once stars switch on,
there is a period of complicated evolution as coupling and temperature
fluctuations become important.  Next, ionization fluctuations become
important culminating in the loss of signal at the end of
reionization.  Thereafter, a weaker signal arises from the remaining
neutral hydrogen in dense clumps that grows as structures continue to
collapse.

Diagonal lines in Figure \ref{fig:pl_fluctuations} show the amplitude
of the mean foreground reduced by a factor ranging from
$10^{-3}-10^{-9}$ (note that it is really fluctuations in the
foregrounds that set the upper limit, but this is less well known than
the mean level).  This gives a measure of the difficulty of removing
the foregrounds and the sensitivity of the instrument required for a
detection.  This diagonal lines serve to identify the most readily
accessible epochs.  The $10^{-5}$ line identifies the level required
to: {\em (i)} access density fluctuations at $z\lesssim4$; {\em (ii)}
access ionization fluctuations at $z\approx6-12$; and {\em (iii)}
detect fluctuations from heating and coupling at $z\approx20$.  It is
perhaps surprising that a similar sensitivity is needed for all three
of these regimes, since the foregrounds grow monotonically with
decreasing frequency. This is explained by the strong evolution in the
amplitude of the signal.  The fourth epoch at $z\approx30-50$ of the
dark ages is considerably more difficult to detect, requiring the
$10^{-7}$ level of foreground removal.

\subsection{Other sources of fluctuations} 
We have focussed on fluctuations in the 21 cm signal that arise from
spatial variation in three main radiation fields - UV, \lya, and
X-rays.  Other sources of fluctuations from the non-linear growth of
structure are possible.

Since the 21 cm signal arises from neutral hydrogen it is interesting
to examine the densest regions of neutral hydrogen.  These occur in
those dense clumps that have achieved the critical density for
collapse, but that lack the mass for efficient cooling of the gas that
would lead to star formation.  This requirement is satisfied for halos
of with virial temperature below $10^4\K$ provided that only atomic
hydrogen is available for cooling or for halos with $T_{\rm
vir}\lesssim300\K$ if molecular hydrogen is present.  These minihalos
should be abundant in the early Universe, although at later times
external ionizing radiation may cause them to evaporate.  Furthermore,
their formation may be prevented with moderate gas heating that raises
the Jean's mass suppressing the collapse of these low mass objects.
The high density in these regions implies that collisions can provide
the main coupling mechanism and due to the high temperature of the
virialised gas these minihalos are bright in 21 cm emission
\cite{iliev2002,iliev2003,furlanetto2006mh,shapiro2006}.

When baryons and photons decouple at $z\approx1,100$ there is a sudden
drop in the pressure supporting the baryons against gravitational
collapse and they begin to fall into dark matter overdensities.  It
was recently realised that the relative velocity between the baryons
and dark matter exceeds the local sound speed (which drops
dramatically from the relativistic value of $c/\sqrt{3}$ to $\sim
\sqrt{k_B T/m_p}$) leading to supersonic flows
\cite{tseliakhovich2010}.  These flows have the potential
to suppress the formation of the first gas clouds by preventing the
baryons from collapsing into dark matter halos with low escape
velocities \cite{tseliakhovich2010b}.  This suppression of galaxy
formation in minihalos may have important consequences for the 21 cm
signal during its earliest phases - for example, delaying the onset of
\lya coupling.  It has even been suggested that the increased
modulation of the earliest galaxies might boost the \lya coupling
fluctuations significantly \cite{dalal2010}.  Ultimately, it appears
that as the higher mass halos required for atomic cooling begin to
collapse, the memory of this effect is greatly reduced
\cite{stacy2011,maio2011}.  It is possible that by suppressing the
building blocks of larger galaxies mild echos of this effect might be
detectable in late galaxies in a similar way to the effects of
inhomogeneous reionization \cite{babich2006,pritchard2007bub,wyithe2007}.

In the local Universe, shocks are known to be an important mechanism for IGM heating.  Shocks heat the gas directly converting bulk motion into thermal energy. If magnetic fields are present in the shock, strong shocks can also accelerate charged particles.  This can lead to radiative emission of photons at energies from radio to X-ray energies due to inverse Compton scattering of CMB photons from the charged particle.  The resulting X-rays can further heat the IGM.  At high redshift, shocks can be broadly divided into two categories: strong shocks around large scale structure and weak shocks in the diffuse IGM due to the low sound speed of the gas.  

Large scale structure shocks occur as gas surrounding an overdense region decouples from the Hubble flow and undergoes turnaround.  As the gas begins to collapse inwards derivations from spherical symmetry will lead to shell crossing and shocking of the gas.  These shocks have been discussed both analytically \cite{furlanetto2004} and observed in simulations \cite{kuhlen2006,shapiro2006}.  The temperature distribution of shocked gas can be estimated using a Press-Schecter formalism under the assumption that all gas that has
undergone turn around has shocked and estimating the shock temperature from the characteristic peculiar velocity of the collapsing overdense region.  These models show that, in the absence of X-ray heating, the thermal energy of the IGM is dominated by large scale shocks.  Since these shocks trace the collapse of structure, they only become significant at redshifts $z\lesssim20$. These shocks cannot play an important role in ionizing the IGM, since only a small fraction of the baryons participate in them and they generate $\lesssim 1$eV per participating baryon at $z\gtrsim 10$ \cite{dopita2011}.

Finally, we note that there is one final radiation background that
could in principle lead to fluctuations in the 21 cm signal: the
diffuse radio background.  In our calculations, we have assumed that
the ambient 21 cm radiation field is dominated by the CMB, so that
$T_\gamma=T_{\rm CMB}$.  This is likely to be a good assumption, but
one can imagine a situation in which the 21 cm flux might be
influenced by nearby radio bright sources.  Since 21 cm photons have a
large mean free path a diffuse radio background may build up in the
same way as the diffuse X-ray background grows.  Fluctuations in this
radio background would arise from clustering of the radio sources.  We
note that the only period where these fluctuations would be important
are in the regime where \lya and collisional coupling are unimportant,
so that the spin temperature relaxes to the ambient radiation
temperature $T_\gamma$.

\subsection{Simulation techniques} 

In the previous sections we have focused on analytical descriptions of
calculating the 21 cm signal.  These provide a framework for
understanding the underlying physics that governs the signal.  They
also provide a method for rapidly exploring the dependence of the 21
cm power spectrum on a wide range of astrophysical parameters and
determining their relative importance.  Ultimately, the interpretation
of observations requires comparison of data to theoretical predictions
at a more detailed level.  This necessitates the production of maps
from numerical techniques.  In this section we describe a hierarchy of
different levels of numerical approximations that allow more
quantitative comparisons. A more detailed review can be found in
Ref. \cite{trac2009}.

Closest to the analytic techniques described in this review are
semi-numerical techniques.  These are primarily techniques for
simulating the reionization field based upon an extension of the
excursion set formalism \cite{bond1991,furlanetto2004,zahn2006}.  It
has been realised that the spectrum of ionized fluctuations depends
primarily on a single parameter - the ionized fraction $x_i$
\cite{mcquinn2007}.  Once $x_i$ is fixed the ionization pattern can be
calculated by filtering the density field on progressively smaller
scales and asking if a region is capable of self-ionization.  Only
those regions capable of self-ionization are taken to be ionized; this
amounts to photon counting.  This technique forms the basis of a
number of codes, 21cmFAST \cite{mesinger2010}, SIMFAST21
\cite{santos2010}, which can rapidly calculate the 21 cm signal with
reasonable accuracy.  To add in fluctuations in the \lya and
temperature these codes also evaluate the equivalent of the analytic
calculations from \S\ref{sec:fluctuate}.  The relevant expressions are
convolutions of the density field with a kernel describing the
astrophysics and so may be rapidly evaluated on a numerical grid via
fast Fourier transform (FFT) techniques.  Other semi-numerical
techniques exist such as BEARS \cite{thomas2009}, which is based upon
painting spherically symmetric ionization, heating, or coupling
profiles from a library of 1D radiative transfer calculations
appropriately scaled for a halos star-formation rate and formation
time onto the density field.

Fully numerical simulations are the ultimate destination for these
calculations, but as of yet it is difficult to achieve the required
simulation volume and dynamic range required for the full calculation
of the 21 cm signal.  Numerical simulations can be broadly split into
two classes: those that fully account for gas hydrodynamics and those
that do not.  The latter are essentially dark matter N-body
calculations with a prescription for painting baryons onto the
simulation in a post-processing step.  Radiative transfer
prescriptions are then applied in a further post-processing step to
calculate the evolution of the ionized regions.  Such calculations
\cite{mellema2006,mcquinn2007,trac2007,partl2011} are of great utility
in describing the large scale properties of the 21 cm signal.  Full
hydrodynamic calculations \cite{trac2008} 
are typically restricted to smaller volumes making them
unrepresentative of the cosmic volume.  They have the advantage of
self-consistently evolving the dark matter, baryons, and radiation
field.  This allows the evolution of photon sinks, in the form of
mini-halos and dense clumps, to be studied in detail.  As computers
improve these simulations will grow in size.

Finally, most of the simulation work to date has focussed on
reionization and made the assumption that $T_S\gg T_{\rm CMB}$
ignoring the effect of spin-temperature fluctuations.  For predicting
the 21 cm signal it is important that analytic calculations of these
$T_S$ variation be verified numerically.  Work on incorporating \lya
and X-ray propagation into reionization simulations exists, but is at
an early stage of development \cite{semelin2007,baek2009,baek2010}.
These calculations require keeping track of the radiative transfer of
photons in many frequency bins and so becomes numerically expensive.

\subsection{Detectability of the 21 cm signal}

Our focus in this review is on the physics underlying the 21 cm
signal, but it is appropriate to pause for a moment and consider the
instrumental requirements for detecting the signal.  We direct the
interested reader to the review in Ref. \cite{morales2010}, which
covers the near term prospects for detecting the 21 cm signal from
reionization in some detail.

The 21 cm line from the epoch of reionization is redshifted to meter wavelengths requiring radio frequency observations.  While a typical radio telescope is made of a single large dish, an interferometer composed of many dipole anntenae is the preferred design for 21 cm observations.  Cross-correlating the signals from individual dipoles allows a beam to be synthesised on the sky.  Using dipoles allows for arrays with very large collecting areas and a large field of view suitable for surveys.  This comes at the cost of large computational demands and, driven by the long wavelengths, relatively poor angular resolution ($\sim10$ arcmin, which corresponds to the predicted bubble size at the middle of reionization).

The sensitivity of these arrays is determined in part by the
distribution of the elements, with a concentrated core configuration giving the highest sensitivity to the power spectrum \cite{lidz2008}. The desire for longer baselines to boost the angular resolution in order to remove radio point sources places constraints on the compactness achievable in practice.

The variance of a 21 cm power spectrum measurement with a radio array for a single $\mathbf{k}$-mode with line of sight component $k_{||}=\mu k$ is given by \cite{lidz2008}:
\begin{equation}
\sigma_P^2(k,\mu)=\\ \frac{1}{N_{\rm field}}\left[\bar{T}_b^2P_{21}(k,\mu)+T_{\rm sys}^2\frac{1}{B t_{\rm int}}\frac{D^2\Delta D}{n(k_\perp)}\left(\frac{\lambda^2}{A_e}\right)^2\right]^2.
\end{equation}
We restrict our attention to modes in the upper-half plane of the wavevector ${\bf k}$, and include both sample variance and thermal detector noise assuming Gaussian statistics. The first term on the right-hand-side of the above expression provides the contribution from sample variance, while the second describes the thermal noise of the radio telescope.  The thermal noise depends upon the system temperature $T_{\rm sys}$, the survey bandwidth $B$, the total observing time $t_{\rm int}$, the comoving distance $D(z)$ to the center of the survey at redshift $z$, the depth of the survey $\Delta D$, the observed wavelength $\lambda$, and the effective
collecting area of each antennae tile $A_e$.   The effect of the configuration of the antennae is encoded in the number density of baselines $n_\perp(k)$ that observe a mode with
transverse wavenumber $k_\perp$ \cite{mcquinn2006}.  Observing a number of fields $N_{\rm field}$ further reduces the variance.




\subsection{Statistics beyond the power spectrum}
\label{ssec:nongaussian}

In our discussion of the 21 cm fluctuations, we have focussed on the power spectrum as a statistical quantity that can be measured from maps of the sky.  When experiments lack the sensitivity to image the 21 cm fluctuations at high signal to noise it is important to compress the information into statistical quantities that can be accurately measured.  If 21 cm fluctuations were Gaussian then the power spectrum would contain all information about the signal.  However, the 21 cm signal is highly non-Gaussian since the presence of ionized bubbles or spheres that have been heated imprints definite features in space.  It is therefore important to think about non-Gaussian statistics that can be used to extract information that is not contained in the power spectrum from 21 cm observations. 

The simplest form of non-Gaussianity that arises in the 21 cm signal is the primordial non-Gaussianity induced in the density field during inflation.  This is often characterised by the $f_{\rm NL}$ parameter defined by assuming a quadratic correction to the inflaton potential so that schematically $\phi=\phi_{\rm G}+f_{\rm NL}\phi_{\rm G}^2$, with $\phi$ the full inflaton potential and $\phi_{\rm G}$ the Gaussian potential.  This form of non-Gaussianity shows up clearly in the bispectrum, the Fourier transform of the 3pt correlation function \cite{maldacena2003}. Measuring $f_{\rm NL}$ is an important goal of cosmology since it can effectively distinguish between different classes of inflationary potential.  Unfortunately, $f_{\rm NL}$ is expected to be small ($\sim (1-n_s)\sim0.05$ for slow roll inflation models) and constraints from the CMB and galaxy surveys are relatively weak \cite{komatsu2009}.  One long term hope is that observations of the pristine 21 cm signal at $z>30$ could lead to very stringent non-Gaussianity constraints, since such surveys can probe very large volumes of space \cite{cooray2006ng,pillepich2007}.

In the near term, studies of 21 cm non-Gaussianity will be most useful for learning about astrophysics in more detail.  Ionized bubbles during reionization will imprint a particular form of non-Gaussianity on the 21 cm signal that should contain useful information about the sizes and topology of ionized regions.  The challenge is to develop statistics matched to that form of non-Gaussianity, which is currently an unsolved problem.  Some examples of statistics that have been explored are the 1 pt function (or probability distribution function) of the 21 cm brightness field, which develops a skewness as reionization leads to many pixels with zero signal \cite{furlanetto2004,harker2009}; the Euler characteristic \cite{friedrich2011}, which determines the number of holes in a connected surface; and there are many other possibilities including the bispectrum, wavelets, and threshold statistics \cite{lee2011} that have yet to be properly explored.  In addition to new statistics, it is possible for signatures of non-Gaussianity to show up as a modification to the shape of the power spectrum \cite{joudaki2011}.

\subsection{Prospects for cosmology}
\label{ssec:cosmology}

In this review, we have mostly focused on the ability of the 21 cm signal to constrain different aspects of astrophysics and the physics that needs to be understood in order to do so.  In the first instance the effects of the astrophysics of galaxy formation needs to be understood before fundamental physics can be addressed.  Once this is understood there is considerable potential for addressing cosmology.
21 cm observations represent one of the only ways of accessing the large volumes and linear modes present at high redshifts.  Studies show that 21 cm experiments have considerable potential to improve on measurements of cosmological parameters \cite{mcquinn2005,mao2008}, on
models of inflation \cite{barger2009,adshead2011}, and on measurements of the neutrino mass \cite{pritchard2008nu}.

We have touched upon the possibility of constraining exotic heating and structure formation in the early Universe via the 21 cm global signal.  These experiments are primarily sensitive to key transitions such as the onset of \lya coupling, the onset of heating, and the reionization of the Universe.  These key moments can be affected by different heating mechanisms in similar ways and so most likely will provide upper limits on the contribution of exotic physics.  Such
elements as heating by dark matter decay or annihilation
\cite{furlanetto2006dm,valdes2007,belikov2009,slatyer2009},
evaporating primordial blackholes \cite{mack2008}, or other such physics might all be constrained.


Ideally one would target the cosmic dark ages, where the 21 cm signal depends on well-understood linear physics.  In this regime, one would have the possibility of making precision measurements similar to the CMB, but at many different redshifts.  This represents the long term goal of 21 cm observations as a tool of cosmology.  Unfortunately, many technical challenges need to be overcome before this is realistic.  Foregrounds become much larger relative to the signal at the low frequencies relevant for the cosmic dark ages.  This requires large experiments to achieve the required sensitivity.  Since the sensitivity of an interferometer to the 21 cm power spectrum typically decreases on smaller angular scales as a result of fewer long baselines, large angular scales are easier to access. It was shown in \cite{gordon2009} that an experiment with 10 km$^2$ collecting area could detect a constrained isocurvature mode via its signature on large angular scales in the 21 cm power spectrum at $z\approx30$.  In contrast, detecting the running of the primordial power spectrum via the 21 cm power spectrum on scales $k\gtrsim1\,\iMpc$ requires experiments with $\gtrsim$10 km$^2$ to achieve the necessary sensitivity \cite{adshead2011}. 

These low frequencies will probably require escaping the Earth's ionosphere to space or the lunar surface \cite{jester2009}. The payoff would be unprecedented precision on cosmological parameters and a precise picture of the earliest stages of structure formation.  This could lend insight into fundamental physics of phenomena like variations in the fine-structure constant \cite{khatri2007} and the presence of cosmic superstrings \cite{khatri2008}. 

More accessible in the near future, is the EoR where the astrophysics of star formation must be disentangled to get at cosmology.  This will likely involve the use of redshift space distortions and astrophysical modeling as a way of removing the astrophysics leaving the cosmological signal behind.  In the future, with sensitivity sufficient to image the 21 cm fluctuations one can imagine developing sophisticated algorithms to exploit the full 3D information in the maps to separate astrophysics, much as one currently removes
foregrounds in the CMB.  This has not been studied in great detail, but preliminary work on modelling out the contribution of reionization appears promising.

It has been shown in simulations that just one parameter - the ionized fraction $x_i$ - does a good job of describing the ionization power spectrum at the few percent level.  This arises from well understood physics and amounts to a simple photon counting argument.  Thus, at first order we might hope to be able to predict the ionization contribution to the 21 cm signal during reionization and marginalise over it in the same manner as the bias of galaxies in galaxy surveys. At higher levels of precision, accounting for the modifications from the clustering of the sources or sinks of ionizing photons will become important.  Modelling of the ionization power spectrum need only degrade cosmological constraints by factors of a few \cite{mao2008}.

\section{Intensity mapping in atomic and molecular lines}
\label{sec:intensity}

\subsection{21 cm intensity mapping and dark energy}

Reionization eliminated most of the neutral hydrogen in the Universe,
but not all.  In overdense regions the column depth can be sufficient
for self-shielding to preserve neutral hydrogen.  Examples of such
pockets of neutral hydrogen are seen as damped Ly$\alpha$ systems in
quasar spectra.  Whereas the pre-reionization 21 cm signal provides
information about the topology of ionized region, the post-reionization
signal describes the clustering of collapsed halos based on the
underlying density field of matter.

The notion of intensity mapping of 21 cm bright galaxies was
introduced in several papers \cite{loeb2008im,wyithe2008,chang2008}, and similar ideas were discussed by \cite{battye2004}.
The starting point was the many experimental efforts to detect the 21
cm signal from the epoch of reionization when the IGM is close to
fully neutral.  After reionization approximately 1\% of the baryons
are contained in neutral hydrogen ($\Omega_{\rm HI}\approx0.01$).
Although this reduces the signal significantly relative to a fully
neutral IGM, the amplitude of Galactic synchrotron emission (which
provides the dominant foreground) is several orders of magnitude less
at the frequencies corresponding to 21 cm emission at redshifts
$z=1-3$.  Consequently, the signal to noise ratio for radio
experiments targeted at intensity mapping might be expected to be
comparable to experiments targeted at the epoch of reionization (EoR).
In both cases, the new technologies driven by developments in
computing power makes possible previous unattempted observations.

\begin{figure} [ht]
\begin{center}
\includegraphics[width=3.2in]{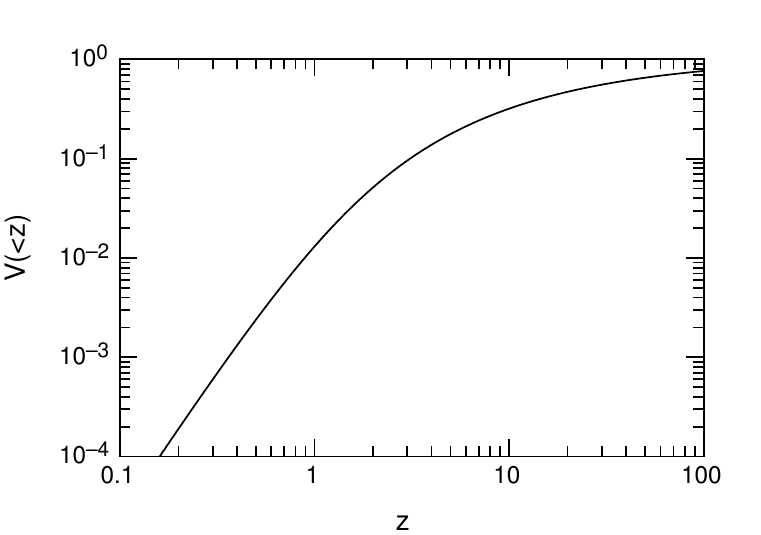}
\caption{The fraction of the total comoving volume of the observable
Universe that is available up to a redshift $z$ \cite{loeb2008im}.}
\label{fig:covolume}
\end{center}
\end{figure}
Such observations would be extremely important for our understanding of the Universe.  It is perhaps humbling to realise that existing observations from the CMB and galaxy surveys fill only a small fraction of the potentially observable Universe.  Figure \ref{fig:covolume}a shows the available comoving volume out to a given redshift.  At present, the deepest sky survey over a considerable fraction of the sky is the first Sloan Digital Sky Survey (SDSS), whose luminous red galaxies (LRG) sample has a mean redshift of $z\approx0.3$.  By measuring the 21 cm intensity fluctuations at $z=1-3$ the comoving volume probed by experiment would be increased by two orders of magnitude. This is particularly important since our ability to constrain cosmological parameters depends critically on the survey volume.  In general, constraints scale with volume $V$ as $\sigma\propto V_{\rm eff}^{-1/2}$ since a larger volume means that more independent Fourier modes can be constrained.  For example, one might improve constraints on quantities such as the neutrino mass and the running of the primordial power spectrum \cite{visbal2009}.

Perhaps the greatest utility for intensity mapping (IM) surveys is in
constraining dark energy via measurements of baryon acoustic
oscillations (BAOs) in the galaxy power spectrum.  BAO arise from the
same physics that produces the spectacular peak structure in the CMB.
These oscillations have a characteristic wavelength set by the sound
horizon at recombination and so provide a ``standard ruler".
Measurements of this distance scale can be used to probe the geometry
of the Universe and to constrain its matter content.  Measurements of
the oscillations in the CMB at $z\approx1100$ and more locally in the
BAO at $z\lesssim3$ would provide exquisite tests of the flatness of
the Universe and of the equation of state of the dark energy.
Redshifts in the range $z\approx1-3$ are of great interest since they
cover the regime in which dark energy begins to dominate the energy
budget of the Universe.  While galaxy surveys will begin to probe this
range in the next decade, covering sufficiently large areas to the
required depth is very challenging for galaxy surveys.  This has been
considered in depth by the Dark Energy Task Force \cite{albrecht2006}.

To get a sense of the observational requirements of BAO observations
with 21 cm intensity mapping, let us consider some numbers.  Beyond the
third peak of the BAO non-linear evolution begins to wash out the
signal.  The peak of the third BAO peak has a comoving wavelength of
35$h^{-1}{\rm\,Mpc}$, which for adequate samples requires pixels
perhaps half that size.  At $z=1.5$ the angular scale corresponding to
18$h^{-1}{\rm\,Mpc}$ is 20 arcmin, which sets the required angular
resolution of the instrument.  Estimating the HI mass enclosed in such
a volume is uncertain, but based on the observed value of $\Omega_{\rm
HI}\sim10^{-3}$ is $\sim2\times10^{12}M_\odot$.

\begin{figure}
\begin{center}
\includegraphics[scale=0.3]{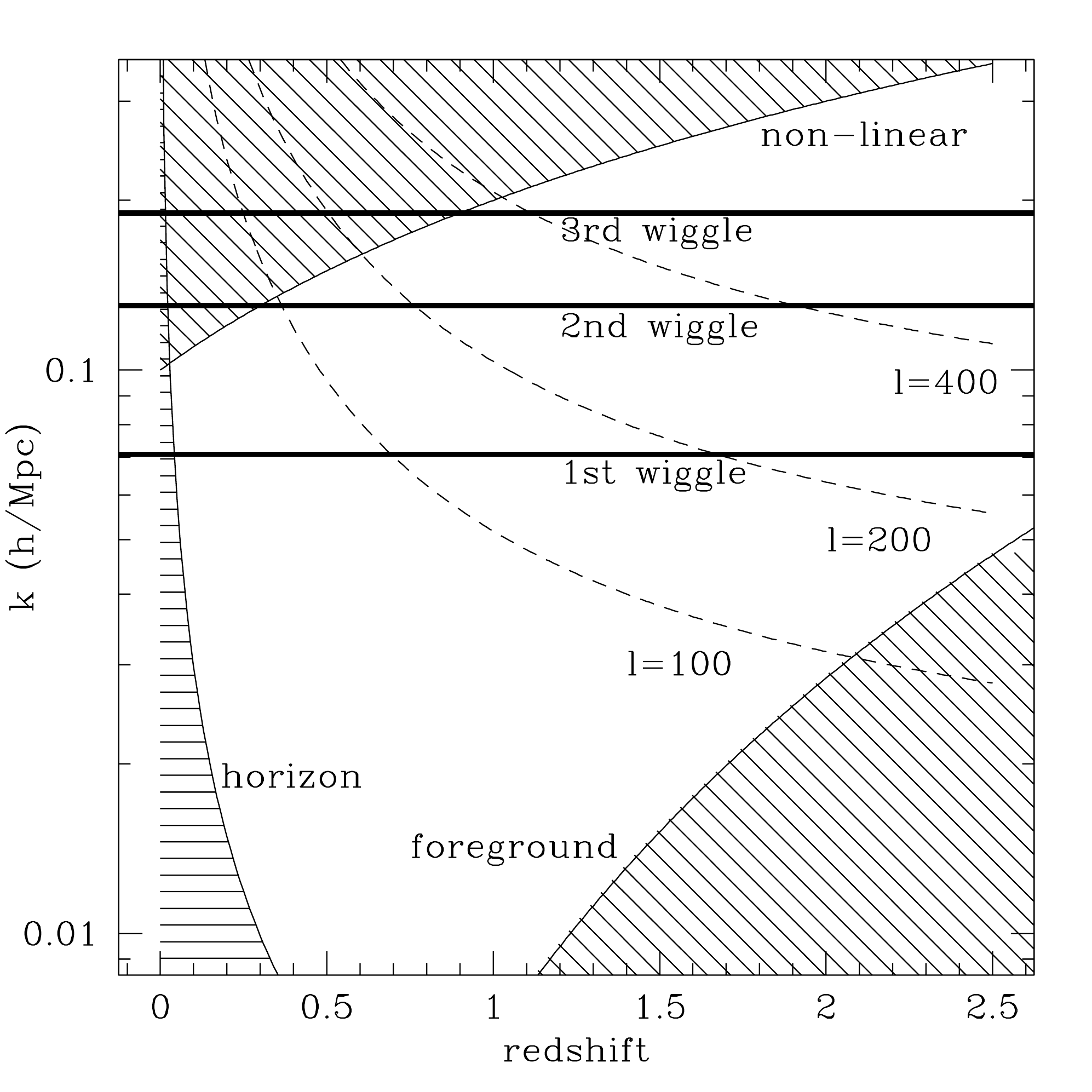}
\caption{The observable parameter space in redshift and in scale ($k$) for BAO studies. The shaded regions are observationally inaccessible. The horizontal lines indicate the scale of the first three BAO wiggles, and the dashed lines show contours of constant spherical harmonic order $\ell$ \cite{chang2008}.}
\label{fig:chang_foreground}
\end{center}
\end{figure}

The mean brightness temperature $T_b$ of the 21 cm line can be estimated using
\begin{equation}
T_b=0.3(1+\delta)\left(\frac{\Omega_{\rm HI}}{10^{-3}}\right)\left(\frac{\Omega_m+a^3\Omega_\Lambda}{0.29}\right)^{-1/2}\left(\frac{1+z}{2.5}\right)^{1/2}{\,\rm mK},
\end{equation}
where $1+\delta=\rho_g/\bar{\rho}_g$ is the normalised neutral gas density and $a=(1+z)^{-1}$ is the scale factor.  The amplitude of the signal is therefore considerably smaller than that of the 21 cm signal during reionization.  However, the foregrounds are reduced by a factor of $[(1+8)/(1+1.5)]^{-2.6}\sim30$, redeeming the situation.
Diffuse foregrounds may be removed in a similar fashion to the 21 cm signal at high redshift.  Figure \ref{fig:chang_foreground} shows the relevant region of the BAO that could be detected.  The top excluded region accounts for the non-linear growth of structures, which erases the baryon wiggles in the power spectrum.  The left exclusion region comes from the finite volume of the Universe, which limits the largest wavelength mode that fits within the survey.  The right region is a rough estimate of the limit from foregrounds based upon a simple differencing of neighbouring frequency channels to remove the correlated foreground component.  The white accessible region demonstrates that the first three BAO peaks could be accessible with intensity mapping experiments.

\begin{figure}[htbp]
\begin{center}
\includegraphics[scale=0.3]{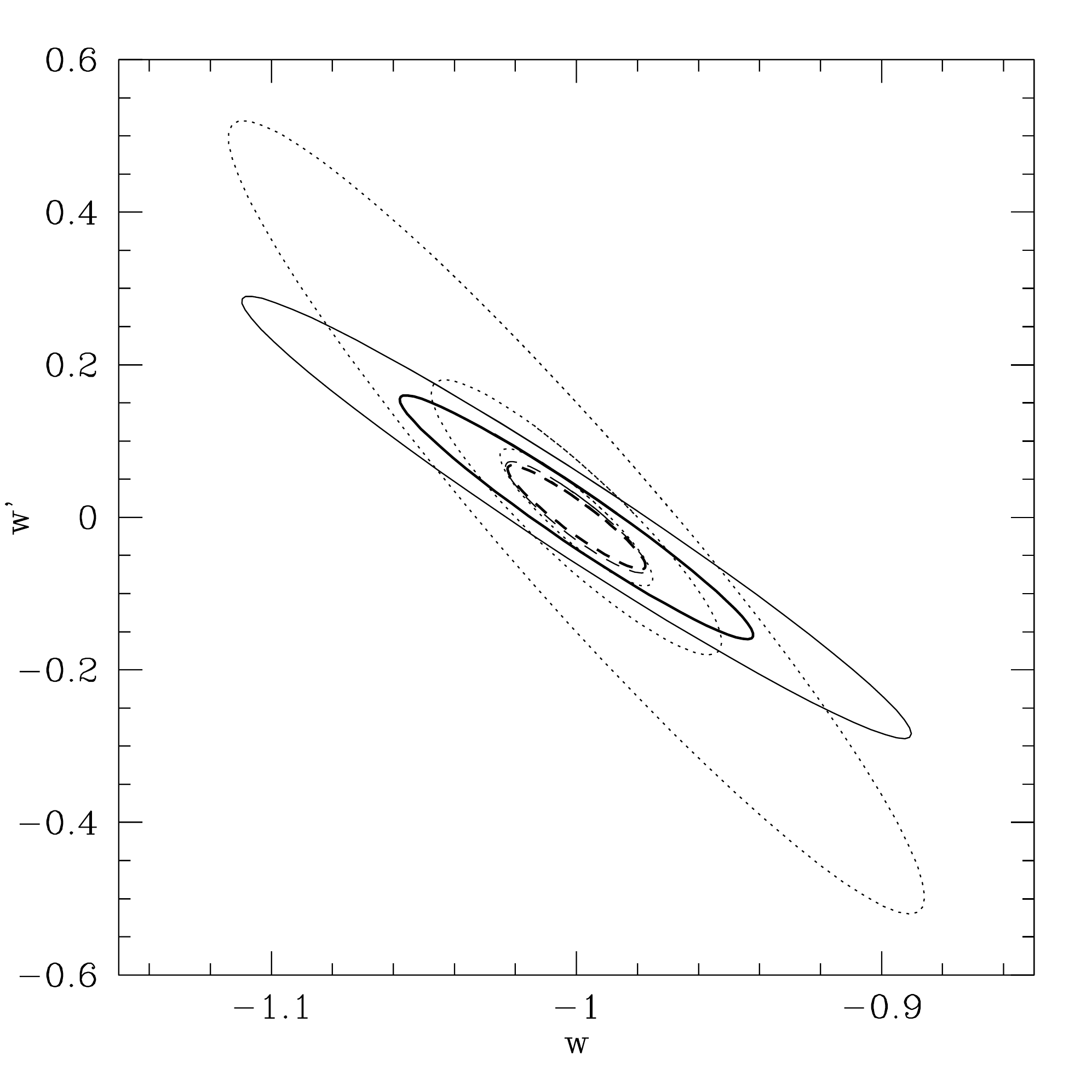}
\caption{Constraints on the equation of state parameter $w=p/\rho c^2$
(horizontal axis) and its redshift variation for an equation of state $w=w_p+(a_p-a)w'$ (vertical
axis) at $z=0$.  The 1-$\sigma$ contour for IM combined with Planck (inner
thick solid for baseline model, outer thin solid for worst case), the
Dark Energy Task Force stage I projects with Planck (outer dotted),
the stage I and III projects with Planck (intermediate dotted), the
stage I, III, and IV projects with Planck (inner dotted), and all
above experiments combined (dashed, again thick for baseline, thin for
worst case; the two contours are nearly
indistinguishable). \cite{chang2008}}
\label{fig:chang_w}
\end{center}
\end{figure}

Observations of the BAO can be used to constrain the dark energy equation of state $w$, which may be parametrised as $w=w_p+(a_p-a)w'$, where $w_p$ and $a_p$ are the value of the equation of state at a pivot redshift where the errors in $w$ and $w'$ are uncorrelated. Figure \ref{fig:chang_w} shows the ability of an IM telescope covering a square aperture of size 200 m by 200 m subdivided into 16 cylindrical sections each 12.5 m wide and 200 m long.  After integrating for 100 days with bandwidth 3MHz, this experiment is roughly comparable to the stage IV milestone set by the Dark Energy Task Force (DETF) committee \cite{albrecht2006},  and provides very good dark energy constraints.  The power of such an experiment makes it a potential alternative to more traditional galaxy surveys \cite{wyithe2008BAO,visbal2009,wyithe2009}.

Reaching the thermal noise required by a signal of $\sim$300 $\mu$K at
frequencies of 600 MHz is within the range of existing telescopes.  As
a first step towards demonstrating that such observations are
possible, Ref. \cite{chang2010} made use of the Green Bank Telescope
(GBT) to observe radio spectra corresponding to 21 cm emission at
$0.53<z<1.12$ over 2 square degrees on the sky.  This raw radio data
was processed by removing a smooth component, dominated by the
foregrounds, to leave a map of intensity fluctuations.
Cross-correlating these intensity fluctuations with a catalog of DEEP2
galaxies in the same field showed a distinct correlation in agreement
with theory. These experiments are continuing with the next step to
understand instrumental systematics at the level required to cleanly
measure the 21 cm intensity auto-power spectrum.  They constitute an
important step towards demonstrating the feasibility of intensity
mapping with the 21 cm line.

21 cm intensity mapping probes the neutral hydrogen contained within
galaxies and one might reasonably worry about how robust a tracer this
is of the density field.  It is believed that damped \lya absorbers
(DLAs), which contain the bulk of neutral hydrogen, are relatively
low-mass systems and so should have relatively low bias (when compared
to the highly biased bright galaxies seen in high redshift galaxy
surveys) \cite{wyithe2008}.  On top of this, however, one might worry
that fluctuations in the ionizing background could lead to spatial
variation in the neutral hydrogen content.  But after reionization the
mean free path for ionizing photons increases and the ionizing
background should become fairly uniform with modulation only at the
percent level \cite{wyithe2009}.

\subsection{Intensity mapping in other lines}

So far, we have described the possibilities for 21 cm intensity
mapping.  The 21 cm line has a number of nice qualities - it is
typically optically thin, it is a line well separated in frequency
from other atomic lines, and hydrogen is ubiquitous in the Universe. It is however only one line out of many that have been observed in local galaxies, which begs the question ``is 21 cm the best line to use and what additional information might be gained by looking at intensity mapping in other lines?"  The analysis of the impact of intensity mapping using lines other than the 21 cm line has not yet been fully explored.  Because the physical conditions leading to emission by these species are quite varied the cross-species joint analysis of intensity maps is a complex topic that the community has only begun to examine.  Because of this variation, the cross-analysis is potentially a rich source of information on conditions at high redshift.

One area where intensity mapping in lines other than 21 cm would be
particularly interesting is during the epoch of reionization.  One of
the challenges for understanding the first galaxies is the difficulty
of placing the galaxies seen in the Hubble Ultradeep Field (HUDF) into
a proper context.  By focusing on a small patch of sky, the HUDF sees very faint galaxies, but it is unclear how representative this patch is of the whole Universe at that time.  For comparison, the full HUDF is approximately 3$\times$3 arcmin in size comparable with the size of an individual ionized bubble, expected to be $\sim$ ten arcminutes in diameter during the middle stages of reionization.  Moreover, it is apparent that the galaxies seen in the HUDF are the brightest galaxies and that fainter, as yet unseen, galaxies contribute significantly to star formation and reionization.  The {\em James Webb Space Telescope} (JWST) will see even fainter galaxies and transform our view of the galaxy population at $z\sim10$, but there will still be a substantial level of star formation that it will not be able to resolve \cite{barkana2000}.

\begin{figure}[htbp]
\begin{center}
\includegraphics[scale=0.35]{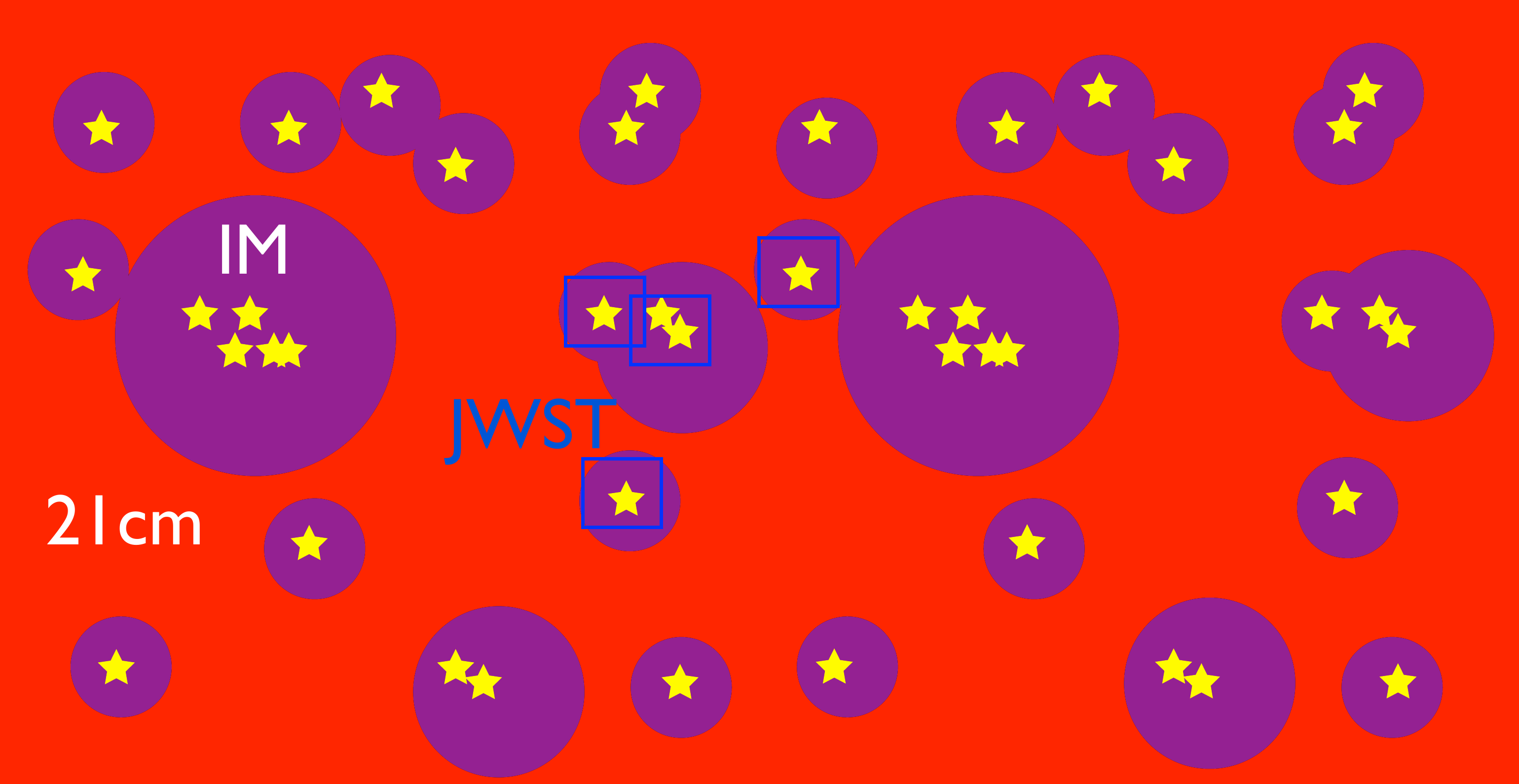}
\caption{Cartoon of the role intensity mapping would play in
understanding galaxy formation.  Deep galaxy surveys with HST and JWST image the properties of individual galaxies in small fields (blue boxes).  21 cm tomography (red filled region) provides a ``negative space" view of the Universe by determining the properties of the neutral gas surrounding groups of galaxies.  Intensity mapping (purple filled regions) fills in the gaps providing information about the collective properties of groups of galaxies.  Together the three would
give a complete view of the early generation of galaxies in the infant universe.}
\label{fig:IMcartoon}
\end{center}
\end{figure}

Combining these galaxy surveys with 21 cm observations and intensity
mapping would allow a powerful synergy between three independent types
of observations directed at understanding the first galaxies and the
epoch of reionization (illustrated in Figure \ref{fig:IMcartoon}).
Deep galaxy surveys with HST and JWST would inform us of the detailed
properties of small numbers of galaxies during the EoR.  21 cm
tomography provides information about the neutral hydrogen gas
surrounding groups of galaxies. Intensity mapping fills in the gaps
providing information about the total emission and clustering of the
full population of galaxies, even those below the sensitivity
threshold of the JWST. Together these three techniques would provide a
highly complete view of galaxies at high redshift and transform our
understanding of the origins of galaxy formation.

The first steps towards understanding intensity mapping in molecular
lines were made by Righi, Hern\'andez-Monteagudo \& Sunyaev
\cite{righi2008}, who considered the possibility that redshifted
emission from CO rotational lines might contribute a foreground to CMB
experiments. They showed that cross-correlating CMB maps of differing
spectral resolution at frequencies of $\nu\approx30{\rm\,GHz}$ could
be used to constrain the CO emission of galaxies at $z\approx3-8$.
This might be done, for example, by correlating CMB maps from CBI with
those of the {\em Planck} LFI.  These ideas were focussed on the
notion of 2D maps, but contain the seed of later work.

\begin{figure}[htbp]
\begin{center}
\includegraphics[scale=0.5]{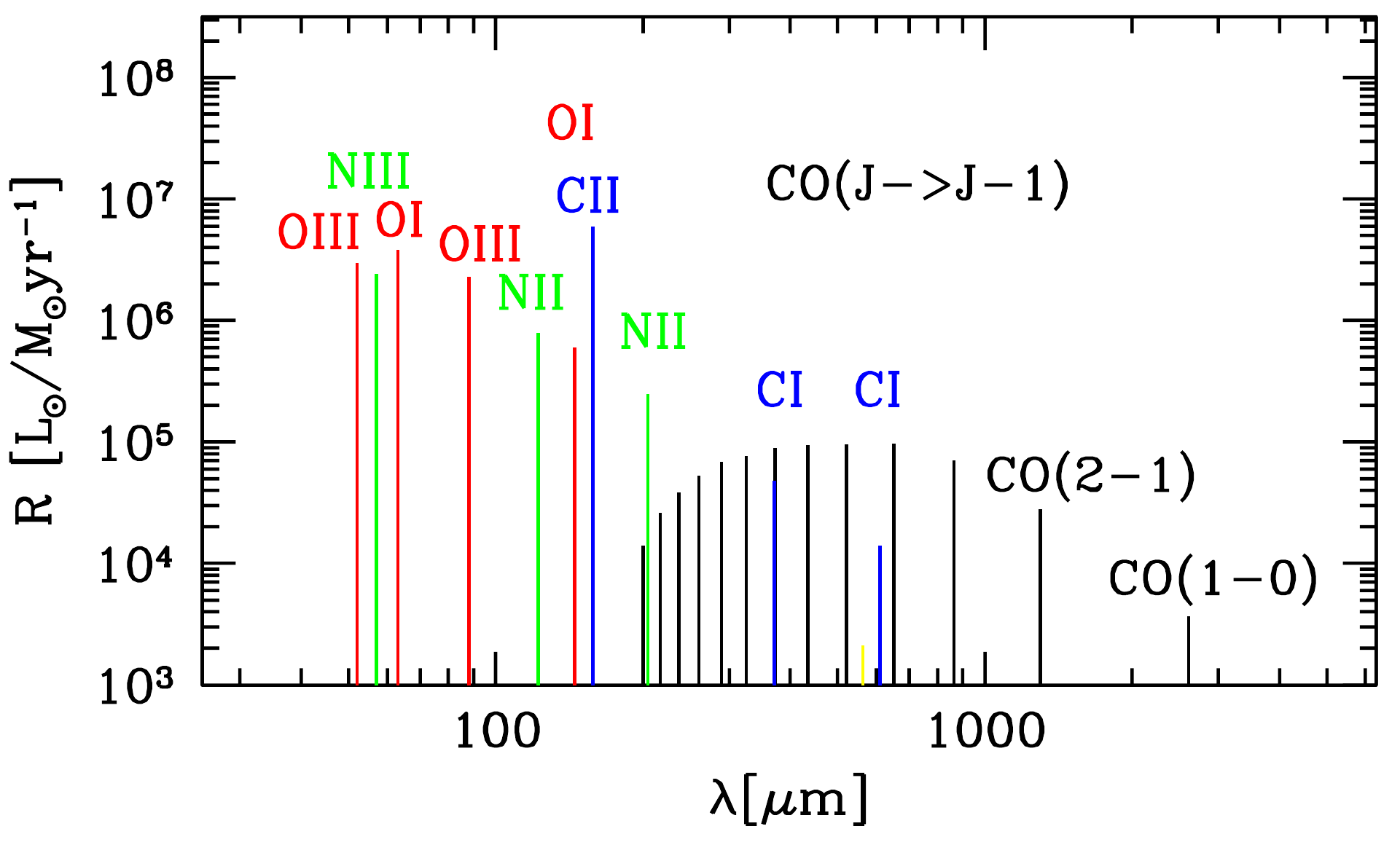}
\caption{Ratio between line luminosity, $L$, and star formation rate,
$\dot{M}_*$, for various lines observed in galaxies and taken from
Table 1 of Ref. \cite{visbal2011}.  For the first 7 lines this ratio is
measured from a sample of low redshift galaxies. The other lines have
been calibrated based on the galaxy M82.  Some weaker lines, for
example for HCN, have been omitted for clarity.}
\label{fig:linelist}
\end{center}
\end{figure}

Ref. \cite{visbal2010} explored for the first time the
notion of looking at a variety of lines in 3D intensity maps. Figure
\ref{fig:linelist} shows the major lines that appear in the emission
spectrum of M82.  By making a map at the appropriate frequency any of
these lines could be studied by intensity mapping.  Two major
challenges arise.  The first is that continuum foregrounds are
typically larger than the signal from these lines by 2-3 orders of
magnitude.  This is an identical, although more tractable problem, as
occurs in the case of 21 cm studies of the EoR and has been studied in
considerable detail.  Studies \cite{mcquinn2006,wang2006,liu2011} show
that, provided the continuum foreground is spectrally smooth in
individual sky pixels, it can be removed leaving very little residuals
in the cleaned signal.

Potentially more challenging is the issue of line confusion.  If we
look for the CO(1-0) line (rest frame frequency of 115 GHz) in a map
made at 23 GHz (corresponding to emission by CO at $z=4$) then our map
will additionally consider emission from other lines in galaxies at
other redshifts (e.g. CO(2-1) from galaxies at $z=9$).  However, the
contaminating emission arises from different galaxies which opens the
possibility of combining maps at different frequencies corresponding
to different lines from the same galaxies as a way of isolating a
particular redshift.  The emission from lines in the same galaxies
will correlate, while emission from lines in galaxies at different
redshifts will not\footnote{This is similar to the suggestion of
using the 21 cm map as a template to detect the deuterium hyperfine
line \cite{sigurdson2006}.}.

\begin{figure}
\includegraphics[width=2.5in,height=2.5in]{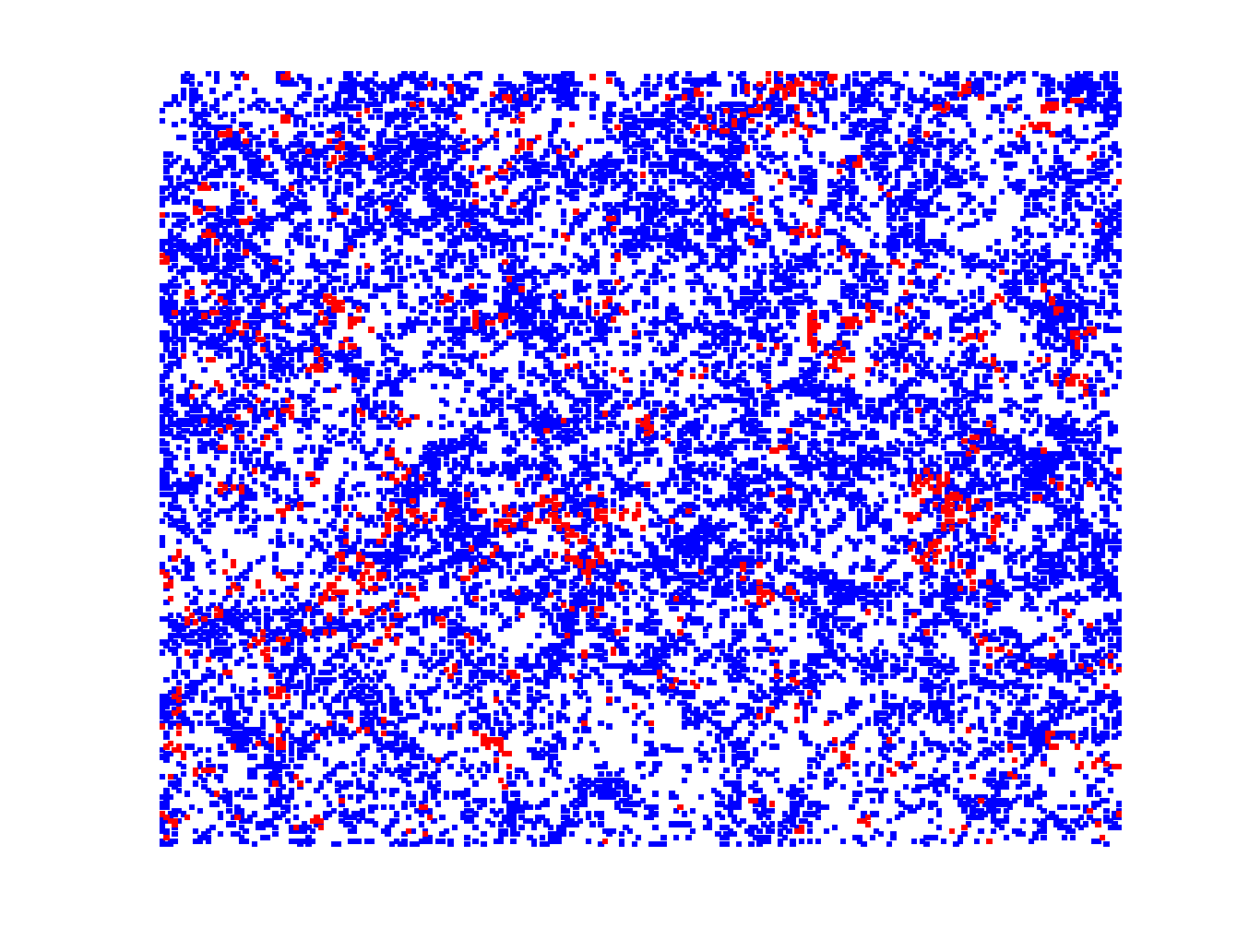}
\includegraphics[width=2.5in,height=2.5in]{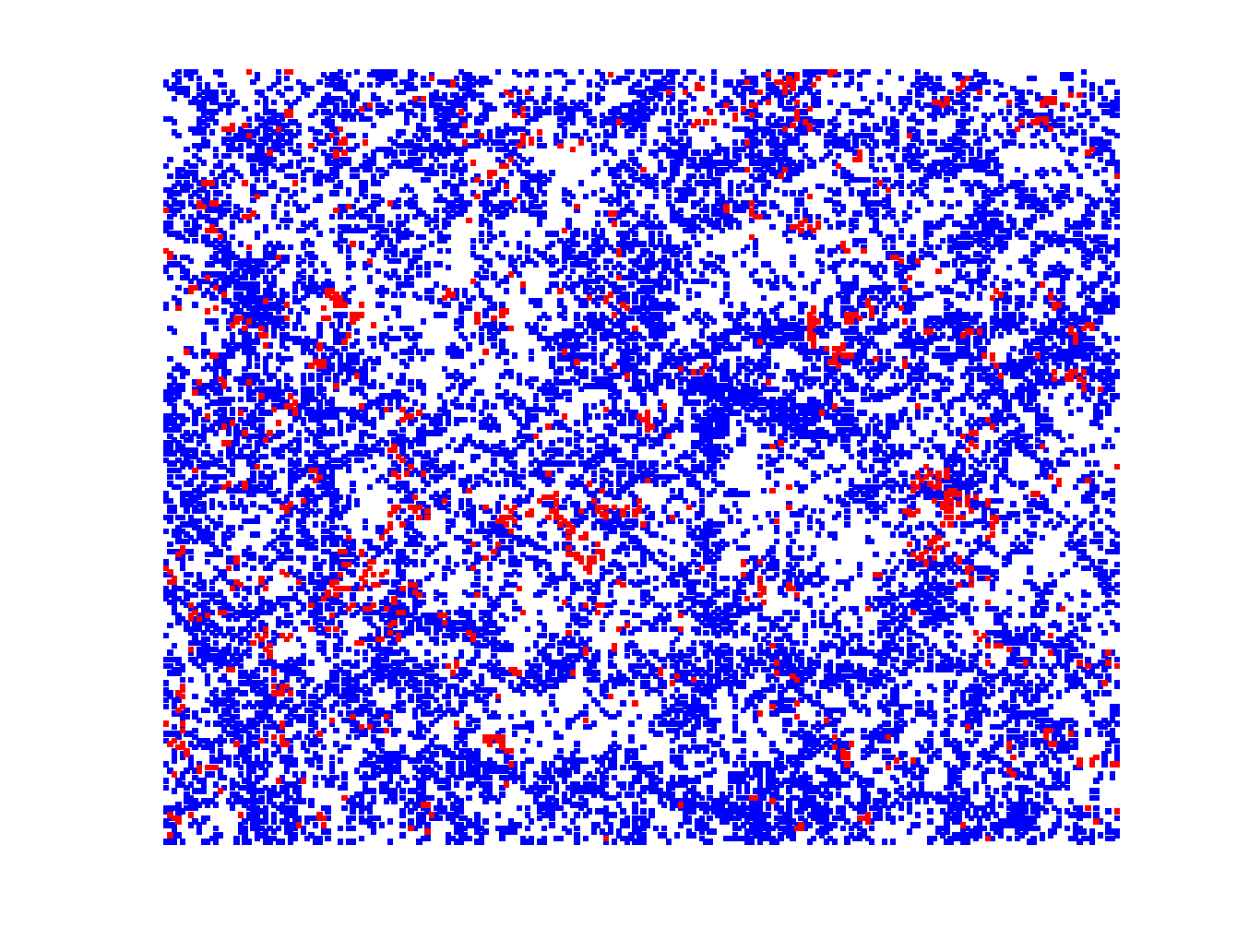}
\caption{ A slice from a simulated realization of line emission from
galaxies at an observed wavelength of 441$\mu$m (left) and 364$\mu$m
(right) \cite{visbal2011}.  The slice is in the plane of the sky and
spans $250\times250$ comoving Mpc$^2$ with a depth of $\Delta \nu /
\nu = 0.001$. The colored squares indicate pixels which have line
emission greater than $200 {\rm Jy/Sr}$ for the left panel and $250
{\rm Jy/Sr} $ for the right panel.  The emission from OI(63$\mu$m) and
OIII(52$\mu$m) is shown in red on the left and right panels,
respectively, originating from the same galaxies at $z=6$.  All of the
lines illustrated in Figure \ref{fig:linelist} are included and
plotted in blue.  Cross correlating data at these two observed
wavelengths would reveal the emission in OI and OIII from $z=6$ with
the other emission lines being essentially uncorrelated. }
\label{fig:badlines}
\end{figure}

Fortunately, it is possible to statistically isolate the fluctuations
from a particular redshift by cross correlating the emission in two
different lines \cite{visbal2010,visbal2011}.  If one compares the
fluctuations at two different wavelengths, which correspond to the
same redshift for two different emission lines, the fluctuations will
be strongly correlated. However, the signal from any other lines
arises from galaxies at different redshifts which are very far apart
and thus will have much weaker correlation (see Figure
\ref{fig:badlines}).  In this way, one can measure either the
two-point correlation function or power spectrum of galaxies at some
target redshift weighted by the total emission in the spectral lines
being cross correlated.

\begin{figure}
\begin{center}
\includegraphics[width=9.2cm]{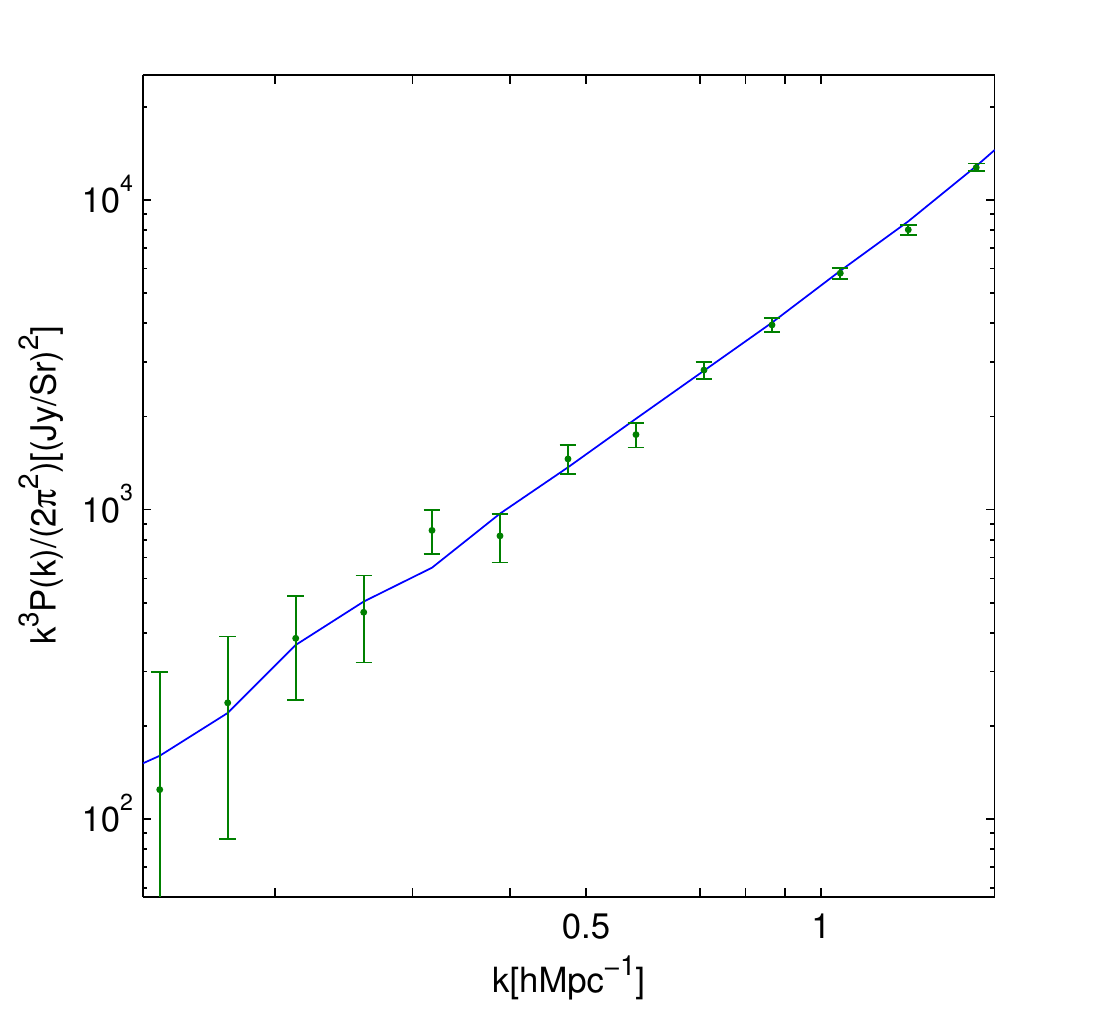}
\caption{\label{spicaplot} The cross power spectrum of OI(63 $\mu$m)
and OIII(52 $\mu$m) at $z=6$ measured from mock simulation data for a
hypothetical infrared space telescope similar to SPICA \cite{visbal2011}.  The solid
line is the cross power spectrum measured when only line emission from
galaxies in the target lines is included.  The points are the
recovered power spectrum when detector noise, contaminating line
emission, galaxy continuum emission, and dust in our galaxy and the
CMB are included.  The error bars follow Eq.~(\ref{ccerror}) with
$P_{\rm1total}$ and $P_{\rm2total}$ calculated from the simulated
data, including detector noise, contaminating line emission and sample
variance. }
\label{fig:SPICA}
\end{center}
\end{figure}

The cross power spectrum at a wavenumber $k$ can be written as,
\begin{equation}
P_{1,2}(\vec{k}) = \bar{S}_{1}\bar{S}_{2}\bar{b}^2 P(\vec{k}) + P_{\rm
shot},
\end{equation}
where $\bar{S}_{1}$ and $\bar{S}_{2}$ are the average fluxes in lines
$1$ and $2$ respectively, $\bar{b}$ is the average bias factor of the
galactic sources, $P(\vec{k})$ is the matter power spectrum, and
$P_{\rm shot}$ is the shot-noise power spectrum due to the discrete
nature of galaxies.  The
root-mean-square error in measuring the cross power spectrum at a
particular $k$-mode is given by \cite{visbal2010},
\begin{equation}
\delta P^2_{1,2}=\frac{1}{2}(P_{1,2}^2 + P_{\rm1total}P_{\rm2total}),
\label{ccerror}
\end{equation}
where $P_{\rm1total}$ and $P_{\rm2total}$ are the total power spectra
corresponding to the first line and second line being cross
correlated.  Each of these includes a term for the power spectrum of
contaminating lines, the target line, and detector noise. Figure
\ref{fig:SPICA} shows the expected errors in the determination of the
cross power spectrum using the OI(63 $\mu$m) and OIII(52 $\mu$m) lines
at a redshift $z=6$ for an optimized spectrometer on a 3.5 meter
space-borne infrared telescope similar to SPICA, providing background
limited sensitivity for 100 diffraction limited beams covering a
square on the sky which is $1.7^\circ$ across (corresponding to 250
comoving Mpc) and a redshift range of $\Delta z=0.6$ (280 Mpc) with a
spectral resolution of $(\Delta \nu/\nu)=10^{-3}$ and a total
integration time of $2\times 10^6$ seconds.

We emphasize that one can measure the line cross power spectrum from
galaxies which are too faint to be seen individually over detector
noise.  Hence, a measurement of the line cross power spectrum can
provide information about the total line emission from all of the
galaxies which are too faint to be directly detected.  One possible
application of this technique would be to measure the evolution of
line emission over cosmic time to better understand galaxy evolution
and the sources that reionized the Universe.  Changes in the minimum
mass of galaxies due to photoionization heating of the intergalactic
medium during reionization could also potentially be measured
\cite{visbal2010}.

\begin{figure}
\begin{center}
\includegraphics[width=9.2cm]{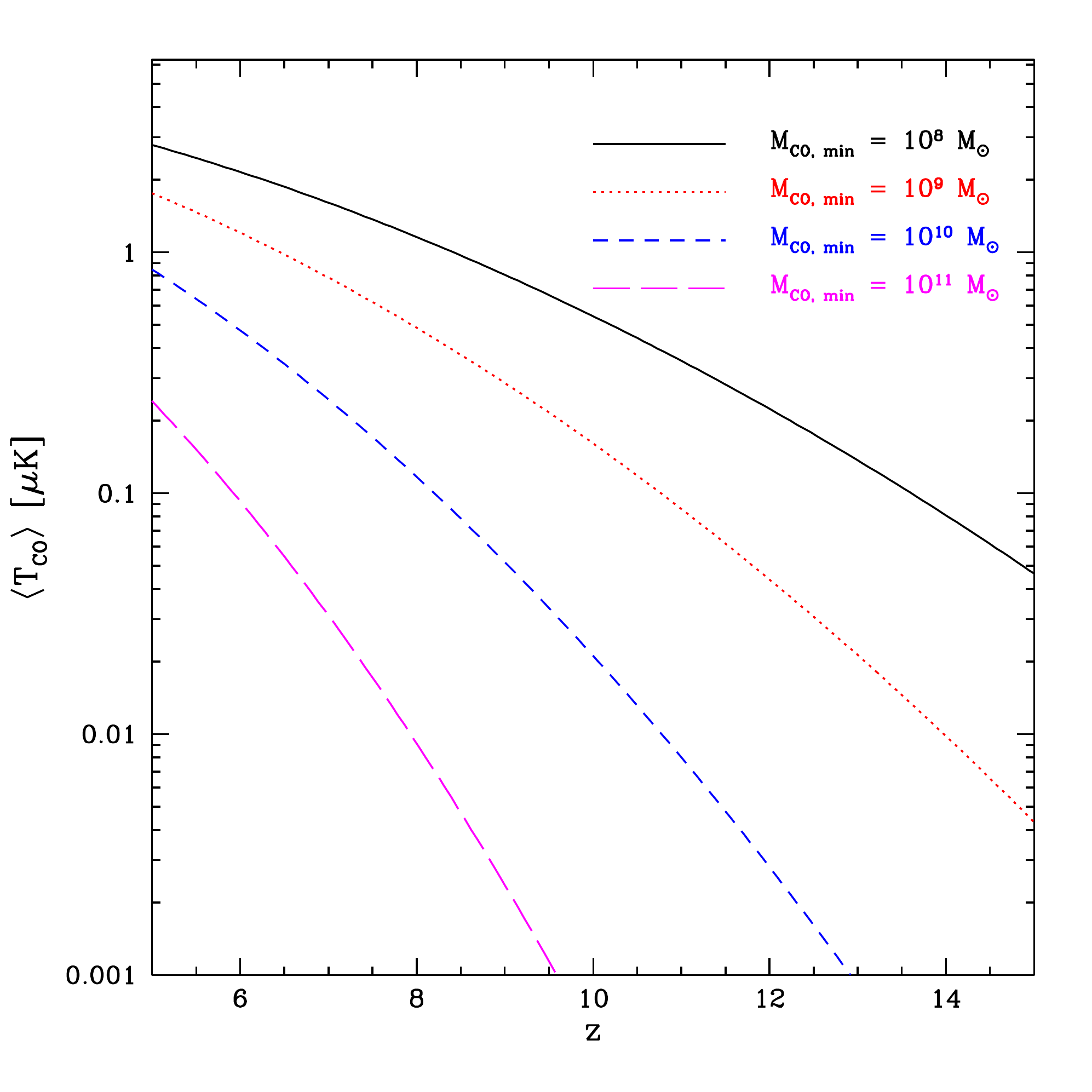}
\caption{The global mean CO brightness temperature as a function of
redshift \cite{lidz2011}.  The curves show the volume-averaged CO
brightness temperature for several different values of $M_{\rm co,
min}$, the minimum halo mass hosting a CO luminous galaxy. In all
cases, the minimum host halo mass of star forming galaxies is the
atomic cooling mass, $M_{\rm sf, min} \approx 10^8 M_\odot$, and so
the curves with larger $M_{\rm co, min}$ describe models in which
galaxies with low star formation rates (SFRs) are not CO luminous. If
we had instead varied $M_{\rm sf, min}$, while fixing both $M_{\rm co,
min} = M_{\rm sf, min}$ and the SFR density at a given redshift, the
model variations would be considerably smaller.}
\label{fig:meanTCO}
\end{center}
\end{figure}

As a definite example of the sort of information that might be
gathered in molecular line intensity mapping, we will discuss CO, the
only line other than the 21 cm line to have been somewhat investigated
\cite{carilli2011,lidz2011,gong2011}.  CO emission is widely used as a
tracer for the mass of cold molecular gas in a galaxy.  This cold
molecular gas provides the fuel for star formation and so determining
its abundance provides a key constraint on models of star formation
and galaxy evolution.  Indeed one of the key science goals for ALMA is
to directly image CO emission in individual galaxies at $z\approx7$.
Figure \ref{fig:meanTCO} shows the redshift evolution of the mean
brightness temperature (a measure of the radio intensity) for the
CO(2-1) line as a function of the minimum mass required for a galaxy
to be CO bright.  These calculations identify a signal strength of
$0.01-1{\rm \,\mu K}$ at $z=8$ as the observational target. This
represents an order of magnitude increase in sensitivity over existing
CMB experiments like CBI and might be enabled by improved techniques
with dishes exploiting focal plane arrays.

\begin{figure}
\begin{center}
\includegraphics[width=9.2cm]{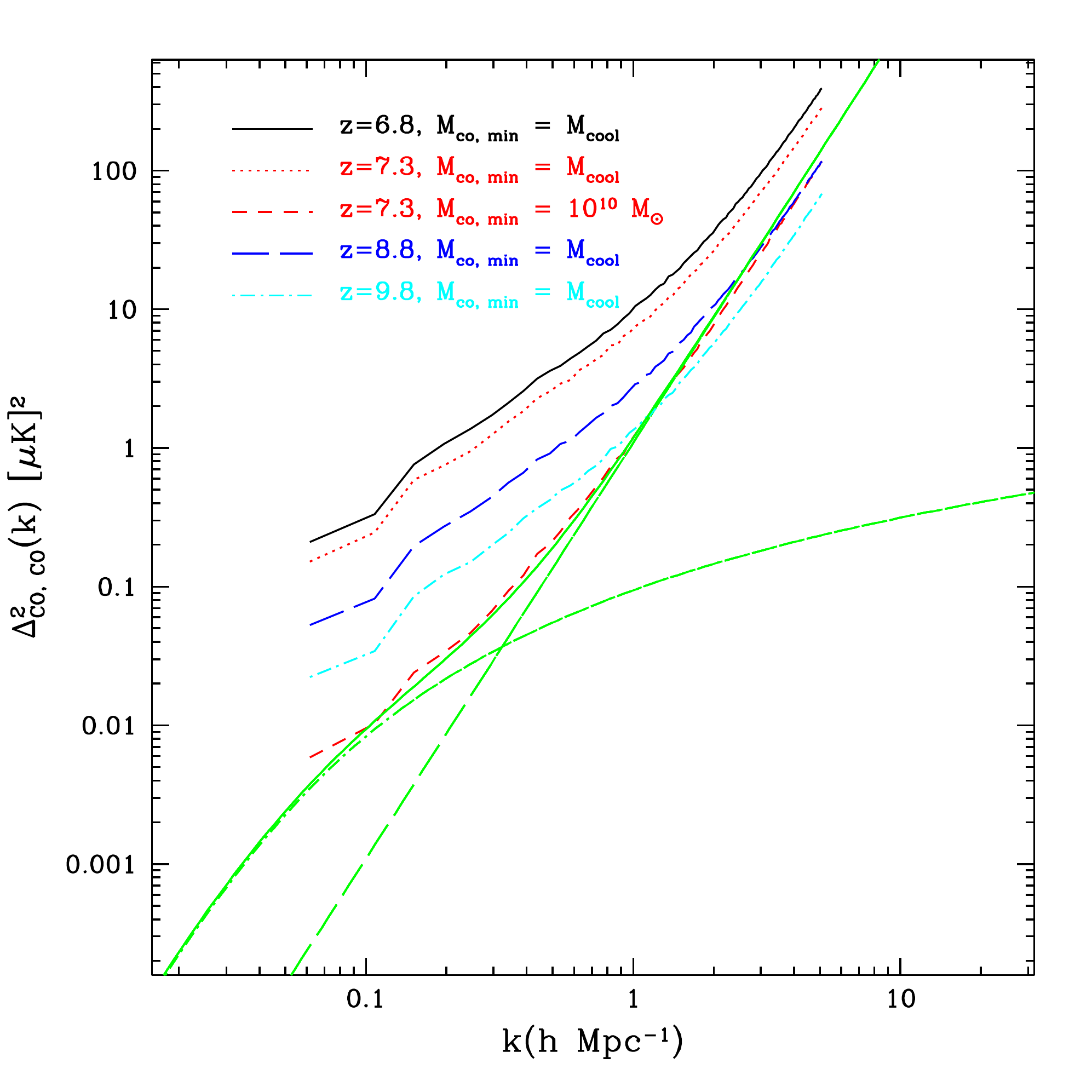}
\caption{The auto power spectrum of CO brightness temperature
fluctuations \cite{lidz2011}. The black solid, red dotted, red dashed,
blue dashed, and cyan dot-dashed curves show simulated CO power
spectra at different redshifts for various values of $M_{\rm co, min}$
and fixed duty cycle. The green solid line is the halo model
prediction for the signal for $z=7.3$, $M_{\rm co, min} = 10^{10}
M_\odot$, and $f_{\rm duty} = 0.1$. The green dashed line shows the
Poisson term, while the green dot-dashed curve represents the
clustering term.  }
\label{fig:autoCO}
\end{center}
\end{figure}

Figure \ref{fig:autoCO} shows the evolution of the auto-power spectrum
of the CO intensity fluctuations.  This signal evolves strongly with
redshift due to its dependence on $\langle T_{\rm CO}\rangle$ and has
a shape that depends on the relative contribution of the clustering of
CO bright galaxies and a Poisson shot noise term, which is determined
by the distribution of CO galaxy masses.  Measurement of the shape and
evolution of this power spectrum would therefore give considerable
information about the properties of early CO galaxies.

\subsection{Cross-correlation of molecular and 21 cm intensity maps}

\begin{figure}[htbp]
\begin{center}
\includegraphics[scale=0.5]{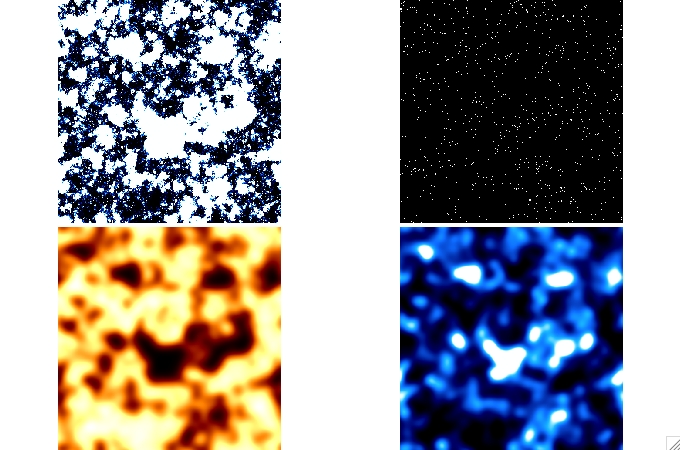}
\caption{Redshift slices at $z\approx7$ taken from the simulation of
Ref. \cite{lidz2011} and illustrating the galaxy distribution (top
right), ionization field (top left), 21 cm fluctuations (bottom left),
and CO intensity fluctuations (bottom right).  These last two fields
have been smoothed to an angular resolution of 6 arcmin.  The CO
bright galaxies seen in the CO intensity map are also responsible for
ionizing the neutral gas.  This can be seen by the correlation between
bright CO regions and dark (ionized) patches in the 21 cm map. The
associated signal provides a quantitative version of the cartoon in
Figure \ref{fig:IMcartoon}. \cite{lidz2011}}
\label{fig:21cm_co}
\end{center}
\end{figure}

Earlier, we described 21 cm tomography observations with instruments like MWA, LOFAR, and PAPER that hope to map the 21 cm emission from neutral hydrogen during the EoR. These maps trace out the ionization field, which is determined by the distribution of galaxies.  Regions with many galaxies will tend to have ionized the surrounding IGM leading to a ``hole" in the 21 cm signal.  In contrast, since those same galaxies will have undergone considerable star formation and so produced metals, regions with galaxies will appear CO bright.  This is clearly seen in the bottom two panels of Figure \ref{fig:21cm_co} - blue CO bright regions
correspond to black patches with no signal in the 21 cm map.  This
anti-correlation can be detected statistically and provides a very
direct constraint on the size of the ionized regions.  Maps made in
lines from different atoms or molecules might be used in a related
way.  Each atomic species will reflect different properties of the
host galaxies and combining maps may enable a detailed understanding
of the host metallicity and internal properties to be achieved.

The model used for these calculations was based upon an empirical
calibration of the CO luminosity function to observations of galaxies
at $z\lesssim3$ \cite{lidz2011}.  This was then extrapolated to higher
redshifts as a best guess at how CO galaxies might evolve.  More
satisfying would be to connect the CO luminosity to the intrinsic
properties of the galaxy interstellar medium (ISM).  Such modeling of
the detailed properties of CO emission from galaxies has been
attempted \cite{obreschkow2009}.  CO emission in local galaxies is
known to originate in giant molecular clouds (GMCs) whose typical size
is $\sim$10 pc.  These clouds are observed to be optically thick to CO
emission and so the empirically determined scaline of the CO
luminosity with molecular hydrogen mass, $L_{\rm CO}\propto M_{H_2}$,
is best explained by the non-overlap of these clouds in angle and
velocity.  The relationship then results from the CO luminosity being
proportinal to the number of clouds.  In addition, the detailed
physics of the heating of these clouds by stars and active galactic
nuclei (AGN), which determines the excitation state of the different
CO rotational levels, must be accounted for.  There is increased
interest in numerical simulation of the chemistry and properties of
GMCs \cite{shetty2011}, which can inform semi-analytic modelling.
Connecting the global properties of CO emission to such ISM details is
challenging, but would be greatly rewarding in providing a new way of
learning about the ISM of galaxies at high redshift.  Importantly the
chemistry that determines the CO abundance may change dramatically in
the denser metal poor ISM of early galaxies \cite{sternberg2011}.

Clearly, there is interesting information to be harvested from
measurements of intensity fluctuations in line emission from galaxies
at high redshifts.  In the above discussion we have focussed on
preliminary results from the study of CO, but similar results apply to
the array of other atomic and molecular lines.  An important question
is how the signal might best be observed and which lines are most
important.  Some lines, like CO, can be targeted from the ground but
others, such as OI(63 $\mu$m), lie in the infra-red and must be
observed from space. The detailed design of the optimal instrument and
the requirements for sensitivity and angular resolution are not yet
well understood.  Furthermore, the statistical tools for removing
continuum and line contamination efficiently need to be developed.

A future space borne infrared telescope such as SPICA could be ideal
for intensity mapping.  As illustrated in Fig. \ref{fig:SPICA}
\cite{visbal2011}, an optimized instrument on SPICA could measure the
OI(63 $\mu$m) and OIII(52 $\mu$m) line signals very accurately out to
high redshifts.  Such an instrument would require the capability to
take many medium resolution spectra at adjacent locations on the sky.
However, it may be possible to sacrifice angular resolution since the
IM technique does not involve resolving individual sources.

\section{21 cm forest}
\label{sec:forest}

While most interest in the redshifted 21 cm line has focused on the
case where the CMB forms a backlight, an important alternative is the
case where the 21 cm absorption is imprinted on the light from radio
loud quasars.  Historically, the first attempts to detect the
extra-galactic 21 cm line were of this nature \cite{field1959obs}. The
resulting one dimensional absorption spectra are expected to show a
forest of lines originating from absorption in clumps of neutral
hydrogen along the line of sight and have been called the 21 cm forest
by analogy to the Ly$\alpha$ forest.

Whereas the Ly$\alpha$ forest is primarily visible at redshifts
$z\lesssim6$, when the Universe is mostly ionized, the 21 cm forest is
strongest at higher redshifts where there is a considerable neutral
fraction.  Additionally, since the 21 cm line is optically thin and
does not saturate, the 21 cm forest can contain detailed information
about the properties of the IGM as well as collapsed gaseous
halos. There is a clear complementarity to the two forests for probing
the evolution of the IGM over a wide range of redshifts.

The 21 cm forest differs from the emission signal of the diffuse
medium in a number of important ways.  First, because the brightness
temperature of radio loud quasars is typically $\sim10^{11-12}\,\K$
the 21 cm line is always seen in absorption against the source. The
strength of the absorption feature depends upon the 21 cm optical
depth
\begin{eqnarray}\label{forest_optical_depth}
\tau_{\nu_0}(z)&=&\frac{3}{32\pi}\frac{h_pc^3 A_{10}}{k_B\nu_0^2}\frac{x_{\rm HI} n_{\rm H}(z)}{T_S(1+z)(\ud v_\parallel/\ud r_\parallel)},\\
&\approx& 0.009(1+\delta)(1+z)^{3/2}\frac{x_{\rm HI}}{T_S}.\nonumber
\end{eqnarray}
We see from this expression that the decrement depends primarily upon
the neutral fraction and spin temperature.  In contrast to the case
where the CMB is the backlight, there is no saturation regime at large
values of $T_S$.  The decrement is maximised for a fully neutral and
cold IGM - heating or ionizing the gas will reduce the observable
signature.

\begin{figure}[htbp]
\begin{center}
\includegraphics[scale=0.3]{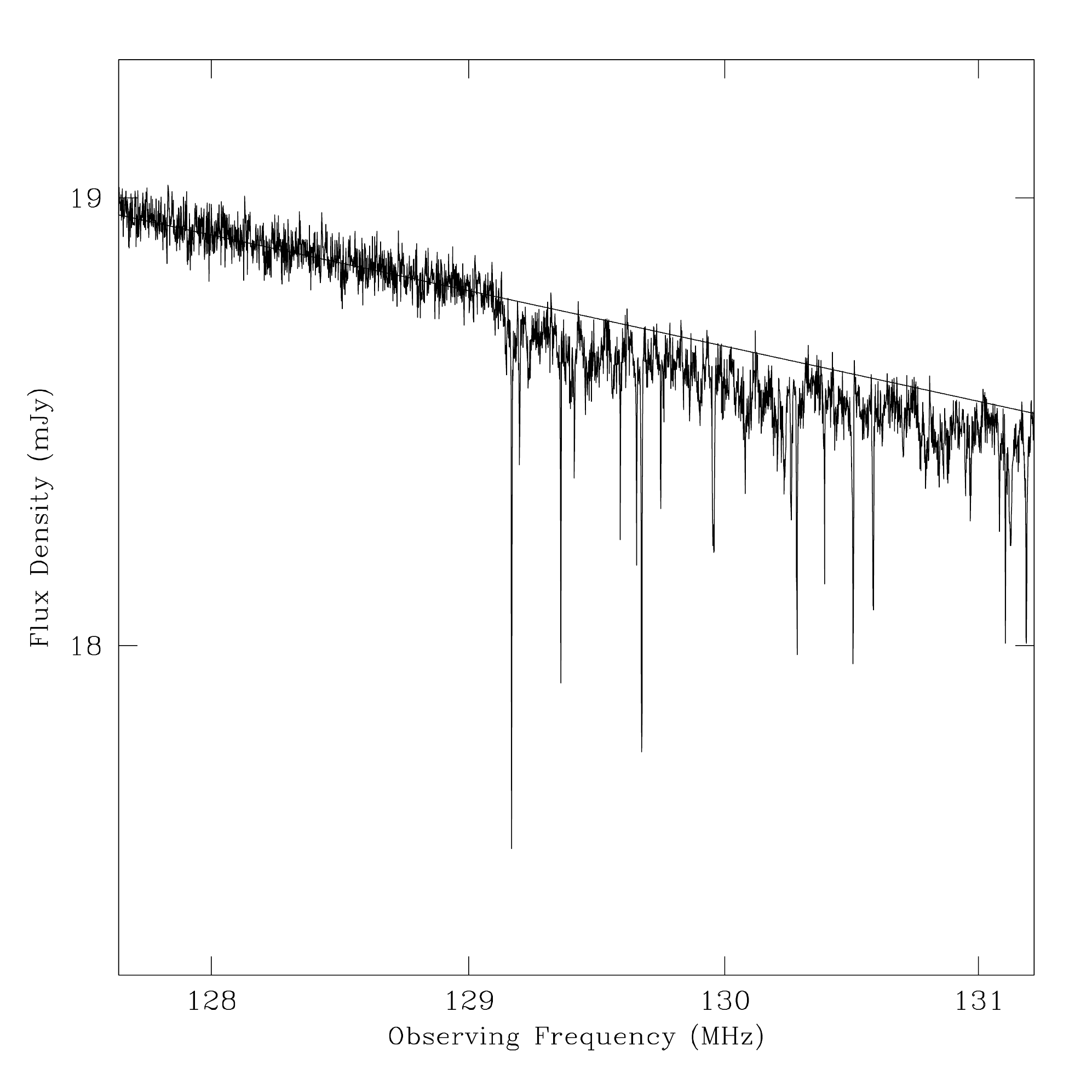}
\caption{ Simulated spectrum from 128 to 131 MHz of a source with
brightness $S(120{\,\rm MHz}) = 20 {\,\rm mJy}$ at $z = 10$ using a
model spectrum based on that of Cygnus A and assuming HI 21 cm
absorption by the IGM. Thermal noise has been added using the
specifications of the square kilometer array (SKA) 
and assuming 10 days integration with 1 kHz wide spectral channels.
The solid line is the model spectrum without noise or absorption
\cite{carilli2002}.}
\label{fig:carilli_forest}
\end{center}
\end{figure}

This signature may show several distinct features: {\em (i)} A mean
intensity decrement blueward of the 21 cm restframe frequency whose
depth depends on the mean IGM optical depth at that redshift; {\em
(ii)} small-scale variations in the intensity due to fluctuations in
the density, neutral fraction and temperature of the IGM at different
points along the line of sight; ({\em iii}) transmission windows due
to photoionized bubbles along the line of sight; and {\em (iv)} deep
absorption features arising from the dense neutral hydrogen clouds in
dwarf galaxies and minihalos \cite{carilli2002,furlanetto2002}.
Figure \ref{fig:carilli_forest} shows an example of a 21 cm forest
spectrum in which a number of these features can be seen.

\begin{figure}[htbp]
\begin{center}
\includegraphics[scale=0.3,angle=270]{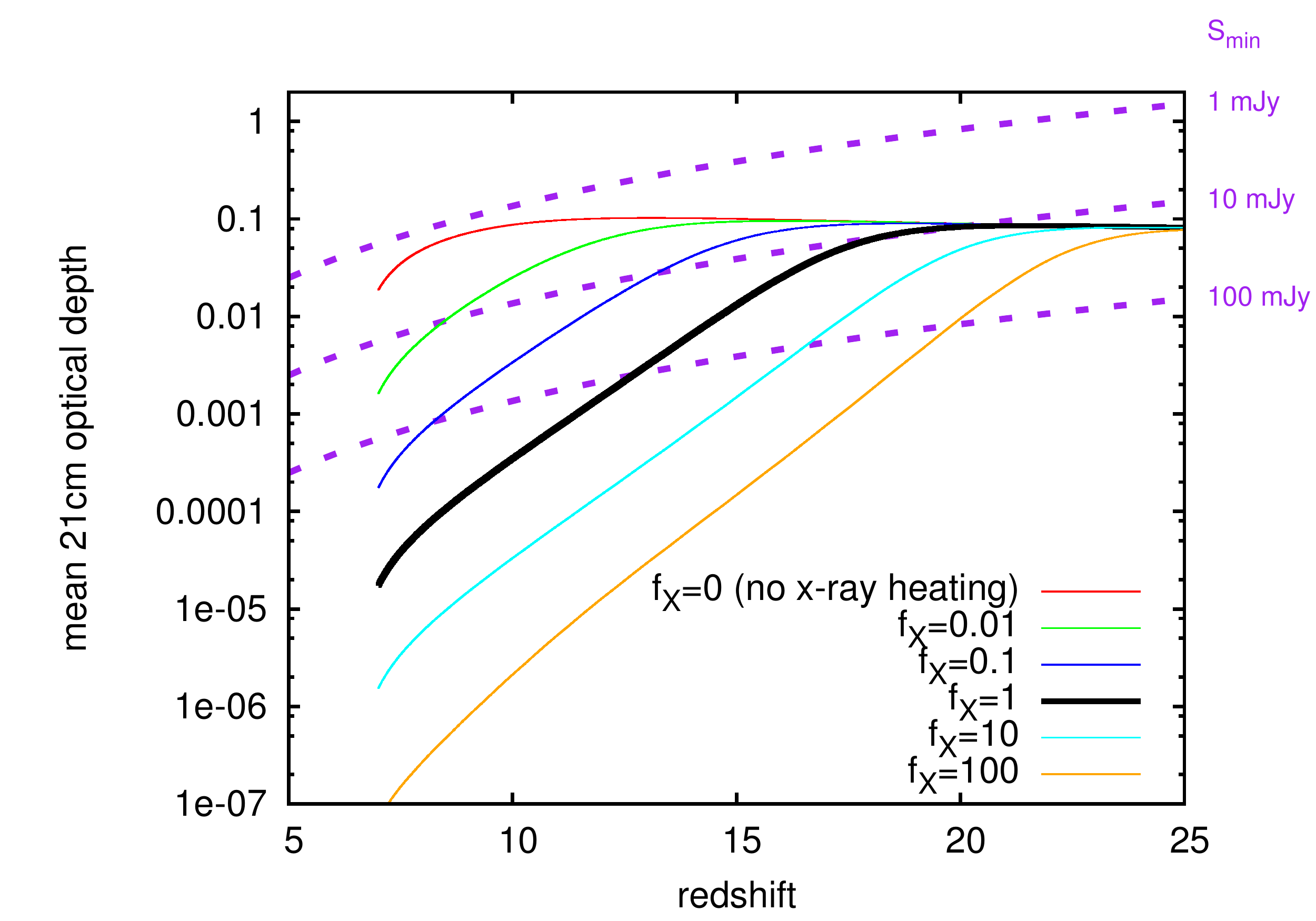}
\caption{Mean 21cm optical depth as a function of redshift, for
varying $f_X$ values: from top to bottom, $f_X$ = 0 (no x-ray
heating), 0.01, 0.1, 1 (thick black line), 10, and 100. Also included
are lines of $S_{\rm min}$ (dashed purple lines) as defined in \eref{smin_eqn}.
These lines indicate, for each value of $\tau$ on the left-hand
axis, the minimum observed flux density of a source that would allow a
detection at signal-to-noise of 5 of the flux decrement due to
absorption, assuming an array with effective area $A_{\rm eff} = 10^6
{\rm m^2}$, frequency resolution $\Delta\nu_{\rm ch}$ = 1 kHz, system
temperature $T_{\rm sys} = 100 {\rm K}(\nu/200{\rm\,MHz})^{-2.8}$, and integration time $t_{\rm int}$
= 1 week. When the mean optical depth lines (solid) cross above the
$S_{\rm min}$ lines (dashed), the flux decrement is detectable at the
corresponding redshift \cite{mack2011}.}
\label{fig:mack_forest}
\end{center}
\end{figure}

Since the evolution of the optical depth depends on the mean neutral
fraction and the spin temperature, we can understand the evolution of
the 21 cm forest based on Figure \ref{fig:pl_global}.  At early times,
the IGM is fully neutral and the evolution is dictated by the spin
temperature.  $T_S$ tracks the gas kinetic temperature and rises from
tens of K to thousands of K by the end of reionization, causing
$\tau_{\nu_0}$ to fall by several orders of magnitude due to heating
alone.  Then, as reionization takes place and the neutral fraction
drops from $x_H=1$ to the $x_H\sim10^{-4}$ seen in the \lya forest,
the optical depth drops even further.  Tracing the evolution of the
mean optical depth would provide a useful constraint on the thermal
evolution of the IGM and give a clear indication of the end of
reionization.  Figure \ref{fig:mack_forest} shows the evolution of
$\tau_{\nu_0}$ in a model where $T_S=T_K$ showing the very significant
evolution in the optical depth with redshift.

Detecting the mean decrement in the 21 cm forest may be challenging
since it is relatively weak and requires detailed fitting of the
unabsorbed continuum level. A potentially more robust method is to
exploit the statistics of individual features.  As reionization
occurs, the appearance of ionized bubbles will show up in the forest
as an increasing number of windows of near total transparency.  If
these lines could be resolved, the distribution of equivalent widths
would give a measure of the process of reionization
\cite{carilli2002,furlanetto2002,xu2009}.

It is possible that the forest does not have to be fully resolved and could be analysed statistically instead.  For example,
the appearance of ionized regions will lead to a change in the
variance of the signal in different frequency bins. This has been shown to be an effective discriminant of the end of reionization \cite{mack2011} and has the advantage of not requiring narrow features to be resolved in frequency.

The signal from the diffuse IGM leads to a relatively low optical depth, as seen in \eref{forest_optical_depth}.  The signal from minihalos can be considerably stronger
\cite{furlanetto2002,iliev2002,xu2011}.  These are
structures that have collapsed and virialised, but because of their low mass to not reach the temperature of $10^4{\,\rm K}$ required for atomic hydrogen cooling.  Their high density contrast and relatively low temperature can lead to optical depths as high as $\tau\sim0.1$ within a frequency width of $\Delta\nu\sim 2{\rm\,kHz}$ \cite{furlanetto2006forest}.  The mean overdensity required to give an optical depth $\tau$ is given by
\begin{equation}
(1+\delta)\sim40\left(\frac{\tau}{0.01}\right)^2\left(\frac{T_S}{100{\rm\,K}}\right)^2\left(\frac{10}{1+z}\right)^3.
\end{equation}
The number density of minihalos is sensitive to the thermal properties
of the IGM, which if heated may raise the Jeans mass sufficiently to
prevent the collapse of the least massive minihalos that dominate the
signal in the forest.  Detailed descriptions of line profiles and their
statistics have been considered based upon analytic models of the
density distribution around halos
\cite{furlanetto2002,xu2010,xu2011,meiksin2011} and in high resolution
simulations \cite{shapiro2006}.  Even a single line of sight showing
such structures could be used to provide information on early halo
formation.

Unfortunately, the sensitivity required to detect the 21 cm forest is
challenging.  We can consider the thermal noise required to directly
detect the flux decrement arising from the mean absorption.  Using the
relationship $S_\nu=2k_B T_a/A_e$ to relate the antennae temperature
to the flux sensitivity for a single polarisation and the radiometer
equation gives the minimum source brightness for the decrement to be
visible
\begin{equation}\label{smin_eqn}
S_{\rm min}=16{\,\rm mJy}\left[\frac{S/N}{5}\frac{0.01}{\tau}\frac{10^6{\,\rm m^2}}{A_{\rm eff}} \frac{T_{\rm sys}}{400{\rm\,K}}\right]\sqrt{\frac{1{\rm\,kHz}}{\Delta\nu}\frac{1{\rm\,week}}{t_{\rm int}}},
\end{equation}
where we have assumed that both polarisation states are measured.  The
value of $A_{\rm eff}/T_{\rm sys}$ used is appropriate for SKA giving
a sense of the challenge in detecting the forest.  The system
temperature is typically set by the sky temperature and is determined by galactic foregrounds.

The major uncertainty in the utility of the 21 cm forest is the
existence of radio loud sources at high redshifts.  The most promising
set of sources are radio loud quasars.  We are currently ignorant of
the number density of radio-loud quasars at high redshifts. Quasars
have currently been observed to redshifts as high as $z=7.1$
\cite{mortlock2011}, but their number density drops rapidly at
redshifts $z\gtrsim3$ \cite{hopkins2007}.  Furthermore, the radio loud
fraction of quasars is poorly understood and may evolve at higher
redshifts \cite{jiang2007}.  Models calibrated to the number of counts
at low redshifts lead to predictions of $\sim2000$ sources across the
sky with $S>6{\rm\,Jy}$ at $8<z<12$ \cite{haiman2004}. Many of these
sources would show up in existing radio surveys, but it is currently
impossible to distinguish these high-z radio bright quasars from low-z
radio galaxies.  Future large area surveys with NIR photometry, such
as EUCLID or WFIRST could enable the identification of these high-z
quasars providing targets for the SKA \cite{willott2010}.  Gamma ray
bursts (GRBs) and hypernovae may provide alternative high-z sources
\cite{ioka2005,souza2011,campisi2011}, but are believed to have a lower maximum
brightness at the low frequencies of interest because of synchrotron
self-absorption.  GRBs do however have the virtue of already having
been observed in the desired redshift range.

This rapid decline in radio-bright quasars is unfortunate since the
signal is most strong early on before ionization and heating have
begun significantly.  In any case, it seems likely that only a few
sight lines to radio bright objects will become available for 21 cm
forest observations.  Since these observations are affected by totally
different systematics than 21 cm tomography, they could be very important
for identifying the properties of the IGM in the early Universe.

\section{Conclusions and outlook}
\label{sec:conclusions}

Cosmology in the 21st century is engaged in the process of pushing the boundaries of sensitivity and completeness of our understanding of the Universe.  Observations of the redshifted 21 cm line offer a new window into the properties of the Universe at redshifts $z=1-150$ filling in a crucial gap in observations of the period where the first structures and stars had formed.  The 21 cm signal depends upon physics on different scales from the atomic physics of hydrogen, through the astrophysics of star formation, to the physics of cosmological structures.  Through this dependence the 21 cm signal has enormous potential to improve our understanding of the Universe.  Over the next decade, radio experiments will begin the process of exploiting 21 cm observations and will reveal how much of this potential can be realised.

In this review, we have explored four key ways in which 21 cm observations might be exploited to learn about cosmology and astrophysics.  Low angular resolution observations with small numbers of radio dipoles can measure the sky averaged ``global" 21 cm signal providing information about key milestones in the history of the Universe. Understanding these moments when the first radiation backgrounds illuminated intergalactic space will place constraints on the properties of the first galaxies and limit the contribution of more exotic physics.  Experiments involving just a handful of people, such as EDGES and CoRE, have taken the first steps along this path. There is currently a growing effort to using different radio dipole designs and attempting to combine the benefits of interferometers with individual absolutely calibrated dipoles.  There is a significant challenge in separating a relatively smooth 21 cm signal from smooth galactic foregrounds and much hard work ahead before these experiments yield robust results.

More ambitiously, collections of thousands of dipoles, digitally combined into radio interferometers will have the raw sensitivity to measure fluctuations in the 21 cm signal, providing detailed information about how the Universe was heated and ionized.  The first generation of such instruments - LOFAR, MWA, PAPER, and 21CMA - have or soon will be completed and data is starting to become available.  These first instruments will peer through a hazy window of low sensitivity, but should determine when the epoch of reionization took place.  Looking further ahead, the SKA will see this period clearly mapping the ionized structures and providing information about heating and coupling of the IGM during the formation of the first galaxies. The combination of 21 cm observations with observations of the CMB and with surveys of high redshift galaxies and information from the Lyman alpha forest should enable the reionization history and details of reionization to be inferred \cite{pritchard2010combo,mitra2011}. 

21 cm observations also have the potential to constrain fundamental cosmology and represent a path to a new level of precision cosmology.  Exploiting 21 cm observations for fundamental cosmology will be a challenging task and it is unclear how much can realistically be hoped for.  During the EoR astrophysics mixes with cosmology in the 21 cm fluctuations and until the performance of instruments is better understood it is not clear whether these two can be separated.  In the further future, 21 cm observations of the cosmic dark ages provide a long term hope that most of the volume of the Universe could one day be mapped and used for cosmology.

Applying similar techniques at lower redshifts allows for intensity mapping of the neutral hydrogen in galaxies, providing an alternative to traditional redshift surveys of galaxies in the optical-infrared bands. In contrast to the traditional approach, only the cumulative line emission from many galaxies over large volumes is being mapped, and galaxies are not being resolved individually.  This intensity mapping technique could help shed light on the properties of
dark energy through the use of baryon acoustic oscillations as a standard ruler to measure the expansion of the Universe.  Further, intensity mapping in other molecular and atomic lines can complement 21 cm tomography allowing a complete picture of the formation of stars and metals to be developed.

Finally, observations of individual radio bright sources, such as quasars, might allow observations of the 21 cm forest.  This provides another window into the properties of the IGM at the end of reionization that depends on very different systematics than 21 cm tomography.  The 21 cm forest would allow the small scale properties of the IGM to be studied in great detail and so constrain the properties of dark matter.  This is a powerful technique and the main uncertainty is the abundance of target sources.  Large area NIR surveys should begin to answer this question soon, as more high redshift quasars are found and followed up to determine the radio loud fraction.  Just a handful of sufficiently bright radio sources at $z\gtrsim7$ would make this a very valuable probe of reionization and the IGM.

21 cm cosmology is a new field with much yet to be discovered, but it is one with great potential for the future.  Much of this potential stems from breakthroughs in computing power which make ``digital radio astronomy" feasible and allow for extremely large radio interferometers.  It is to be hoped that in the decades to come 21 cm observations will transform our understanding of the cosmic dawn and the epoch of reionization pushing farther our detailed understanding of the cosmos.

\section*{Acknowledgments}
AL
acknowledges funding from NSF grant AST-0907890 and NASA grants
NNX08AL43G and NNA09DB30A.

\section*{References}
\bibliographystyle{unsrt}
\bibliography{review}{}

\begin{thebibliography}{100}

\bibitem{santos2008}
M.~G. {Santos}, A.~{Amblard}, J.~{Pritchard}, H.~{Trac}, R.~{Cen}, and
  A.~{Cooray}.
\newblock {Cosmic Reionization and the 21 cm Signal: Comparison between an
  Analytical Model and a Simulation}.
\newblock {\em \apj}, 689:1--16, December 2008.

\bibitem{komatsu2011}
E.~{Komatsu} et~al.
\newblock {Seven-year Wilkinson Microwave Anisotropy Probe (WMAP) Observations:
  Cosmological Interpretation}.
\newblock {\em \apjs}, 192:18, February 2011.

\bibitem{fob}
S.~R. {Furlanetto}, S.~P. {Oh}, and F.~H. {Briggs}.
\newblock {Cosmology at low frequencies: The 21 cm transition and the
  high-redshift Universe}.
\newblock {\em \physrep}, 433:181--301, October 2006.

\bibitem{morales2010}
M.~F. {Morales} and J.~S.~B. {Wyithe}.
\newblock {Reionization and Cosmology with 21-cm Fluctuations}.
\newblock {\em \araa}, 48:127--171, September 2010.

\bibitem{goldenberg1960}
H.~M. Goldenberg, D.~Kleppner, and N.~F. Ramsey.
\newblock Atomic hydrogen maser.
\newblock {\em Phys. Rev. Lett.}, 5(8):361--362, Oct 1960.

\bibitem{vandehulst1945}
H.C. {van de Hulst}.
\newblock {Radiogolven uit het Wereldruim (Radio waves from space)}.
\newblock {\em Ned. Tijdschr. Natuurk.}, 11:210--221, 1945.

\bibitem{ewen1951}
H.~I. {Ewen} and E.~M. {Purcell}.
\newblock {Observation of a Line in the Galactic Radio Spectrum: Radiation from
  Galactic Hydrogen at 1,420 Mc./sec.}
\newblock {\em \nat}, 168:356, September 1951.

\bibitem{kanekar2007}
N.~{Kanekar}, J.~N. {Chengalur}, and W.~M. {Lane}.
\newblock {HI 21-cm absorption at $z\sim3.39$ towards PKS 0201+113}.
\newblock {\em \mnras}, 375:1528--1536, March 2007.

\bibitem{srianand2010}
R.~{Srianand}, N.~{Gupta}, P.~{Petitjean}, P.~{Noterdaeme}, and C.~{Ledoux}.
\newblock {Detection of 21-cm, H$_{2}$ and deuterium absorption at $z>3$ along
  the line of sight to J1337+3152}.
\newblock {\em \mnras}, 405:1888--1900, July 2010.

\bibitem{bagla2009}
J.~S. {Bagla} and A.~{Loeb}.
\newblock {The hyperfine transition of $^3$He$^+$ as a probe of the
  intergalactic medium}.
\newblock {\em ArXiv e-prints}, May 2009.

\bibitem{mcquinn2009}
M.~{McQuinn} and E.~R. {Switzer}.
\newblock {Redshifted intergalactic $^3$He$^+$ 8.7 GHz hyperfine absorption}.
\newblock {\em \prd}, 80(6):063010, September 2009.

\bibitem{sigurdson2006}
K.~{Sigurdson} and S.~R. {Furlanetto}.
\newblock {Measuring the Primordial Deuterium Abundance during the Cosmic Dark
  Ages}.
\newblock {\em Physical Review Letters}, 97(9):091301, September 2006.

\bibitem{rybicki1986}
G.~B. {Rybicki} and A.~P. {Lightman}.
\newblock {\em {Radiative Processes in Astrophysics}}.
\newblock Wiley-VCH, June 1986.

\bibitem{sobolev1957}
V.~V. {Sobolev}.
\newblock {The Diffusion of L{$\alpha$} Radiation in Nebulae and Stellar
  Envelopes.}
\newblock {\em \sovast}, 1:678, October 1957.

\bibitem{field1958}
G.~B. {Field}.
\newblock {Excitation of the Hydrogen 21cm line}.
\newblock {\em Proc.~I.~R.~E.}, 46:240, 1958.

\bibitem{allison1969}
A.~C. {Allison} and A.~{Dalgarno}.
\newblock {Spin Change in Collisions of Hydrogen Atoms}.
\newblock {\em \apj}, 158:423, October 1969.

\bibitem{zygelman2005}
B.~{Zygelman}.
\newblock {Hyperfine Level-changing Collisions of Hydrogen Atoms and Tomography
  of the Dark Age Universe}.
\newblock {\em \apj}, 622:1356--1362, April 2005.

\bibitem{furlanettobros}
S.~R. {Furlanetto} and M.~R. {Furlanetto}.
\newblock {Spin-exchange rates in electron-hydrogen collisions}.
\newblock {\em \mnras}, 374:547--555, January 2007.

\bibitem{furlanettobros_pH}
S.~R. {Furlanetto} and M.~R. {Furlanetto}.
\newblock {Spin exchange rates in proton-hydrogen collisions}.
\newblock {\em \mnras}, 379:130--134, July 2007.

\bibitem{kuhlen2006}
M.~{Kuhlen}, P.~{Madau}, and R.~{Montgomery}.
\newblock {The Spin Temperature and 21 cm Brightness of the Intergalactic
  Medium in the Pre-Reionization era}.
\newblock {\em \apjl}, 637:L1--L4, January 2006.

\bibitem{liszt2001}
H.~{Liszt}.
\newblock {The spin temperature of warm interstellar H I}.
\newblock {\em \aap}, 371:698--707, May 2001.

\bibitem{hirata2006col}
C.~M. {Hirata} and K.~{Sigurdson}.
\newblock {The spin-resolved atomic velocity distribution and 21-cm line
  profile of dark-age gas}.
\newblock {\em \mnras}, 375:1241--1264, March 2007.

\bibitem{wouth1952}
S.~A. {Wouthuysen}.
\newblock {On the excitation mechanism of the 21-cm (radio-frequency)
  interstellar hydrogen emission line.}
\newblock {\em \aj}, 57:31, 1952.

\bibitem{pritchard2006}
J.~R. {Pritchard} and S.~R. {Furlanetto}.
\newblock {Descending from on high: Lyman-series cascades and spin-kinetic
  temperature coupling in the 21-cm line}.
\newblock {\em \mnras}, 367:1057--1066, April 2006.

\bibitem{meiksin2000}
A.~{Meiksin}.
\newblock {Detecting the epoch of first light in 21-CM radiation}.
\newblock In {\em {Perspectives on Radio Astronomy: Science with Large Antennae
  Arrays}}, page~37. M.P. van Harlem (Ed.), 2000.

\bibitem{rybicki2006}
G.~B. {Rybicki}.
\newblock {Improved Fokker-Planck Equation for Resonance-Line Scattering}.
\newblock {\em \apj}, 647:709--718, August 2006.

\bibitem{field1959relax}
G.~B. {Field}.
\newblock {The Time Relaxation of a Resonance-Line Profile.}
\newblock {\em \apj}, 129:551, May 1959.

\bibitem{chen2004}
X.~{Chen} and J.~{Miralda-Escud{\'e}}.
\newblock {The Spin-Kinetic Temperature Coupling and the Heating Rate due to
  Ly{$\alpha$} Scattering before Reionization: Predictions for 21 Centimeter
  Emission and Absorption}.
\newblock {\em \apj}, 602:1--11, February 2004.

\bibitem{hirata2006lya}
C.~M. {Hirata}.
\newblock {Wouthuysen-Field coupling strength and application to high-redshift
  21-cm radiation}.
\newblock {\em \mnras}, 367:259--274, March 2006.

\bibitem{chuzhoy2006heat}
L.~{Chuzhoy} and P.~R. {Shapiro}.
\newblock {Heating and Cooling of the Early Intergalactic Medium by Resonance
  Photons}.
\newblock {\em \apj}, 655:843--846, February 2007.

\bibitem{furlanetto2006heat}
S.~R. {Furlanetto} and J.~R. {Pritchard}.
\newblock {The scattering of Lyman-series photons in the intergalactic medium}.
\newblock {\em \mnras}, 372:1093--1103, November 2006.

\bibitem{loeb_zald2004}
A.~{Loeb} and M.~{Zaldarriaga}.
\newblock {Measuring the Small-Scale Power Spectrum of Cosmic Density
  Fluctuations through 21cm Tomography Prior to the Epoch of Structure
  Formation}.
\newblock {\em Physical Review Letters}, 92(21):211301, May 2004.

\bibitem{bl2005detect}
R.~{Barkana} and A.~{Loeb}.
\newblock {Detecting the Earliest Galaxies through Two New Sources of 21
  Centimeter Fluctuations}.
\newblock {\em \apj}, 626:1--11, June 2005.

\bibitem{chen2006}
X.~{Chen} and J.~{Miralda-Escude}.
\newblock {The 21cm Signature of the First Stars}.
\newblock {\em \apj}, submitted (astro-ph/0605439), May 2006.

\bibitem{santos2006}
M.~G. {Santos} and A.~{Cooray}.
\newblock {Cosmological and astrophysical parameter measurements with 21-cm
  anisotropies during the era of reionization}.
\newblock {\em \prd}, 74(8):083517, October 2006.

\bibitem{fzh2004}
S.~R. {Furlanetto}, M.~{Zaldarriaga}, and L.~{Hernquist}.
\newblock {The Growth of H II Regions During Reionization}.
\newblock {\em \apj}, 613:1--15, September 2004.

\bibitem{nasser2005}
A.~{Nusser}.
\newblock {The spin temperature of neutral hydrogen during cosmic
  pre-reionization}.
\newblock {\em \mnras}, 359:183--190, May 2005.

\bibitem{kuhlen2005}
M.~{Kuhlen} and P.~{Madau}.
\newblock {The first miniquasar}.
\newblock {\em \mnras}, 363:1069--1082, November 2005.

\bibitem{furlanetto2004}
S.~R. {Furlanetto} and A.~{Loeb}.
\newblock {Large-Scale Structure Shocks at Low and High Redshifts}.
\newblock {\em \apj}, 611:642--654, August 2004.

\bibitem{geil2009}
P.~M. {Geil} and J.~S.~B. {Wyithe}.
\newblock {Modification of the 21-cm power spectrum by quasars during the epoch
  of reionization}.
\newblock {\em \mnras}, 399:1877--1887, November 2009.

\bibitem{seager1999}
S.~{Seager}, D.~D. {Sasselov}, and D.~{Scott}.
\newblock {A New Calculation of the Recombination Epoch}.
\newblock {\em \apjl}, 523:L1--L5, September 1999.

\bibitem{me2000}
J.~{Miralda-Escud{\'e}}, M.~{Haehnelt}, and M.~J. {Rees}.
\newblock {Reionization of the Inhomogeneous Universe}.
\newblock {\em \apj}, 530:1--16, February 2000.

\bibitem{bromm2001}
V.~{Bromm}, R.~P. {Kudritzki}, and A.~{Loeb}.
\newblock {Generic Spectrum and Ionization Efficiency of a Heavy Initial Mass
  Function for the First Stars}.
\newblock {\em \apj}, 552:464--472, May 2001.

\bibitem{zackrisson2011}
E.~{Zackrisson}, C.-E. {Rydberg}, D.~{Schaerer}, G.~{{\"O}stlin}, and
  M.~{Tuli}.
\newblock {The Spectral Evolution of the First Galaxies. I. James Webb Space
  Telescope Detection Limits and Color Criteria for Population III Galaxies}.
\newblock {\em \apj}, 740:13, October 2011.

\bibitem{bl2001}
R.~{Barkana} and A.~{Loeb}.
\newblock {In the beginning: the first sources of light and the reionization of
  the universe}.
\newblock {\em \physrep}, 349:125--238, July 2001.

\bibitem{sheth1999mfn}
R.~K. {Sheth} and G.~{Tormen}.
\newblock {Large-scale bias and the peak background split}.
\newblock {\em \mnras}, 308:119--126, September 1999.

\bibitem{pawlik2009}
A.~H. {Pawlik}, J.~{Schaye}, and E.~{van Scherpenzeel}.
\newblock {Keeping the Universe ionized: photoheating and the clumping factor
  of the high-redshift intergalactic medium}.
\newblock {\em \mnras}, 394:1812--1824, April 2009.

\bibitem{bolton2009}
J.~S. {Bolton} and G.~D. {Becker}.
\newblock {Resolving the high redshift Ly{$\alpha$} forest in smoothed particle
  hydrodynamics simulations}.
\newblock {\em \mnras}, 398:L26--L30, September 2009.

\bibitem{furlanetto2005tax}
S.~R. {Furlanetto} and S.~P. {Oh}.
\newblock {Taxing the rich: recombinations and bubble growth during
  reionization}.
\newblock {\em \mnras}, 363:1031--1048, November 2005.

\bibitem{naoz2005}
S.~{Naoz} and R.~{Barkana}.
\newblock {Growth of linear perturbations before the era of the first
  galaxies}.
\newblock {\em \mnras}, 362:1047--1053, September 2005.

\bibitem{mmr1997}
P.~{Madau}, A.~{Meiksin}, and M.~J. {Rees}.
\newblock {21 Centimeter Tomography of the Intergalactic Medium at High
  Redshift}.
\newblock {\em \apj}, 475:429, February 1997.

\bibitem{venkatesan2001}
A.~{Venkatesan}, M.~L. {Giroux}, and J.~M. {Shull}.
\newblock {Heating and Ionization of the Intergalactic Medium by an Early X-Ray
  Background}.
\newblock {\em \apj}, 563:1--8, December 2001.

\bibitem{pritchard2007xray}
J.~R. {Pritchard} and S.~R. {Furlanetto}.
\newblock {21-cm fluctuations from inhomogeneous X-ray heating before
  reionization}.
\newblock {\em \mnras}, 376:1680--1694, April 2007.

\bibitem{zaroubi2007}
S.~{Zaroubi}, R.~M. {Thomas}, N.~{Sugiyama}, and J.~{Silk}.
\newblock {Heating of the intergalactic medium by primordial miniquasars}.
\newblock {\em \mnras}, 375:1269--1279, March 2007.

\bibitem{ciardi2010}
B.~{Ciardi}, R.~{Salvaterra}, and T.~{Di Matteo}.
\newblock {Ly{$\alpha$} versus X-ray heating in the high-z intergalactic
  medium}.
\newblock {\em \mnras}, 401:2635--2640, February 2010.

\bibitem{SVS1985}
J.~M. {Shull} and M.~E. {van Steenberg}.
\newblock {X-ray secondary heating and ionization in quasar emission-line
  clouds}.
\newblock {\em \apj}, 298:268--274, November 1985.

\bibitem{valdes2008}
M.~{Vald{\'e}s} and A.~{Ferrara}.
\newblock {The energy cascade from warm dark matter decays}.
\newblock {\em \mnras}, 387:L8--L12, June 2008.

\bibitem{furlanetto2010cascade}
S.~R. {Furlanetto} and S.~J. {Stoever}.
\newblock {Secondary ionization and heating by fast electrons}.
\newblock {\em \mnras}, 404:1869--1878, June 2010.

\bibitem{verner1996}
D.~A. {Verner}, G.~J. {Ferland}, K.~T. {Korista}, and D.~G. {Yakovlev}.
\newblock {Atomic Data for Astrophysics. II. New Analytic FITS for
  Photoionization Cross Sections of Atoms and Ions}.
\newblock {\em \apj}, 465:487, July 1996.

\bibitem{oh2001}
S.~P. {Oh}.
\newblock {Reionization by Hard Photons. I. X-Rays from the First Star
  Clusters}.
\newblock {\em \apj}, 553:499--512, June 2001.

\bibitem{glover2003}
S.~C.~O. {Glover} and P.~W.~J.~L. {Brand}.
\newblock {Radiative feedback from an early X-ray background}.
\newblock {\em \mnras}, 340:210--226, March 2003.

\bibitem{furlanetto2006}
S.~R. {Furlanetto}.
\newblock {The global 21-centimeter background from high redshifts}.
\newblock {\em \mnras}, 371:867--878, September 2006.

\bibitem{grimm2003}
{H.-J.} {Grimm}, M.~{Gilfanov}, and R.~{Sunyaev}.
\newblock {High-mass X-ray binaries as a star formation rate indicator in
  distant galaxies}.
\newblock {\em \mnras}, 339:793--809, March 2003.

\bibitem{gilfanov2004}
M.~{Gilfanov}, {H.-J.} {Grimm}, and R.~{Sunyaev}.
\newblock {$L_{X}$-SFR relation in star-forming galaxies}.
\newblock {\em \mnras}, 347:L57--L60, January 2004.

\bibitem{lehmer2010}
B.~D. {Lehmer}, D.~M. {Alexander}, F.~E. {Bauer}, W.~N. {Brandt}, A.~D.
  {Goulding}, L.~P. {Jenkins}, A.~{Ptak}, and T.~P. {Roberts}.
\newblock {A Chandra Perspective on Galaxy-wide X-ray Binary Emission and its
  Correlation with Star Formation Rate and Stellar Mass: New Results from
  Luminous Infrared Galaxies}.
\newblock {\em \apj}, 724:559--571, November 2010.

\bibitem{mineo2010}
S.~{Mineo}, M.~{Gilfanov}, and R.~{Sunyaev}.
\newblock {The collective X-ray luminosity of HMXB as a SFR indicator}.
\newblock {\em Astronomische Nachrichten}, 332:349, May 2011.

\bibitem{dijkstra2011}
M.~{Dijkstra}, M.~{Gilfanov}, A.~{Loeb}, and R.~{Sunyaev}.
\newblock {Constraints on the Redshift Evolution of the L\_X-SFR Relation from
  the Cosmic X-Ray Backgrounds}.
\newblock {\em ArXiv e-prints}, August 2011.

\bibitem{mirabel2011}
I.~F. {Mirabel}, M.~{Dijkstra}, P.~{Laurent}, A.~{Loeb}, and J.~R. {Pritchard}.
\newblock {Stellar black holes at the dawn of the universe}.
\newblock {\em \aap}, 528:A149, April 2011.

\bibitem{madau2004}
P.~{Madau}, M.~J. {Rees}, M.~{Volonteri}, F.~{Haardt}, and S.~P. {Oh}.
\newblock {Early Reionization by Miniquasars}.
\newblock {\em \apj}, 604:484--494, April 2004.

\bibitem{ricotti2004}
M.~{Ricotti} and J.~P. {Ostriker}.
\newblock {X-ray pre-ionization powered by accretion on the first black holes -
  I. A model for the WMAP polarization measurement}.
\newblock {\em \mnras}, 352:547--562, August 2004.

\bibitem{dijkstra2004}
M.~{Dijkstra}, Z.~{Haiman}, and A.~{Loeb}.
\newblock {A Limit from the X-Ray Background on the Contribution of Quasars to
  Reionization}.
\newblock {\em \apj}, 613:646--654, October 2004.

\bibitem{chuzhoy2006}
L.~{Chuzhoy}, M.~A. {Alvarez}, and P.~R. {Shapiro}.
\newblock {Recognizing the First Radiation Sources through Their 21 cm
  Signature}.
\newblock {\em \apjl}, 648:L1--L4, September 2006.

\bibitem{leitherer1999}
C.~{Leitherer}, D.~{Schaerer}, J.~D. {Goldader}, R.~M. {Gonz{\'a}lez Delgado},
  C.~{Robert}, D.~F. {Kune}, D.~F. {de Mello}, D.~{Devost}, and T.~M.
  {Heckman}.
\newblock {Starburst99: Synthesis Models for Galaxies with Active Star
  Formation}.
\newblock {\em \apjs}, 123:3--40, July 1999.

\bibitem{shull1979}
J.~M. {Shull}.
\newblock {Heating and ionization by X-ray photoelectrons}.
\newblock {\em \apj}, 234:761--764, December 1979.

\bibitem{clark2011}
P.~C. {Clark}, S.~C.~O. {Glover}, R.~S. {Klessen}, and V.~{Bromm}.
\newblock {Gravitational Fragmentation in Turbulent Primordial Gas and the
  Initial Mass Function of Population III Stars}.
\newblock {\em \apj}, 727:110, February 2011.

\bibitem{caffau2011}
E.~{Caffau}, P.~{Bonifacio}, P.~{Fran{\c c}ois}, L.~{Sbordone}, L.~{Monaco},
  M.~{Spite}, F.~{Spite}, H.-G. {Ludwig}, R.~{Cayrel}, S.~{Zaggia},
  F.~{Hammer}, S.~{Randich}, P.~{Molaro}, and V.~{Hill}.
\newblock {An extremely primitive star in the Galactic halo}.
\newblock {\em \nat}, 477:67--69, September 2011.

\bibitem{bromm2003}
V.~{Bromm} and A.~{Loeb}.
\newblock {The formation of the first low-mass stars from gas with low carbon
  and oxygen abundances}.
\newblock {\em \nat}, 425:812--814, October 2003.

\bibitem{pritchard2008}
J.~R. {Pritchard} and A.~{Loeb}.
\newblock {Evolution of the 21cm signal throughout cosmic history}.
\newblock {\em \prd}, 78(10):103511, November 2008.

\bibitem{chen2004dm}
X.~{Chen} and M.~{Kamionkowski}.
\newblock {Particle decays during the cosmic dark ages}.
\newblock {\em \prd}, 70(4):043502, August 2004.

\bibitem{furlanetto2006dm}
S.~R. {Furlanetto}, S.~P. {Oh}, and E.~{Pierpaoli}.
\newblock {Effects of dark matter decay and annihilation on the high-redshift
  21cm background}.
\newblock {\em \prd}, 74(10):103502, November 2006.

\bibitem{valdes2007}
M.~{Vald{\'e}s}, A.~{Ferrara}, M.~{Mapelli}, and E.~{Ripamonti}.
\newblock {Constraining dark matter through 21-cm observations}.
\newblock {\em \mnras}, 377:245--252, May 2007.

\bibitem{belikov2009}
A.~V. {Belikov} and D.~{Hooper}.
\newblock {How dark matter reionized the Universe}.
\newblock {\em \prd}, 80(3):035007, August 2009.

\bibitem{slatyer2009}
T.~R. {Slatyer}, N.~{Padmanabhan}, and D.~P. {Finkbeiner}.
\newblock {CMB constraints on WIMP annihilation: Energy absorption during the
  recombination epoch}.
\newblock {\em \prd}, 80(4):043526, August 2009.

\bibitem{cumberbatch2008}
D.~T. {Cumberbatch}, M.~{Lattanzi}, and J.~{Silk}.
\newblock {Signatures of clumpy dark matter in the global 21 cm background
  signal}.
\newblock {\em \prd}, 82(10):103508, November 2010.

\bibitem{finkbeiner2008}
D.~P. {Finkbeiner}, N.~{Padmanabhan}, and N.~{Weiner}.
\newblock {CMB and 21-cm signals for dark matter with a long-lived excited
  state}.
\newblock {\em \prd}, 78(6):063530, September 2008.

\bibitem{mack2008}
K.~J. {Mack} and D.~H. {Wesley}.
\newblock {Primordial black holes in the Dark Ages: Observational prospects for
  future 21cm surveys}.
\newblock {\em ArXiv e-prints}, May 2008.

\bibitem{ricotti2008}
M.~{Ricotti}, J.~P. {Ostriker}, and K.~J. {Mack}.
\newblock {Effect of Primordial Black Holes on the Cosmic Microwave Background
  and Cosmological Parameter Estimates}.
\newblock {\em \apj}, 680:829--845, June 2008.

\bibitem{brandenberger2010}
R.~H. {Brandenberger}, R.~J. {Danos}, O.~F. {Hern{\'a}ndez}, and G.~P.
  {Holder}.
\newblock {The 21 cm signature of cosmic string wakes}.
\newblock {\em \jcap}, 12:28, December 2010.

\bibitem{chippendale2005}
A.~P. {Chippendale}, E.~{Subrahmanyan}, and R.~{Ekers}, 2005.
\newblock New Techniques and Results in Low Frequency Radio Astronomy, Hobart,
  Australia, December 6th-10th.

\bibitem{bowman2008}
J.~D. {Bowman}, A.~E.~E. {Rogers}, and J.~N. {Hewitt}.
\newblock {Toward Empirical Constraints on the Global Redshifted 21 cm
  Brightness Temperature During the Epoch of Reionization}.
\newblock {\em \apj}, 676:1--9, March 2008.

\bibitem{bowman2010}
J.~D. {Bowman} and A.~E.~E. {Rogers}.
\newblock {A lower limit of $\Delta z\gtrsim0.06$ for the duration of the
  reionization epoch}.
\newblock {\em \nat}, 468:796--798, December 2010.

\bibitem{EHT99}
D.~J. {Eisenstein}, W.~{Hu}, and M.~{Tegmark}.
\newblock {Cosmic Complementarity: Joint Parameter Estimation from Cosmic
  Microwave Background Experiments and Redshift Surveys}.
\newblock {\em \apj}, 518:2--23, June 1999.

\bibitem{pritchard2010}
J.~R. {Pritchard} and A.~{Loeb}.
\newblock {Constraining the unexplored period between the dark ages and
  reionization with observations of the global 21 cm signal}.
\newblock {\em \prd}, 82(2):023006, July 2010.

\bibitem{morandi2011}
A.~{Morandi} and R.~{Barkana}.
\newblock {Studying cosmic reionization with observations of the global 21-cm
  signal}.
\newblock {\em ArXiv e-prints}, February 2011.

\bibitem{burns2011}
J.~O. {Burns}, T.~J.~W. {Lazio}, S.~D. {Bale}, J.~D. {Bowman}, R.~F. {Bradley},
  C.~L. {Carilli}, S.~R. {Furlanetto}, G.~J.~A. {Harker}, A.~{Loeb}, and J.~R.
  {Pritchard}.
\newblock {Probing the First Stars and Black Holes in the Early Universe with
  the Dark Ages Radio Explorer (DARE)}.
\newblock {\em ArXiv e-prints}, June 2011.

\bibitem{harker2011}
G.~J.~A. {Harker}, J.~R. {Pritchard}, J.~O. {Burns}, and J.~D. {Bowman}.
\newblock {An MCMC approach to extracting the global 21-cm signal during the
  cosmic dawn from sky-averaged radio observations}.
\newblock {\em \mnras}, 419:1070--1084, January 2012.

\bibitem{morales_hewitt2004}
M.~F. {Morales} and J.~{Hewitt}.
\newblock {Toward Epoch of Reionization Measurements with Wide-Field Radio
  Observations}.
\newblock {\em \apj}, 615:7--18, November 2004.

\bibitem{bharadwaj2004}
S.~{Bharadwaj} and S.~S. {Ali}.
\newblock {The cosmic microwave background radiation fluctuations from HI
  perturbations prior to reionization}.
\newblock {\em \mnras}, 352:142--146, July 2004.

\bibitem{mcquinn2005}
M.~{McQuinn}, O.~{Zahn}, M.~{Zaldarriaga}, L.~{Hernquist}, and S.~R.
  {Furlanetto}.
\newblock {Cosmological Parameter Estimation Using 21 cm Radiation from the
  Epoch of Reionization}.
\newblock {\em \apj}, 653:815--834, December 2006.

\bibitem{bl2005sep}
R.~{Barkana} and A.~{Loeb}.
\newblock {A Method for Separating the Physics from the Astrophysics of
  High-Redshift 21 Centimeter Fluctuations}.
\newblock {\em \apjl}, 624:L65--L68, May 2005.

\bibitem{wang2006hu}
X.~{Wang} and W.~{Hu}.
\newblock {Redshift Space 21 cm Power Spectra from Reionization}.
\newblock {\em \apj}, 643:585--597, June 2006.

\bibitem{mellema2006}
G.~{Mellema}, I.~T. {Iliev}, U.-L. {Pen}, and P.~R. {Shapiro}.
\newblock {Simulating cosmic reionization at large scales - II. The 21-cm
  emission features and statistical signals}.
\newblock {\em \mnras}, 372:679--692, October 2006.

\bibitem{shaw2008}
J.~R. {Shaw} and A.~{Lewis}.
\newblock {Nonlinear redshift-space power spectra}.
\newblock {\em \prd}, 78(10):103512, November 2008.

\bibitem{mao2011}
Y.~{Mao}, P.~R. {Shapiro}, G.~{Mellema}, I.~T. {Iliev}, J.~{Koda}, and
  K.~{Ahn}.
\newblock {Redshift Space Distortion of the 21cm Background from the Epoch of
  Reionization I: Methodology Re-examined}.
\newblock {\em ArXiv e-prints}, April 2011.

\bibitem{barkana2004}
R.~{Barkana} and A.~{Loeb}.
\newblock {Unusually Large Fluctuations in the Statistics of Galaxy Formation
  at High Redshift}.
\newblock {\em \apj}, 609:474--481, July 2004.

\bibitem{ps1974mfn}
W.~H. {Press} and P.~{Schechter}.
\newblock {Formation of Galaxies and Clusters of Galaxies by Self-Similar
  Gravitational Condensation}.
\newblock {\em \apj}, 187:425--438, February 1974.

\bibitem{bond1991}
J.~R. {Bond}, S.~{Cole}, G.~{Efstathiou}, and N.~{Kaiser}.
\newblock {Excursion set mass functions for hierarchical Gaussian
  fluctuations}.
\newblock {\em \apj}, 379:440--460, October 1991.

\bibitem{zahn2006}
O.~{Zahn}, A.~{Lidz}, M.~{McQuinn}, S.~{Dutta}, L.~{Hernquist},
  M.~{Zaldarriaga}, and S.~R. {Furlanetto}.
\newblock {Simulations and Analytic Calculations of Bubble Growth during
  Hydrogen Reionization}.
\newblock {\em \apj}, 654:12--26, January 2007.

\bibitem{zahn2011}
O.~{Zahn}, A.~{Mesinger}, M.~{McQuinn}, H.~{Trac}, R.~{Cen}, and L.~E.
  {Hernquist}.
\newblock {Comparison of reionization models: radiative transfer simulations
  and approximate, seminumeric models}.
\newblock {\em \mnras}, page 532, April 2011.

\bibitem{scannapieco2002}
E.~{Scannapieco} and R.~{Barkana}.
\newblock {An Analytical Approach to Inhomogeneous Structure Formation}.
\newblock {\em \apj}, 571:585--603, June 2002.

\bibitem{barkana2007}
R.~{Barkana}.
\newblock {On correlated random walks and 21-cm fluctuations during cosmic
  reionization}.
\newblock {\em \mnras}, 376:1784--1792, April 2007.

\bibitem{santos2011}
M.~G. {Santos}, M.~B. {Silva}, J.~R. {Pritchard}, R.~{Cen}, and A.~{Cooray}.
\newblock {Probing the first galaxies with the Square Kilometer Array}.
\newblock {\em \aap}, 527:A93, March 2011.

\bibitem{chuzhoy2007}
L.~{Chuzhoy} and Z.~{Zheng}.
\newblock {Radiative Transfer Effect on Ultraviolet Pumping of the 21 cm Line
  in the High-Redshift Universe}.
\newblock {\em \apj}, 670:912--918, December 2007.

\bibitem{semelin2007}
B.~{Semelin}, F.~{Combes}, and S.~{Baek}.
\newblock {Lyman-alpha radiative transfer during the epoch of reionization:
  contribution to 21-cm signal fluctuations}.
\newblock {\em \aap}, 474:365--374, November 2007.

\bibitem{naoz2008}
S.~{Naoz} and R.~{Barkana}.
\newblock {Detecting early galaxies through their 21-cm signature}.
\newblock {\em \mnras}, 385:L63--L67, March 2008.

\bibitem{alvarez2010}
M.~A. {Alvarez}, U.-L. {Pen}, and T.-C. {Chang}.
\newblock {Enhanced Detectability of Pre-reionization 21 cm Structure}.
\newblock {\em \apjl}, 723:L17--L21, November 2010.

\bibitem{bl2005infall}
R.~{Barkana} and A.~{Loeb}.
\newblock {Probing the epoch of early baryonic infall through 21-cm
  fluctuations}.
\newblock {\em \mnras}, 363:L36--L40, October 2005.

\bibitem{iliev2002}
I.~T. {Iliev}, P.~R. {Shapiro}, A.~{Ferrara}, and H.~{Martel}.
\newblock {On the Direct Detectability of the Cosmic Dark Ages: 21 Centimeter
  Emission from Minihalos}.
\newblock {\em \apjl}, 572:L123--L126, June 2002.

\bibitem{iliev2003}
I.~T. {Iliev}, E.~{Scannapieco}, H.~{Martel}, and P.~R. {Shapiro}.
\newblock {Non-linear clustering during the cosmic Dark Ages and its effect on
  the 21-cm background from minihaloes}.
\newblock {\em \mnras}, 341:81--90, May 2003.

\bibitem{furlanetto2006mh}
S.~R. {Furlanetto} and S.~P. {Oh}.
\newblock {Redshifted 21 cm Emission from Minihalos before Reionization}.
\newblock {\em \apj}, 652:849--856, December 2006.

\bibitem{shapiro2006}
P.~R. {Shapiro}, K.~{Ahn}, M.~A. {Alvarez}, I.~T. {Iliev}, H.~{Martel}, and
  D.~{Ryu}.
\newblock {The 21 cm Background from the Cosmic Dark Ages: Minihalos and the
  Intergalactic Medium before Reionization}.
\newblock {\em \apj}, 646:681--690, August 2006.

\bibitem{tseliakhovich2010}
D.~{Tseliakhovich} and C.~{Hirata}.
\newblock {Relative velocity of dark matter and baryonic fluids and the
  formation of the first structures}.
\newblock {\em \prd}, 82(8):083520, October 2010.

\bibitem{tseliakhovich2010b}
D.~{Tseliakhovich}, R.~{Barkana}, and C.~M. {Hirata}.
\newblock {Suppression and spatial variation of early galaxies and minihaloes}.
\newblock {\em \mnras}, 418:906--915, December 2011.

\bibitem{dalal2010}
N.~{Dalal}, U.-L. {Pen}, and U.~{Seljak}.
\newblock {Large-scale BAO signatures of the smallest galaxies}.
\newblock {\em \jcap}, 11:7, November 2010.

\bibitem{stacy2011}
A.~{Stacy}, V.~{Bromm}, and A.~{Loeb}.
\newblock {Effect of Streaming Motion of Baryons Relative to Dark Matter on the
  Formation of the First Stars}.
\newblock {\em \apjl}, 730:L1, March 2011.

\bibitem{maio2011}
U.~{Maio}, L.~V.~E. {Koopmans}, and B.~{Ciardi}.
\newblock {The impact of primordial supersonic flows on early structure
  formation, reionization and the lowest-mass dwarf galaxies}.
\newblock {\em \mnras}, 412:L40--L44, March 2011.

\bibitem{babich2006}
D.~{Babich} and A.~{Loeb}.
\newblock {Imprint of Inhomogeneous Reionization on the Power Spectrum of
  Galaxy Surveys at High Redshifts}.
\newblock {\em \apj}, 640:1--7, March 2006.

\bibitem{pritchard2007bub}
J.~R. {Pritchard}, S.~R. {Furlanetto}, and M.~{Kamionkowski}.
\newblock {Galaxy surveys, inhomogeneous re-ionization and dark energy}.
\newblock {\em \mnras}, 374:159--167, January 2007.

\bibitem{wyithe2007}
J.~S.~B. {Wyithe} and A.~{Loeb}.
\newblock {The imprint of cosmic reionization on galaxy clustering}.
\newblock {\em \mnras}, 382:921--936, December 2007.

\bibitem{dopita2011}
M.~A. {Dopita}, L.~M. {Krauss}, R.~S. {Sutherland}, C.~{Kobayashi}, and C.~H.
  {Lineweaver}.
\newblock {Re-ionizing the universe without stars}.
\newblock {\em \apss}, page 242, July 2011.

\bibitem{trac2009}
H.~{Trac} and N.~Y. {Gnedin}.
\newblock {Computer Simulations of Cosmic Reionization}.
\newblock {\em ArXiv e-prints}, June 2009.

\bibitem{mcquinn2007}
M.~{McQuinn}, A.~{Lidz}, O.~{Zahn}, S.~{Dutta}, L.~{Hernquist}, and
  M.~{Zaldarriaga}.
\newblock {The morphology of HII regions during reionization}.
\newblock {\em \mnras}, 377:1043--1063, May 2007.

\bibitem{mesinger2010}
A.~{Mesinger}, S.~{Furlanetto}, and R.~{Cen}.
\newblock {21CMFAST: a fast, seminumerical simulation of the high-redshift
  21-cm signal}.
\newblock {\em \mnras}, 411:955--972, February 2011.

\bibitem{santos2010}
M.~G. {Santos}, L.~{Ferramacho}, M.~B. {Silva}, A.~{Amblard}, and A.~{Cooray}.
\newblock {Fast large volume simulations of the 21-cm signal from the
  reionization and pre-reionization epochs}.
\newblock {\em \mnras}, page 864, June 2010.

\bibitem{thomas2009}
R.~M. {Thomas}, S.~{Zaroubi}, B.~{Ciardi}, A.~H. {Pawlik}, P.~{Labropoulos},
  V.~{Jeli{\'c}}, G.~{Bernardi}, M.~A. {Brentjens}, A.~G. {de Bruyn}, G.~J.~A.
  {Harker}, L.~V.~E. {Koopmans}, G.~{Mellema}, V.~N. {Pandey}, J.~{Schaye}, and
  S.~{Yatawatta}.
\newblock {Fast large-scale reionization simulations}.
\newblock {\em \mnras}, 393:32--48, February 2009.

\bibitem{trac2007}
H.~{Trac} and R.~{Cen}.
\newblock {Radiative Transfer Simulations of Cosmic Reionization. I.
  Methodology and Initial Results}.
\newblock {\em \apj}, 671:1--13, December 2007.

\bibitem{partl2011}
A.~M. {Partl}, A.~{Maselli}, B.~{Ciardi}, A.~{Ferrara}, and V.~{M{\"u}ller}.
\newblock {Enabling parallel computing in CRASH}.
\newblock {\em \mnras}, 414:428--444, June 2011.

\bibitem{trac2008}
H.~{Trac}, R.~{Cen}, and A.~{Loeb}.
\newblock {Imprint of Inhomogeneous Hydrogen Reionization on the Temperature
  Distribution of the Intergalactic Medium}.
\newblock {\em \apjl}, 689:L81--L84, December 2008.

\bibitem{baek2009}
S.~{Baek}, P.~{Di Matteo}, B.~{Semelin}, F.~{Combes}, and Y.~{Revaz}.
\newblock {The simulated 21 cm signal during the epoch of reionization: full
  modeling of the Ly-{$\alpha$} pumping}.
\newblock {\em \aap}, 495:389--405, February 2009.

\bibitem{baek2010}
S.~{Baek}, B.~{Semelin}, P.~{Di Matteo}, Y.~{Revaz}, and F.~{Combes}.
\newblock {Reionization by UV or X-ray sources}.
\newblock {\em \aap}, 523:A4, November 2010.

\bibitem{lidz2008}
A.~{Lidz}, O.~{Zahn}, M.~{McQuinn}, M.~{Zaldarriaga}, and L.~{Hernquist}.
\newblock {Detecting the Rise and Fall of 21 cm Fluctuations with the Murchison
  Widefield Array}.
\newblock {\em \apj}, 680:962--974, June 2008.

\bibitem{mcquinn2006}
M.~{McQuinn}, O.~{Zahn}, M.~{Zaldarriaga}, L.~{Hernquist}, and S.~R.
  {Furlanetto}.
\newblock {Cosmological Parameter Estimation Using 21 cm Radiation from the
  Epoch of Reionization}.
\newblock {\em \apj}, 653:815--834, December 2006.

\bibitem{maldacena2003}
J.~{Maldacena}.
\newblock {Non-gaussian features of primordial fluctuations in single field
  inflationary models}.
\newblock {\em Journal of High Energy Physics}, 5:13, May 2003.

\bibitem{komatsu2009}
E.~{Komatsu}, J.~{Dunkley}, M.~R. {Nolta}, C.~L. {Bennett}, B.~{Gold},
  G.~{Hinshaw}, N.~{Jarosik}, D.~{Larson}, M.~{Limon}, L.~{Page}, D.~N.
  {Spergel}, M.~{Halpern}, R.~S. {Hill}, A.~{Kogut}, S.~S. {Meyer}, G.~S.
  {Tucker}, J.~L. {Weiland}, E.~{Wollack}, and E.~L. {Wright}.
\newblock {Five-Year Wilkinson Microwave Anisotropy Probe Observations:
  Cosmological Interpretation}.
\newblock {\em \apjs}, 180:330--376, February 2009.

\bibitem{cooray2006ng}
A.~{Cooray}.
\newblock {21-cm Background Anisotropies Can Discern Primordial
  Non-Gaussianity}.
\newblock {\em Physical Review Letters}, 97(26):261301, December 2006.

\bibitem{pillepich2007}
A.~{Pillepich}, C.~{Porciani}, and S.~{Matarrese}.
\newblock {The Bispectrum of Redshifted 21 Centimeter Fluctuations from the
  Dark Ages}.
\newblock {\em \apj}, 662:1--14, June 2007.

\bibitem{harker2009}
G.~J.~A. {Harker}, S.~{Zaroubi}, R.~M. {Thomas}, V.~{Jeli{\'c}},
  P.~{Labropoulos}, G.~{Mellema}, I.~T. {Iliev}, G.~{Bernardi}, M.~A.
  {Brentjens}, A.~G. {de Bruyn}, B.~{Ciardi}, L.~V.~E. {Koopmans}, V.~N.
  {Pandey}, A.~H. {Pawlik}, J.~{Schaye}, and S.~{Yatawatta}.
\newblock {Detection and extraction of signals from the epoch of reionization
  using higher-order one-point statistics}.
\newblock {\em \mnras}, 393:1449--1458, March 2009.

\bibitem{friedrich2011}
M.~M. {Friedrich}, G.~{Mellema}, M.~A. {Alvarez}, P.~R. {Shapiro}, and I.~T.
  {Iliev}.
\newblock {Topology and sizes of H II regions during cosmic reionization}.
\newblock {\em \mnras}, 413:1353--1372, May 2011.

\bibitem{lee2011}
K.-G. {Lee} and D.~N. {Spergel}.
\newblock {Threshold Probability Functions and Thermal Inhomogeneities in the
  Ly{$\alpha$} Forest}.
\newblock {\em \apj}, 734:21, June 2011.

\bibitem{joudaki2011}
S.~{Joudaki}, O.~{Dor{\'e}}, L.~{Ferramacho}, M.~{Kaplinghat}, and M.~G.
  {Santos}.
\newblock {Primordial Non-Gaussianity from the 21 cm Power Spectrum during the
  Epoch of Reionization}.
\newblock {\em Physical Review Letters}, 107(13):131304, September 2011.

\bibitem{mao2008}
Y.~{Mao}, M.~{Tegmark}, M.~{McQuinn}, M.~{Zaldarriaga}, and O.~{Zahn}.
\newblock {How accurately can 21cm tomography constrain cosmology?}
\newblock {\em \prd}, 78(2):023529, July 2008.

\bibitem{barger2009}
V.~{Barger}, Y.~{Gao}, Y.~{Mao}, and D.~{Marfatia}.
\newblock {Inflationary potential from 21 cm tomography and Planck}.
\newblock {\em Physics Letters B}, 673:173--178, March 2009.

\bibitem{adshead2011}
P.~{Adshead}, R.~{Easther}, J.~{Pritchard}, and A.~{Loeb}.
\newblock {Inflation and the scale dependent spectral index: prospects and
  strategies}.
\newblock {\em \jcap}, 2:21, February 2011.

\bibitem{pritchard2008nu}
J.~R. {Pritchard} and E.~{Pierpaoli}.
\newblock {Constraining massive neutrinos using cosmological 21cm
  observations}.
\newblock {\em \prd}, 78(6):065009, September 2008.

\bibitem{gordon2009}
C.~{Gordon} and J.~R. {Pritchard}.
\newblock {Forecasted 21cm constraints on compensated isocurvature
  perturbations}.
\newblock {\em \prd}, 80(6):063535, September 2009.

\bibitem{jester2009}
S.~{Jester} and H.~{Falcke}.
\newblock {Science with a lunar low-frequency array: From the dark ages of the
  Universe to nearby exoplanets}.
\newblock {\em \nar}, 53:1--26, May 2009.

\bibitem{khatri2007}
R.~{Khatri} and B.~D. {Wandelt}.
\newblock {21-cm Radiation: A New Probe of Variation in the Fine-Structure
  Constant}.
\newblock {\em Physical Review Letters}, 98(11):111301, March 2007.

\bibitem{khatri2008}
R.~{Khatri} and B.~D. {Wandelt}.
\newblock {Cosmic (Super)String Constraints from 21cm Radiation}.
\newblock {\em Physical Review Letters}, 100(9):091302, March 2008.

\bibitem{loeb2008im}
A.~{Loeb} and J.~S.~B. {Wyithe}.
\newblock {Possibility of Precise Measurement of the Cosmological Power
  Spectrum with a Dedicated Survey of 21cm Emission after Reionization}.
\newblock {\em Physical Review Letters}, 100(16):161301, April 2008.

\bibitem{wyithe2008}
J.~S.~B. {Wyithe} and A.~{Loeb}.
\newblock {Fluctuations in 21-cm emission after reionization}.
\newblock {\em \mnras}, 383:606--614, January 2008.

\bibitem{chang2008}
{T.-C.} {Chang}, {U.-L.} {Pen}, J.~B. {Peterson}, and P.~{McDonald}.
\newblock {Baryon Acoustic Oscillation Intensity Mapping of Dark Energy}.
\newblock {\em Physical Review Letters}, 100(9):091303, March 2008.

\bibitem{battye2004}
R.~A. {Battye}, R.~D. {Davies}, and J.~{Weller}.
\newblock {Neutral hydrogen surveys for high-redshift galaxy clusters and
  protoclusters}.
\newblock {\em \mnras}, 355:1339--1347, December 2004.

\bibitem{visbal2009}
E.~{Visbal}, A.~{Loeb}, and S.~{Wyithe}.
\newblock {Cosmological constraints from 21cm surveys after reionization}.
\newblock {\em \jcap}, 10:30, October 2009.

\bibitem{albrecht2006}
A.~{Albrecht}, G.~{Bernstein}, R.~{Cahn}, W.~L. {Freedman}, J.~{Hewitt},
  W.~{Hu}, J.~{Huth}, M.~{Kamionkowski}, E.~W. {Kolb}, L.~{Knox}, J.~C.
  {Mather}, S.~{Staggs}, and N.~B. {Suntzeff}.
\newblock {Report of the Dark Energy Task Force}.
\newblock {\em ArXiv Astrophysics e-prints}, September 2006.

\bibitem{wyithe2008BAO}
J.~S.~B. {Wyithe}, A.~{Loeb}, and P.~M. {Geil}.
\newblock {Baryonic acoustic oscillations in 21-cm emission: a probe of dark
  energy out to high redshifts}.
\newblock {\em \mnras}, 383:1195--1209, January 2008.

\bibitem{wyithe2009}
J.~S.~B. {Wyithe} and A.~{Loeb}.
\newblock {The 21-cm power spectrum after reionization}.
\newblock {\em \mnras}, 397:1926--1934, August 2009.

\bibitem{chang2010}
{T.-C.} {Chang}, {U.-L.} {Pen}, K.~{Bandura}, and J.~B. {Peterson}.
\newblock {An intensity map of hydrogen 21-cm emission at redshift z\~{}0.8}.
\newblock {\em \nat}, 466:463--465, July 2010.

\bibitem{barkana2000}
R.~{Barkana} and A.~{Loeb}.
\newblock {Identifying the Reionization Redshift from the Cosmic Star Formation
  Rate}.
\newblock {\em \apj}, 539:20--25, August 2000.

\bibitem{righi2008}
M.~{Righi}, C.~{Hern{\'a}ndez-Monteagudo}, and R.~A. {Sunyaev}.
\newblock {Carbon monoxide line emission as a CMB foreground: tomography of the
  star-forming universe with different spectral resolutions}.
\newblock {\em \aap}, 489:489--504, October 2008.

\bibitem{visbal2011}
E.~{Visbal}, H.~{Trac}, and A.~{Loeb}.
\newblock {Demonstrating the feasibility of line intensity mapping using mock
  data of galaxy clustering from simulations}.
\newblock {\em \jcap}, 8:10, August 2011.

\bibitem{visbal2010}
E.~{Visbal} and A.~{Loeb}.
\newblock {Measuring the 3D clustering of undetected galaxies through cross
  correlation of their cumulative flux fluctuations from multiple spectral
  lines}.
\newblock {\em \jcap}, 11:16, November 2010.

\bibitem{wang2006}
X.~{Wang}, M.~{Tegmark}, M.~G. {Santos}, and L.~{Knox}.
\newblock {21 cm Tomography with Foregrounds}.
\newblock {\em \apj}, 650:529--537, October 2006.

\bibitem{liu2011}
A.~{Liu} and M.~{Tegmark}.
\newblock {A method for 21 cm power spectrum estimation in the presence of
  foregrounds}.
\newblock {\em \prd}, 83(10):103006, May 2011.

\bibitem{carilli2011}
C.~L. {Carilli}.
\newblock {Intensity Mapping of Molecular Gas During Cosmic Reionization}.
\newblock {\em \apjl}, 730:L30, April 2011.

\bibitem{lidz2011}
A.~{Lidz}, S.~R. {Furlanetto}, S.~P. {Oh}, J.~{Aguirre}, T.-C. {Chang},
  O.~{Dor{\'e}}, and J.~R. {Pritchard}.
\newblock {Intensity Mapping with Carbon Monoxide Emission Lines and the
  Redshifted 21 cm Line}.
\newblock {\em \apj}, 741:70, November 2011.

\bibitem{gong2011}
Y.~{Gong}, A.~{Cooray}, M.~B. {Silva}, M.~G. {Santos}, and P.~{Lubin}.
\newblock {Probing Reionization with Intensity Mapping of Molecular and
  Fine-structure Lines}.
\newblock {\em \apjl}, 728:L46, February 2011.

\bibitem{obreschkow2009}
D.~{Obreschkow}, I.~{Heywood}, H.-R. {Kl{\"o}ckner}, and S.~{Rawlings}.
\newblock {A Heuristic Prediction of the Cosmic Evolution of the Co-luminosity
  Functions}.
\newblock {\em \apj}, 702:1321--1335, September 2009.

\bibitem{shetty2011}
R.~{Shetty}, S.~C. {Glover}, C.~P. {Dullemond}, and R.~S. {Klessen}.
\newblock {Modelling CO emission - I. CO as a column density tracer and the X
  factor in molecular clouds}.
\newblock {\em \mnras}, 412:1686--1700, April 2011.

\bibitem{sternberg2011}
A.~{Sternberg}, A.~{Dalgarno}, E.~{Herbst}, and Y.~{Pei}.
\newblock {Molecular Clouds at the Reionization Epoch}.
\newblock In {\em IAU Symposium}, volume 280 of {\em IAU Symposium}, May 2011.

\bibitem{field1959obs}
G.~B. {Field}.
\newblock {An Attempt to Observe Neutral Hydrogen Between the Galaxies.}
\newblock {\em \apj}, 129:525, May 1959.

\bibitem{carilli2002}
C.~L. {Carilli}, N.~Y. {Gnedin}, and F.~{Owen}.
\newblock {H I 21 Centimeter Absorption beyond the Epoch of Reionization}.
\newblock {\em \apj}, 577:22--30, September 2002.

\bibitem{furlanetto2002}
S.~R. {Furlanetto} and A.~{Loeb}.
\newblock {The 21 Centimeter Forest: Radio Absorption Spectra as Probes of
  Minihalos before Reionization}.
\newblock {\em \apj}, 579:1--9, November 2002.

\bibitem{mack2011}
K.~J. {Mack} and J.~S.~B. {Wyithe}.
\newblock {Detecting the redshifted 21cm forest during reionization}.
\newblock {\em ArXiv e-prints}, January 2011.

\bibitem{xu2009}
Y.~{Xu}, X.~{Chen}, Z.~{Fan}, H.~{Trac}, and R.~{Cen}.
\newblock {The 21 cm Forest as a Probe of the Reionization and The Temperature
  of the Intergalactic Medium}.
\newblock {\em \apj}, 704:1396--1404, October 2009.

\bibitem{xu2011}
Y.~{Xu}, A.~{Ferrara}, and X.~{Chen}.
\newblock {The earliest galaxies seen in 21 cm line absorption}.
\newblock {\em \mnras}, 410:2025--2042, January 2011.

\bibitem{furlanetto2006forest}
S.~R. {Furlanetto}.
\newblock {The 21-cm forest}.
\newblock {\em \mnras}, 370:1867--1875, August 2006.

\bibitem{xu2010}
Y.~{Xu}, A.~{Ferrara}, F.~S. {Kitaura}, and X.~{Chen}.
\newblock {Searching for the earliest galaxies in the 21 cm forest}.
\newblock {\em Science in China G: Physics and Astronomy}, 53:1124--1129, June
  2010.

\bibitem{meiksin2011}
A.~{Meiksin}.
\newblock {The micro-structure of the intergalactic medium - I. The 21 cm
  signature from dynamical minihaloes}.
\newblock {\em \mnras}, 417:1480--1509, October 2011.

\bibitem{mortlock2011}
D.~J. {Mortlock}, S.~J. {Warren}, B.~P. {Venemans}, M.~{Patel}, P.~C. {Hewett},
  R.~G. {McMahon}, C.~{Simpson}, T.~{Theuns}, E.~A. {Gonz{\'a}les-Solares},
  A.~{Adamson}, S.~{Dye}, N.~C. {Hambly}, P.~{Hirst}, M.~J. {Irwin},
  E.~{Kuiper}, A.~{Lawrence}, and H.~J.~A. {R{\"o}ttgering}.
\newblock {A luminous quasar at a redshift of z = 7.085}.
\newblock {\em \nat}, 474:616--619, June 2011.

\bibitem{hopkins2007}
P.~F. {Hopkins}, G.~T. {Richards}, and L.~{Hernquist}.
\newblock {An Observational Determination of the Bolometric Quasar Luminosity
  Function}.
\newblock {\em \apj}, 654:731--753, January 2007.

\bibitem{jiang2007}
L.~{Jiang}, X.~{Fan}, {\v Z}.~{Ivezi{\'c}}, G.~T. {Richards}, D.~P.
  {Schneider}, M.~A. {Strauss}, and B.~C. {Kelly}.
\newblock {The Radio-Loud Fraction of Quasars is a Strong Function of Redshift
  and Optical Luminosity}.
\newblock {\em \apj}, 656:680--690, February 2007.

\bibitem{haiman2004}
Z.~{Haiman}, E.~{Quataert}, and G.~C. {Bower}.
\newblock {Modeling the Counts of Faint Radio-Loud Quasars: Constraints on the
  Supermassive Black Hole Population and Predictions for High Redshift}.
\newblock {\em \apj}, 612:698--705, September 2004.

\bibitem{willott2010}
C.~J. {Willott}, P.~{Delorme}, C.~{Reyl{\'e}}, L.~{Albert}, J.~{Bergeron},
  D.~{Crampton}, X.~{Delfosse}, T.~{Forveille}, J.~B. {Hutchings}, R.~J.
  {McLure}, A.~{Omont}, and D.~{Schade}.
\newblock {The Canada-France High-z Quasar Survey: Nine New Quasars and the
  Luminosity Function at Redshift 6}.
\newblock {\em \aj}, 139:906--918, March 2010.

\bibitem{ioka2005}
K.~{Ioka} and P.~{M{\'e}sz{\'a}ros}.
\newblock {Radio Afterglows of Gamma-Ray Bursts and Hypernovae at High Redshift
  and Their Potential for 21 Centimeter Absorption Studies}.
\newblock {\em \apj}, 619:684--696, February 2005.

\bibitem{souza2011}
R.~S. {de Souza}, N.~{Yoshida}, and K.~{Ioka}.
\newblock {Populations III.1 and III.2 gamma-ray bursts: constraints on the
  event rate for future radio and X-ray surveys}.
\newblock {\em \aap}, 533:A32, September 2011.

\bibitem{campisi2011}
M.~A. {Campisi}, U.~{Maio}, R.~{Salvaterra}, and B.~{Ciardi}.
\newblock {Population III stars and the long gamma-ray burst rate}.
\newblock {\em \mnras}, 416:2760--2767, October 2011.

\bibitem{pritchard2010combo}
J.~R. {Pritchard}, A.~{Loeb}, and J.~S.~B. {Wyithe}.
\newblock {Constraining reionization using 21-cm observations in combination
  with CMB and Ly{$\alpha$} forest data}.
\newblock {\em \mnras}, 408:57--70, October 2010.

\bibitem{mitra2011}
S.~{Mitra}, T.~R. {Choudhury}, and A.~{Ferrara}.
\newblock {Reionization constraints using principal component analysis}.
\newblock {\em \mnras}, 413:1569--1580, May 2011.

\end{thebibliography}

\end{document}